\documentclass[fleqn,usenatbib]{mnras}

\usepackage{newtxtext,newtxmath}
\usepackage{soul}
\usepackage[T1]{fontenc}
\usepackage{multirow}
\usepackage[table]{xcolor}

\DeclareRobustCommand{\VAN}[3]{#2}
\let\VANthebibliography\thebibliography
\def\thebibliography{\DeclareRobustCommand{\VAN}[3]{##3}\VANthebibliography}

\usepackage{graphicx}	% Including figure files
\usepackage{amsmath}	% Advanced maths commands
\usepackage{float}
\title[Vela individual pulses study]{Vela Pulsar: Single Pulses Analysis with Machine Learning Techniques}

\author[Lousto et al.]{Carlos O. Lousto,$^{1,3}$\thanks{E-mail: colsma@rit.edu (COL)}
Ryan Missel$^{2}$, Harsh Prajapati$^{3}$, Valentina Sosa Fiscella$^{1,4}$, Federico G. L\'opez Armengol$^{1}$,
\newauthor{Prashnna Kumar Gyawali$^{2}$, Linwei Wang$^{2}$, Nathan Cahill$^{3}$, % RIT+Cameron Knight? Carolina Negrelli?
Luciano Combi$^{1,4}$, Santiago del Palacio$^{4}$,}
\newauthor{Jorge A. Combi$^{4,5}$, Guillermo Gancio$^{4}$, Federico Garc\'{i}a$^{4,6}$, Eduardo M. Guti\'errez$^{4}$, Fernando Hauscarriaga$^{4}$}%IAR+Eduardo Gutierrez? G. Mancuso? A. Simaz Bunzel?
\\
$^{1}$Center for Computational Relativity and Gravitation, Rochester Institute of Technology, 85 Lomb Memorial Drive, Rochester,New York 14623, USA\\
$^{2}$Golisano College of Computing and Information Sciences Rochester Institute of Technology Rochester, NY 14623, USA\\
$^{3}$School of Mathematical Sciences, Sciences Rochester Institute of Technology Rochester, NY 14623, USA\\
$^{4}$Instituto Argentino de Radioastronom\'ia (CCT La Plata, CONICET; CICPBA; UNLP), C.C.5, (1894) Villa Elisa, Buenos Aires, Argentina.\\
$^{5}$Facultad de Ciencias Astron\'omicas y Geof\'{\i}sicas, Universidad Nacional de La Plata, Paseo del Bosque, B1900FWA La Plata, Argentina\\
$^{6}$Kapteyn Astronomical Institute, University of Groningen, P.O. BOX 800, 9700 AV Groningen, The Netherlands\\
}

\date{Accepted XXX. Received YYY; in original form ZZZ}

\pubyear{2021}

\graphicspath{{./}{figures/}}

\begin{document}
\label{firstpage}
\pagerange{\pageref{firstpage}--\pageref{lastpage}}
\maketitle

% Abstract of the paper
\begin{abstract}
%The abstract should briefly describe the aims, methods, and main results of the paper.
%It should be a single paragraph not more than 250 words (200 words for Letters).
%No references should appear in the abstract.
We study individual pulses of Vela (PSR\ B0833-45\,/\,J0835-4510) from daily observations of over three hours (around 120,000 pulses per observation), performed simultaneously with the two radio telescopes at the Argentine Institute of Radioastronomy. We select 4 days of observations in January-March 2021 and study their statistical properties with machine learning techniques. We first use density based DBSCAN clustering techniques, associating pulses mainly by amplitudes, and find a correlation between higher amplitudes and earlier arrival times. We also find a weaker (polarization dependent) correlation with the mean width of the pulses. We identify clusters of the so-called mini-giant pulses, with $\sim10\times$ the average pulse amplitude. We then perform an independent study, with Self-Organizing Maps (SOM) clustering techniques. We use Variational AutoEncoder (VAE) reconstruction of the pulses to separate them clearly from the noise and select one of the days of observation to train VAE and apply it to thre rest of the observations. We use SOM to determine 4 clusters of pulses per day per radio telescope and conclude that our main results are robust and self-consistent. These results support models for emitting regions at different heights (separated each by roughly a hundred km) in the pulsar magnetosphere.
We also model the pulses amplitude distribution with interstellar scintillation patterns at the inter-pulses time-scale finding a characterizing exponent $n_{\mathrm{ISS}}\sim7-10$.
In the appendices we discuss independent checks of hardware systematics with the simultaneous use of the two radio telescopes in different one-polarization / two-polarizations configurations. We also provide a detailed analysis of the processes of radio-interferences cleaning and individual pulse folding.
\end{abstract}

% Select between one and six entries from the list of approved keywords.
% Don't make up new ones.
\begin{keywords}
pulsars: Vela -- methods: observational -- methods: statistical
%\Carlos{I didn't find anything better in:https://static.primary.prod.gcms.the-infra.com/static/site/mnras/document/Updated_keyword_list_Jan_2020_.pdf?node=21e8735032e6494cce9b&version=78171:e4aca8bedd490069ed08}
\end{keywords}

%%%%%%%%%%%%%%%%%%%%%%%%%%%%%%%%%%%%%%%%%%%%%%%%%%

%%%%%%%%%%%%%%%%% BODY OF PAPER %%%%%%%%%%%%%%%%%%

\section{Introduction}\label{sec:intro}

In 1967, a rapid pulsating radio-source was discovered in the sky \citep{Hewish+1968}. 
This \emph{pulsar} was soon proved to be a rapidly rotating and higly magnetized neutron star (NS) \citep{Pacini1967, Gold1968, Goldreich+1969, Radhakrishnan+1969a, Radhakrishnan+1969b, Sturrock1971}. 
In the standard model, pulsars have an external magnetic field with a strong dipole moment and, around the magnetic poles of the star,  charged particles accelerate and emit a collimated beam of light. 
The misalignment of the magnetic moment with the rotation axis of the star generates 
a \textit{lighthouse} effect that allows us to detect such emission only when the magnetic moment coincides with our line of sight, once every rotational period.
Since the first discovery, more than a thousand pulsars have been found \citep[see][for a general review]{Lorimer2008}.

The radio emission from pulsars is rather weak, varying from a few $\mu \mathrm{Jy}$ to $1\,\mathrm{Jy}$, at $1.4\,\mathrm{GHz}$.
For this reason, to obtain a clear profile over the background noise, the observation of pulsars proceeds by integrating over thousands of pulses, with a technique usually called \textit{folding}. 
This technique is feasible because of the remarkable regularity of the pulses over time.
The detection of each individual pulse, however, is achievable for very bright pulsars, and the variability of these single-pulses reveal details of the structure and emission processes in these objects.
Particularly, the pulsar of Vela (PSR B0833$-$45\,/\,J0835$-$4510), discovered over 50~years ago \citep{1968Natur.220..340L}, has a flux of $S_{1400} = 1.1 \mathrm{Jy}$ and has been the focus of many single-pulses studies. The pioneer work of \cite{Downs1983} discovered that stronger single-pulses arrive earlier if compared with the mean pulse. Further, to explain deviations of the observed polarization from the dipole model, they proposed that single-pulses are composed of four independent components that originate at  different heights in the magnetosphere. \cite{Johnston+2001} and \cite{Kramer:2002us} increased the resolution in time and focused in the variability of the micro-structure, arguing that single-pulses are composed of multiple sweeping beams of light. They also claimed the existence of sporadic giant micro-pulses and a bump in the leading and trailing parts of the profile, respectively, that present a different energy distribution than the rest of the pulse, suggesting a different origin \citep[see also][]{Cairns+2001}. %\FedeLA{Still to review:} \cite{Palfreyman2011, Palfreyman+2016}.

At the same time, the pulsar of Vela is particularly interesting because, being young ($\sim 10^4\,\mathrm{yr}$), it is prone to have regular \emph{glitches}, 
sudden speeds up in its rotation frequency $\nu$, with $\Delta \nu / \nu \propto 10^{-9}$, approximately once every three years \citep{Radhakrishnan+1969c, Reichley+1969, Dodson+2007}.
The Vela PSR also manifests several \emph{micro-glitches} over a year \citep{Cordes+1988}.
These events are thought to be produced by the sudden coupling of a fast rotating superfluid core with the crust of the pulsar, but the details of such coupling have been challenged \citep{Andersson:2012iu,Chamel:2012ae,Piekarewicz:2014lba} and are a matter of current research \citep[see][for a review on models of pulsar glitches]{Haskell:2015jra}.
The analysis of single-pulses can bring further insight into the details of the physical mechanism behind these glitches.
Remarkably, in 2016, Vela was observed to glitch \emph{live}  \citep{2018Natur.556..219P}. 
This observation showed that the glitch process took under five seconds, and that the pulsar did not pulse for one period, with the prior pulse to the null being very broad and the two following pulses featuring low linear polarization. %\Carlos{Updated info here with new glitch:}
The latest glitch (so far) had occurred in 2019, around MJD~58515, and was reported by \cite{atel_newglitch_vela}. 
The radio timing observations performed at the Argentine Institute of Radioastronomy (IAR) of this event were first summarized in \cite{atel_vela} and then expanded in \cite{Gancio2020}.
A new glitch has just occurred on July 21 2021 (MJD 59417.6) that was first reported in \cite{2021ATel14806....1S}.

Given the high rotational frequency of Vela PSR ($\sim 11\,\mathrm{Hz}$), an observation lasting a few hours involves a large amount of single-pulses ($\sim 10^5$).
Even more, a daily observational campaign of these single-pulses represents a challenge to our analysis techniques.
The last five years have witnessed a notable increase of the machine learning techniques in the analysis of astrophysical data. %Notably because of the large amount of data generated by new observations. 
In particular, for pulsar data analysis and statistical methods, their main application has been on search and detection of new pulsars in surveys \citep{2020MNRAS.493.1842L,2019MNRAS.490.5424G,2019SCPMA..6259507W,2019MNRAS.483.3673M,2016MNRAS.459.1519D,2014MNRAS.443.1651M,2014ApJ...781..117Z}. The techniques are also applied to fast radio bursts (FRBs) searches \citep{2020A&A...642A..26Z}
and specialize in extracting pulsars over radio-frequency interference (RFI)/noise \citep{,2020MNRAS.497.1661A,2018ApJ...866..149Z,2018A&C....23...15B}. 
Also interesting here are the techniques that deal with single pulses \citep{2018MNRAS.480.3457M,2018MNRAS.480.3302P}.
The idea of classifying those individual pulses in terms of amplitude
and width, to a first approach, and then in terms of other particular features using machine learning,
has already been expressed briefly at the end of Sections 4.3 (in relation
to magnetars) and 4.5 (in relation to FRBs) of \cite{Gancio2020}. Once we have a robust classification and characterization of families of single pulses, we can study their statistical distributions over different time scales, from a single day, through weeks, or up to years and possibly assess whether there is a connection between the single pulses distribution and the glitching activity.

%The observation of pulsars proceeds by averaging over thousands of pulsed to obtain a clear pulse over the background noise (of both instrumental and astrophysical origin, the later intrinsic to the pulsar and due to interstellar media).
%The single pulse observations require very bright pulsars (such as Vela in the south or Crab in the northern hemisphere). In particular for Vela, a young glitching pulsar, we expect variations in shape from pulse to pulse. 
%High-resolution single-pulse studies of the Vela pulsar have been performed in \citep{Kramer:2002us}.
%The idea of classifying those individual pulses in terms of first, amplitude,
%and width, and then other particular features using machine learning algorithms
%has already been expressed briefly at the end of sections 4.3 (in relation
%to magnetars) and 4.5 (in relation to FRB) of our paper \cite{Gancio2020}. Once we have a robust classification and characterization of families of single pulses we can study its statistical distributions over a single day (or subsets of) observation and between different days up to years, i.e. at all possible temporal scales.

%\FedeLA{Consider to remove the following paragraph:}
%It was also raised the question of using single pulse (possibly of a high S/N pulsar subset) for timing (ToAs) to improve or assess this as an alternative or additional method, and apply the techniques here developed to J0437-4715 \citep{Kerr:2015dla}. This will be developed as a parallel project by the authors and compared to the standard timing process as in Ref.~\citep{Fiscella:2020jey}.

The use of machine learning algorithms is well suited for the classification of single pulses in at least two ways: Clustering and Supervised learning. 
In Sec.~\ref{sec:math} we group pulses with similar features by applying the technique of Density-Based Spatial Clustering of Applications with Noise (DBSCAN) \citep{DBSCAN}, using the criterium of density in data space, setting a threshold and a number of expected clusters, involving an iterative optimization process.
% In Sec.~\ref{sec:CS} we treat first the data with a variational autoencoder (VAE) reconstruction method \citep{kingma2014autoencoding} with a training algorithm \citep{2006Sci...313..504H} based on the observation with the best signal-to-noise rate (S/N), and then perform the self-organizing maps (SOM) clustering algorithm \citep{teuvo1988som}. 
% \FedeLA{Rewritting last sentence, please check the idea holds:} 
%{Prashnna: Removed the citation for \citep{2006Sci...313..504H}. I don't think its appropriate for VAE, or its training setup.}
In Sec.~\ref{sec:CS} we train a Variational AutoEncoder (VAE) method \citep{kingma2014autoencoding} to reconstruct the observation of the highest signal-to-noise ratio (S/N), and then apply the trained method to the rest of our data. Based on such reconstructions, we apply an algorithm of Self-Organizing Maps (SOM) clustering \citep{teuvo1988som}. In Sec.~\ref{sec:scintillations} we model the individual pulse amplitude distribution assuming it is due entirely to insterstellar scintillation, while in Sec.~\ref{sec:magneto} we model the SOM clusters in terms of magnetosphere strata. We close the paper with a discussion in Sec.~\ref{sec:conclusions} of future applications of this technique to other bright pulsars that allow single pulse analysis and in particular to Vela close to one of its large glitches.

%\FedeLA{TO DO: Review the rest of the Sections, in order.}\Carlos{Kind of completed by Carlos. Check}

%Since our data naturally separates in days (and Radio telescope) of observation, we can use one of these subsets of data to perform a supervised learning and obtain trained data that in turn can be used to 
%potentially determine unforeseen classes in the other days data. The statistical quality of this process can be evaluated via a generalization error (Empirical risk) \citep{}.

\section{Observations}\label{sec:Observations}

The main observational program of the Argentine Institute of Radioastronomy consists on the follow up of a set of targeted pulsars, with daily observations of up $\sim 3.5$ hours.
In 2019, the radio telescope Dr. Carlos Varsavsky (A1) was updated to match the same parameters and components as the second radio telescope, Dr. Esteban Bajaja (A2). In this direction, the low noise amplifiers, radio frequency filters, among other components where upgraded. As a result, now both radio telescopes have almost identical front end's and the second polarization of A1 has been recovered, in contrast with the receiver presented in \citep{Gancio2020}.
In the case of the very bright pulsar of Vela, IAR's antennas sensitivity allows the detection of single pulses with a time resolution of 73~$\mu$s (from a 56~MHz sampling averaged over $N_\mathrm{av}=4096$ cycles).
The observations used in this paper have been performed simultaneously with A1 and A2. A1 has been set in a single polarization mode that allows for a total 112~MHz bandwidth, while A2 adds its two (circular)  polarizations, each with a 56~MHz bandwidth. Both receivers have been chosen to be centered at 1400~MHz and share the same band pass filter (ZX75BP-1280-S+) \citep[see Fig.~3 in][]{Gancio2020}. 
These different settings have been chosen to see the effects of bandwidth and total power on our analysis as well as the different (local) RFI content of each radio telescope due to their different location (150 meters apart) in the IAR campus.

We analize six observations on January, 21th, 24th and 28th, 2021, performed with both radio telescopes for over three hours. An additional set of observations was performed on March 29th, 2021, in which we changed the A1 configuration from a single polarization at 112~MHz bandwidth, into the sum of two polarizations of 56~MHz bandwidth each, matching A2 configuration, which remained unchanged.
The number of single-pulses in each observation is given in Table~\ref{tab:observations}.

\begin{table}
	\centering
	\caption{Date of each observation, the corresponding number of single pulses, instantaneous period at the beginning of the observation, and the MJD at the beginning and at the end of each.}
	\label{tab:observations}
	\begin{tabular}{l c c l l} % four columns, alignment for each
		\hline
		Day 2021& initial MJD & final MJD & \# pulses & $P_{\mathrm{inst}}$ [ms]\\
		\hline
		Jan. 21 & 59235.128553 & 59235.254277 & 121495 & 89.407366\\
		Jan. 24 & 59239.117013 & 59239.240620 & 119448 & 89.407676\\
		Jan. 28 & 59242.088680 & 59242.222170 & 128999 & 89.407915\\
		Mar. 29 & 59302.943356 & 59303.076850 & 128999 & 89.413017\\
		\hline
	\end{tabular}
\end{table}

All observations have been cleaned from radio frequency interferences using the code RFIClean \citep{Maan2020} with protection of the fundamental frequency of Vela (11.184~Hz) at each of the days of observations, 
as described in the appendix \ref{sec:RFIClean}.
%\Carlos{@Valentina: Include the specific values of the rotational frequency for each day in the table \ref{tab:observations} if they differ significantly, otherwise give just the common significant number in this text} \Valentina{Done!} 

\section{Analysis Methods}\label{sec:analysis}

In this section we report the analysis of the observations pulse by pulse via two independent approaches followed by independent teams in a study to compare and interpret results in an unbiased fashion. The large amount of
data (120,000 -- 130,000 per observation per antenna, totalizing nearly a million pulses) is well suited for statistical and machine learning studies. Our first approach has been carried out with DBSCAN, focusing on the peak amplitude, width, and location (time of arrival) of the strongest single-pulses. The second approach has been carried out using a combination of the Variational AutoEncoder reconstruction (VAE) and the self-organizing maps clustering (SOM) techniques. This analysis focuses on a global clustering of the pulses into four main components for which we obtain the mean pulses and its statistical properties. This analysis allow us to identify mini-giant pulses, interstellar scintillation at sub-seconds time-scales,
and to model the different emission regions in the pulsar magnetosphere.

In the following subsections we will describe the results of the two approaches for each of the observations in the three days of January 2021, and present March 29th cross-check observations in an appendix \ref{sec:2021-03-29}. 

\subsection{Density based (DBSCAN) techniques and results}\label{sec:math}

%\Carlos{Blurb description of DBSCAN techniques: Harsh and Nathan, please check!}
%\Carlos{Description Results; Reconstruction pulses. Clustering. Correlations Width-Peak-Amplitude-Skew}

We start by characterizing the individual pulses by extracting three basic properties from them: the peak amplitude, the peak position and the width of the peak in time. While we have 1220 bins in time, we focus on a relatively small window (less than 100 bins) where we expect the pulse to arrive, based on the total average pulse. Since the peak in the data looks jagged, we smooth it doing a least squares second order polynomial fit taking five points around the maximum, what provides us with the peak amplitude and peak position. The full width is computed when the pulse arrives to half the maximum amplitude. We also experimented with denoising techniques, but found the above least squares fit was providing more robust results.

\begin{figure}\begin{center}
	\includegraphics[width=0.9\columnwidth]{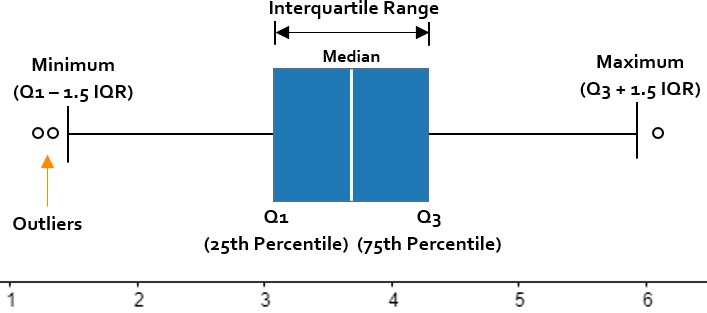}
    \caption{Explanation of the representations in the box plots below.}
    \label{fig:box}
\end{center}
\end{figure}

We first provide box-plots for the pulses amplitude distributions versus its peak location and versus its width for each of the 8 observations. This follows the nomenclature described in Fig.~\ref{fig:box} for its median and 50\% around the maximum in blue blocks. Bars in black cover three times this blocks maximum around the 50\% width and finally those points lying outside the bars, denoted by circles, are labeled as outliers.
The linear box-plots are supplemented by insets with the (log of the) number of pulses in each amplitude slice.

Now that we have extracted and analyzed those three properties of the individual pulses, we feed those as features
to the clustering algorithm. As such, each pulse is represented as a three dimensional vector and we can thus
classify those points in a 3-dimensional space based on similarities indices.
Here we will focus on the Density-Based Spatial Clustering of Applications with Noise (DBSCAN) clustering method \cite{DBSCAN,ester1996densitybased} because this algorithm (compared to the classic K-Means) is better at identifying
higher density regions and is better at identifying arbitrarily shaped clusters rather than being biased towards spherical clusters.
This clustering analysis is an unsupervised learning method that separates the data points into several specific groups, 
such that the data points in the same group are close in their similarity indices,
and data points in different groups have different distance properties.
DBSCAN can discover clusters of different shapes and sizes from a large amount of data containing noise and outliers.
DBSCAN does not require to specify the number of clusters to use it, but a function is required to calculate the distance 
between data points and some criteria for the relative scale of distance in that space.
For this, we use a default value of 0.6 for the maximum distance between two samples. This is the most important DBSCAN parameter to choose appropriately. Additionally, a minimum number of samples is required for a point to be considered in a neighborhood. Here, we choose 3, including the core point itself, where a core point is one that has at least the minimum number of points required at the maximum distance.
% We have also to choose the minimum number of samples (here 3, including the core point itself) for a point to be considered in a neighborhood. (Core points are those that have at least a minimun number of points at the maximum distance). 
Finally, we will consider an Euclidean metric for the distance measures.

DBSCAN has limitations in its performance when inappropriate criterion values and measurements are used, meaning either parameter searches or prior domain knowledge is required to get optimal results.
% DBSCAN has some limitations when an appropriate distance of neighborhood value is not passed and would require further domain knowledge. It also cannot generalize well to clusters with much different densities given the fixed initial combination of maximum distance-minimum number of points.
Another limitation is in its ability to generalize to clusters containing highly different densities from the initial ratio of maximum distance to number of points. OPTICS \cite{optics99} addresses this issue providing a similar clustering at multiple values of the maximum distance with optimized memory usage.

\begin{figure*}
	\includegraphics[width=\linewidth]{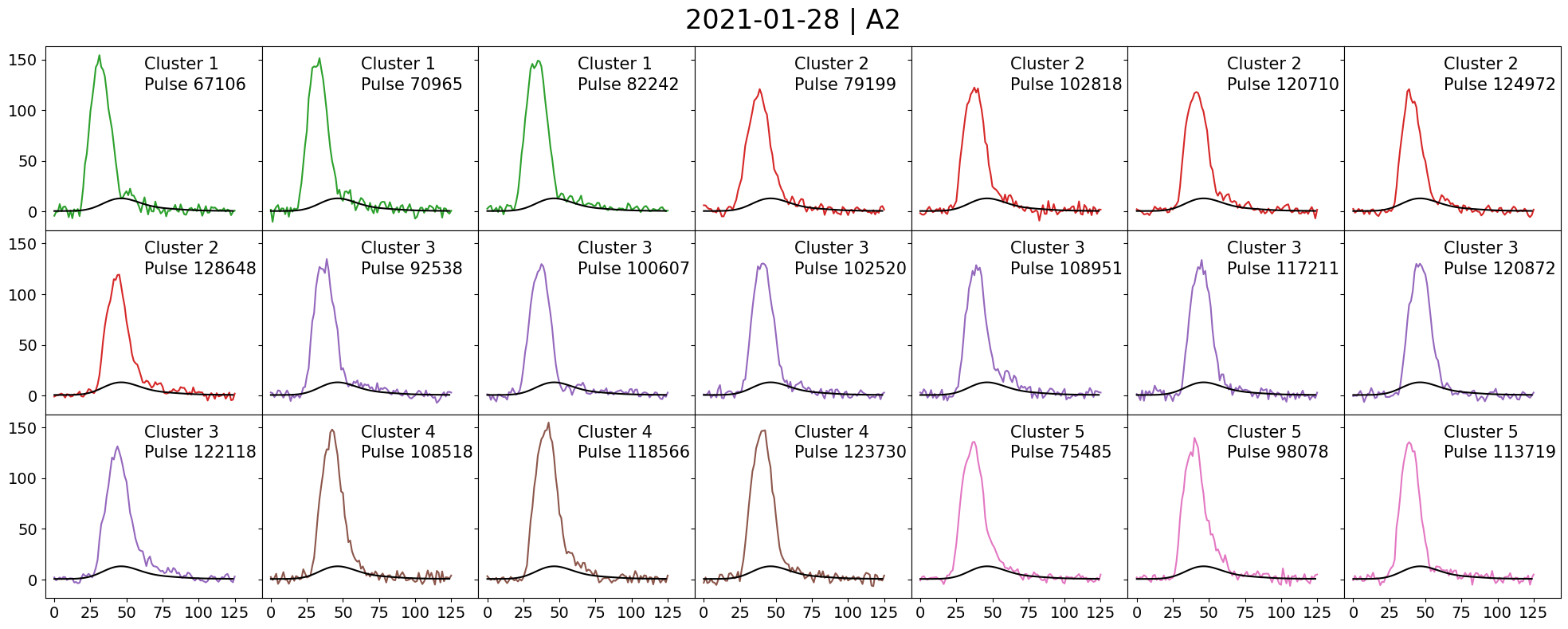}
    \caption{Single pulses in each of the top clusters obtained using DBSCAN. The black line corresponds to the average pulse.}
    \label{fig:all_sequences}
\end{figure*}

Figure \ref{fig:all_sequences} displays the individual pulses of the top amplitude clusters as determined by the DBSCAN clustering algorithm for the 2021-01-28 observation with antenna A2. Those display factors above $\times10$ the average pulse amplitude and tend to appear earlier than the average. The clustering has different properties of peak amplitude and location, and width to separate them.

In Section \ref{sec:CS} we will use a different (to Density-based clustering DBSCAN) approach to perform clustering tasks, a hierarchical clustering algorithm based on learned data features, the Self-Organizing Map (SOM).

\subsubsection{Observations 2021-01-21}\label{sec:2021-01-21}

\begin{figure*}
	% To include a figure from a file named example.*
	% Allowable file formats are eps or ps if compiling using latex
	% or pdf, png, jpg if compiling using pdflatex
	\includegraphics[width=.95\columnwidth]{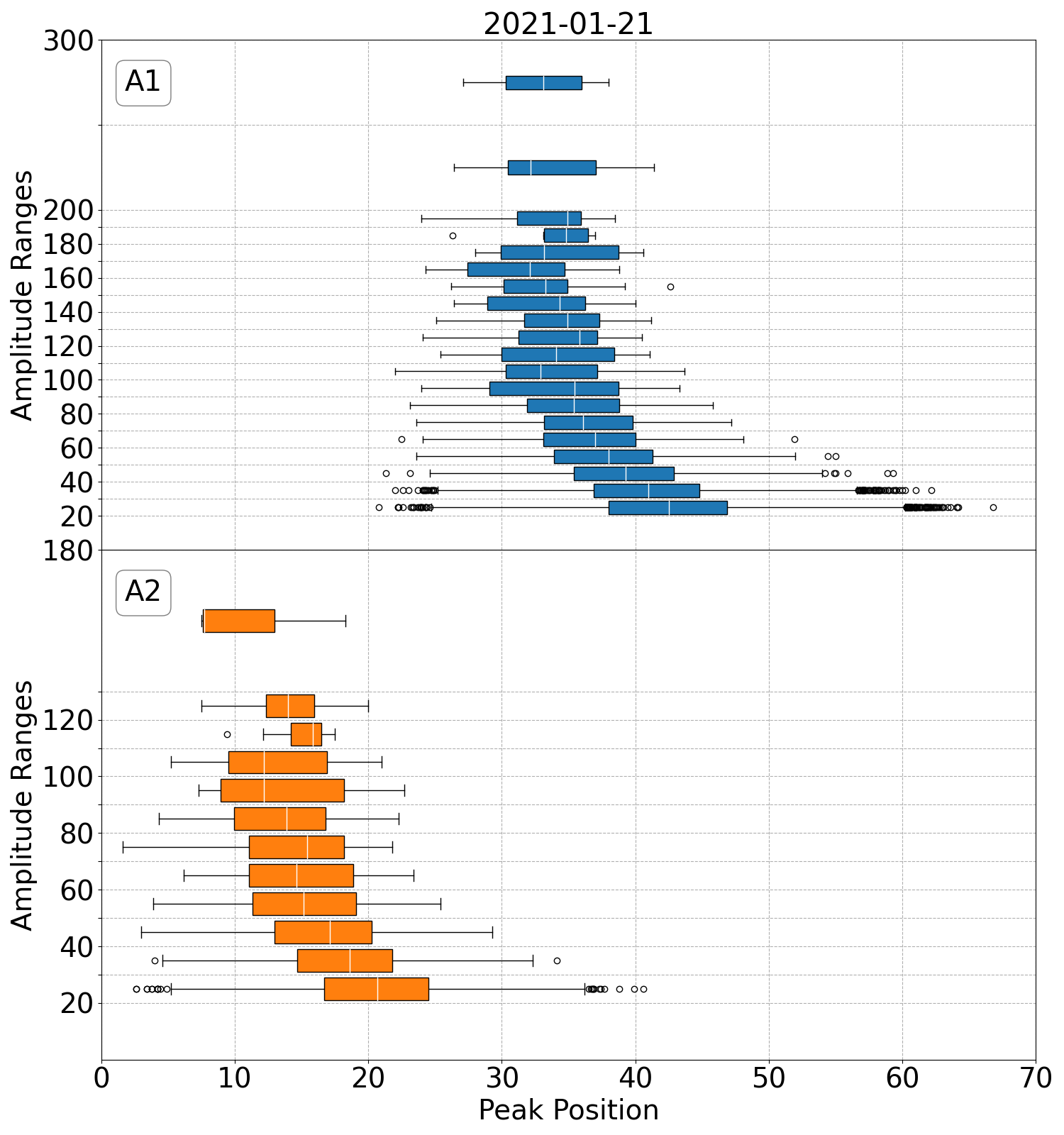}
	\includegraphics[width=.95\columnwidth]{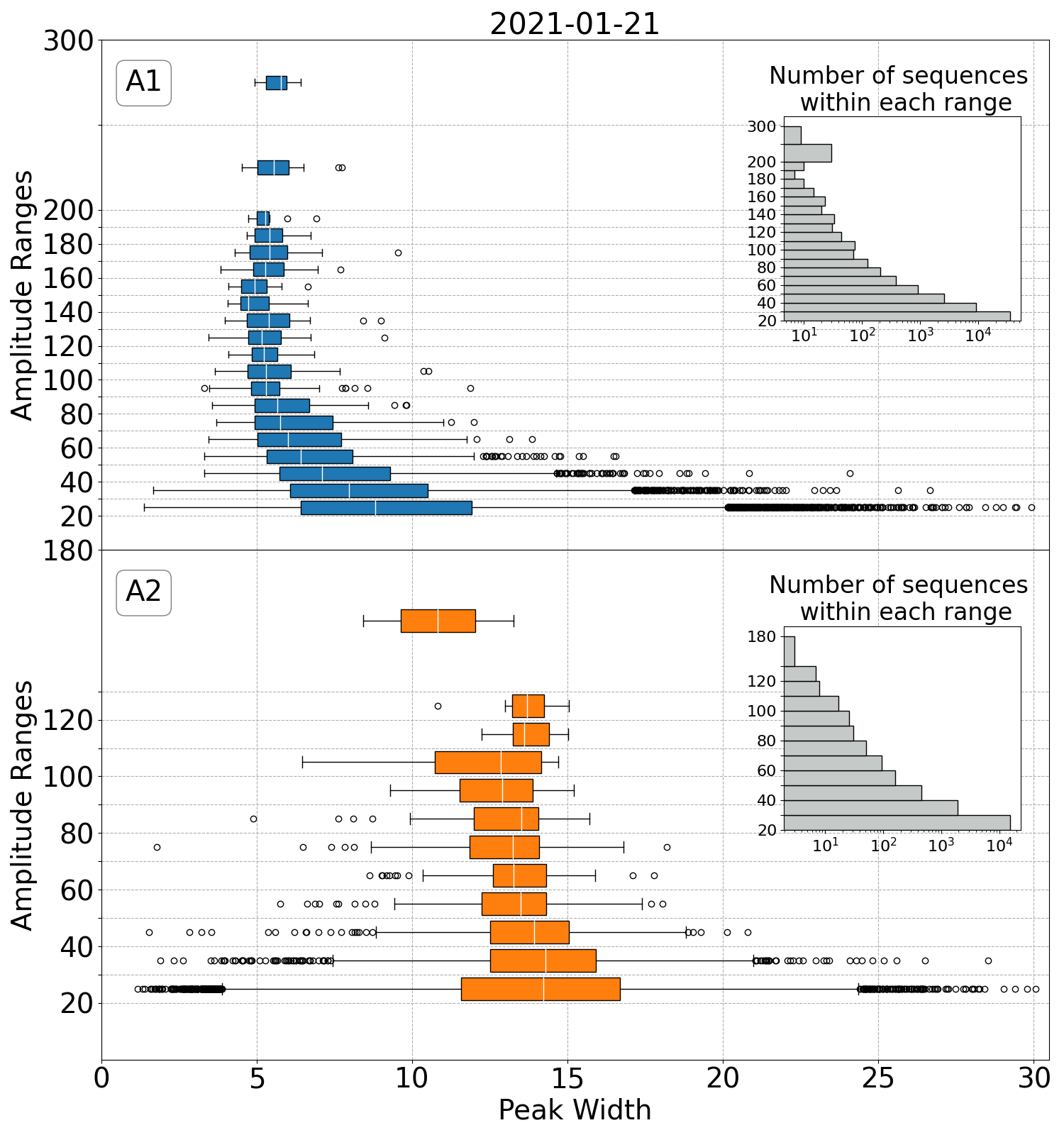}
    \caption{On the left, the distribution of the pulse amplitudes versus arrival time as measured by the position of their peaks for the 2021-01-21 observations. Observations with A1 are on top and with A2 on bottom.
    On the right, the distribution of the pulses amplitudes versus its mean widths.}
    \label{fig:12A12}
\end{figure*}

Fig.~\ref{fig:12A12} displays, on the left, the distribution of the pulses amplitudes versus arrival time as measured by the position of their peaks in units of the instantaneous period over 1220 tick marks (see Appendix~\ref{sec:sp}). Insets show the histograms of pulse amplitudes for each observation. There is a trend for peaks with larger amplitudes to appear earlier that pulses with lower amplitudes for both radio telescopes observations.
On the right column, Fig.~\ref{fig:12A12} displays the distribution of the pulses amplitudes versus its width, measured by the half amplitude of their peaks. There is a trend for peaks with larger amplitudes to appear thinner that pulses with lower amplitude in Radio telescope A1 observations that is not quite confirmed by the corresponding observation with A2, indicating a potential dependence of this effect on the single polarization observations with A1.

%indicating its generation takes place at a potentially upper location in the neutron star magnetosphere.
%\Carlos{Quantification of the effect: }
%A displacement of -10/1200 over a period of 0.089 seconds gives a time difference that at the speed of light leads to nearly 223 km ( for each mark tick $D=0.0892s/1200*299792km/s= 22.3km$). Comparable to the size of a typical magnetosphere \citep{lorimer2012handbook}. [Note that the plots are restricted to a few tens of ticks marks out of the 1200 we divided every single period.]  

%\Carlos{Note that to assign a physical meaning to the neutron star magnetosphere, we should first discard that all those effect are due to scintillation (that would be interesting in itself, but had to be described independently). In particular if we confirm there are actually two (or more) components (average+high amplitude pulse).}

\begin{figure*}
	% To include a figure from a file named example.*
	% Allowable file formats are eps or ps if compiling using latex
	% or pdf, png, jpg if compiling using pdflatex
	\includegraphics[width=1.15\columnwidth]{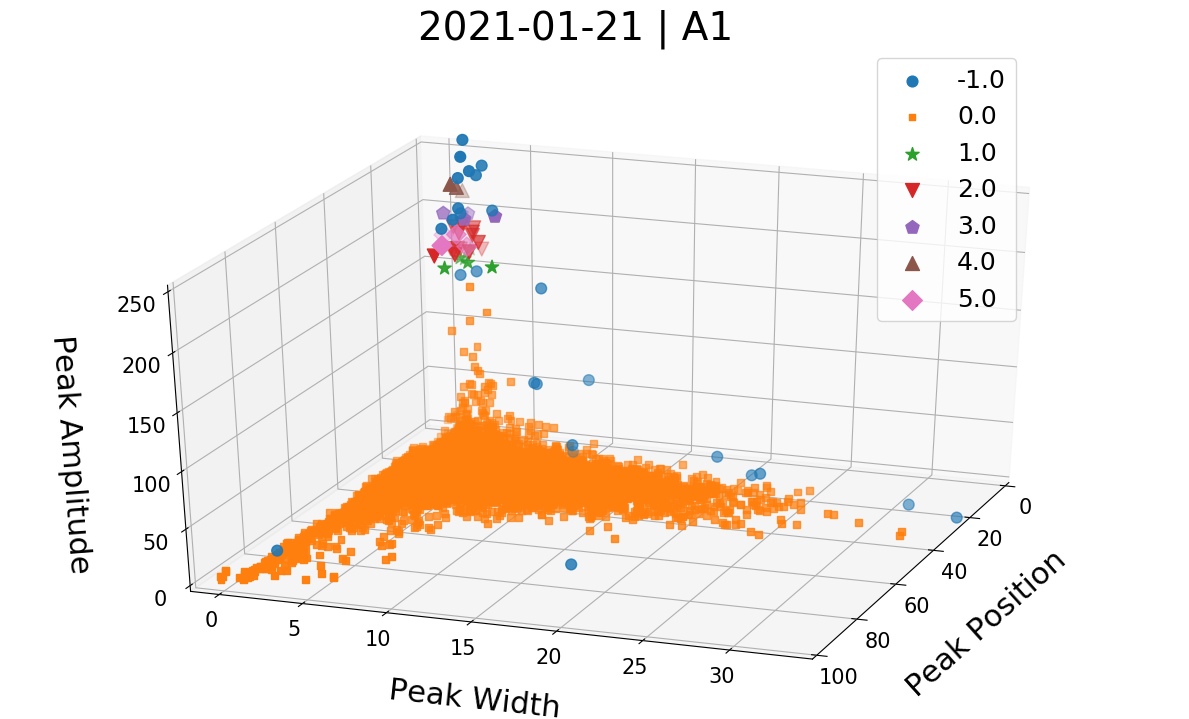}
	\includegraphics[width=0.75\columnwidth]{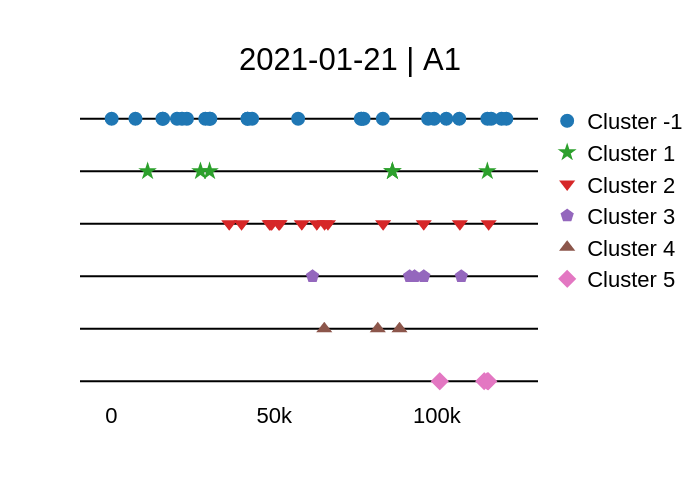}\\
	\includegraphics[width=1.15\columnwidth]{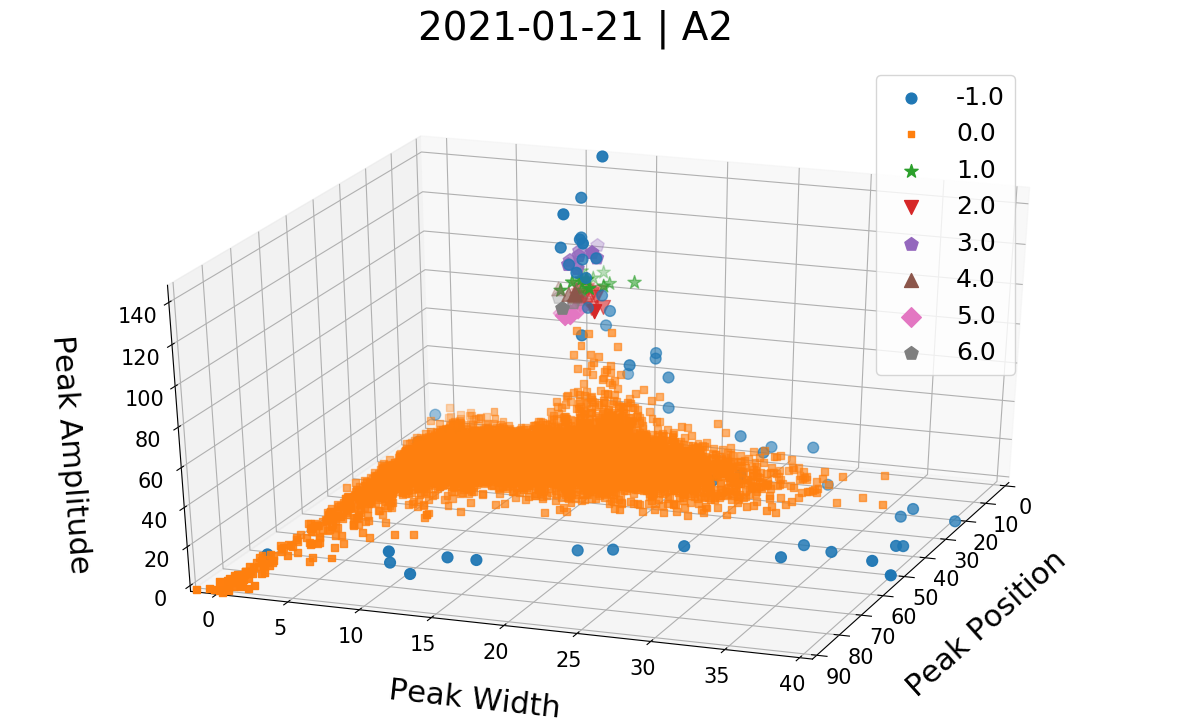}
	\includegraphics[width=0.75\columnwidth]{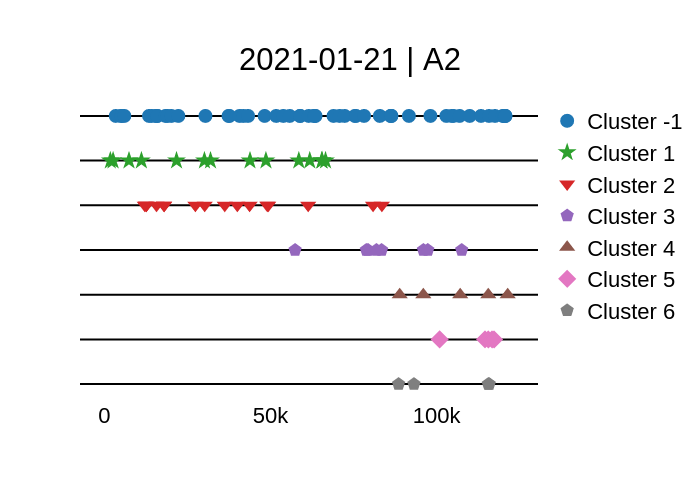}
    \caption{(on the left) 3D distribution of the pulses peak amplitudes, position, and widths for the 2021-01-21 observations. Different colors represent different clusters according to density based scan criteria.
    (On the right) Distribution of the pulses over the duration of the observations 2021-01-21. Upper figures are for Radio telescope A1, lower figures are for Radio telescope A2 on bottom. Different colors represent different clusters according to DBSCAN criteria.}
    \label{fig:3A12}
\end{figure*}

We next performed a Density Based Scan analysis of the pulses clustering for each radio telescope's observation as displayed in Fig.~\ref{fig:3A12}, which seems to preferentially select the different clusters by strata of increasing amplitudes with an spread in the peak location and width. It also selects a baseline cluster (in orange, labeled as 0) and an enveloping outlier (in light blue, labeled as $-1$). We also display the detail of each pulse in the top amplitude DBSCAN clusters over the duration of the observation, labeled by the pulse index number. There seems to be a preference of large pulses amplitudes towards the second half of the observations from both radio telescopes.

%In Fig.~\ref{fig:1A12pi} we display the detail of each pulse in the top amplitude DBSCAN clusters over the duration of the observation, labeled by the pulse index number.
%
%\begin{figure}
%	\includegraphics[width=\columnwidth]{1A1pulse_index.png}
%	\includegraphics[width=\columnwidth]{1A2pulse_index.png}
%    \caption{Distribution of the pulses over the duration of the observations 2021-01-21. Upper figures as for Radio telescope A1, lower figures are for %Radio telescope A2 on. Different colors represent different clusters according to density based scan criteria.}
%    \label{fig:1A12pi}
%\end{figure}

%In Fig.~\ref{fig:4A12} we display the detail of each pulse in the top amplitude DBSCAN clusters in comparison with the total average pulse in blue.

%\begin{figure}
	% To include a figure from a file named example.*
	% Allowable file formats are eps or ps if compiling using latex
	% or pdf, png, jpg if compiling using pdflatex
%	\includegraphics[width=\columnwidth]{4A1a.png}
%	\includegraphics[width=\columnwidth]{4A1b.png}
%	\includegraphics[width=\columnwidth]{4A2a.png}
%	\includegraphics[width=\columnwidth]{4A2b.png}
%    \caption{Distribution of the pulses peak amplitudes, position, and widths. Upper figures as for Radio telescope A1, lower figures are for Radio telescope A2 on. Different colors represent different clusters according to density based scan criteria. The total average pulse in depicted in blue.}
%    \label{fig:4A12}
%\end{figure}

\subsubsection{Observations 2021-01-24}\label{sec:2021-01-24}

\begin{figure*}
	% To include a figure from a file named example.*
	% Allowable file formats are eps or ps if compiling using latex
	% or pdf, png, jpg if compiling using pdflatex
	\includegraphics[width=.95\columnwidth]{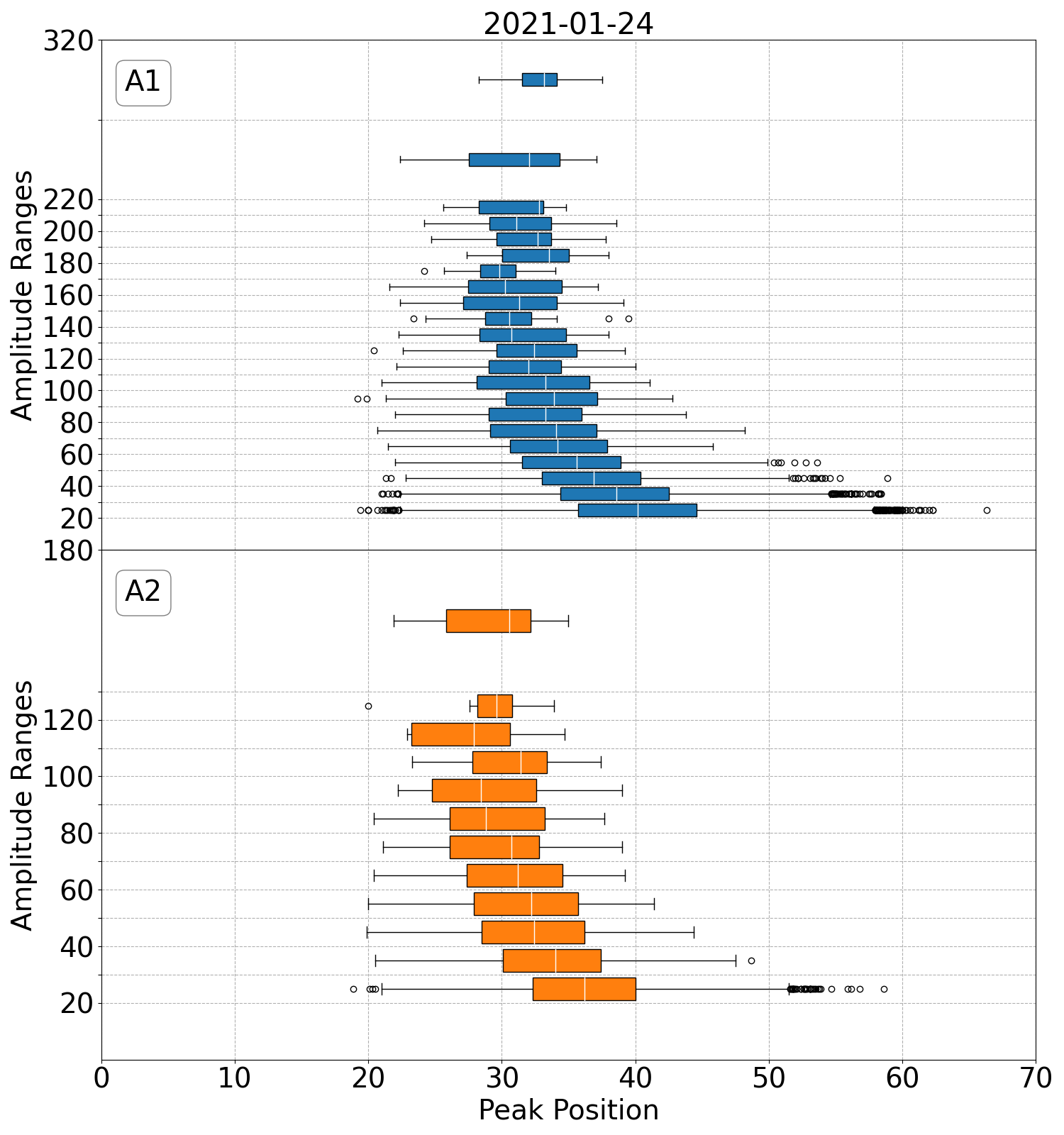}
	\includegraphics[width=.95\columnwidth]{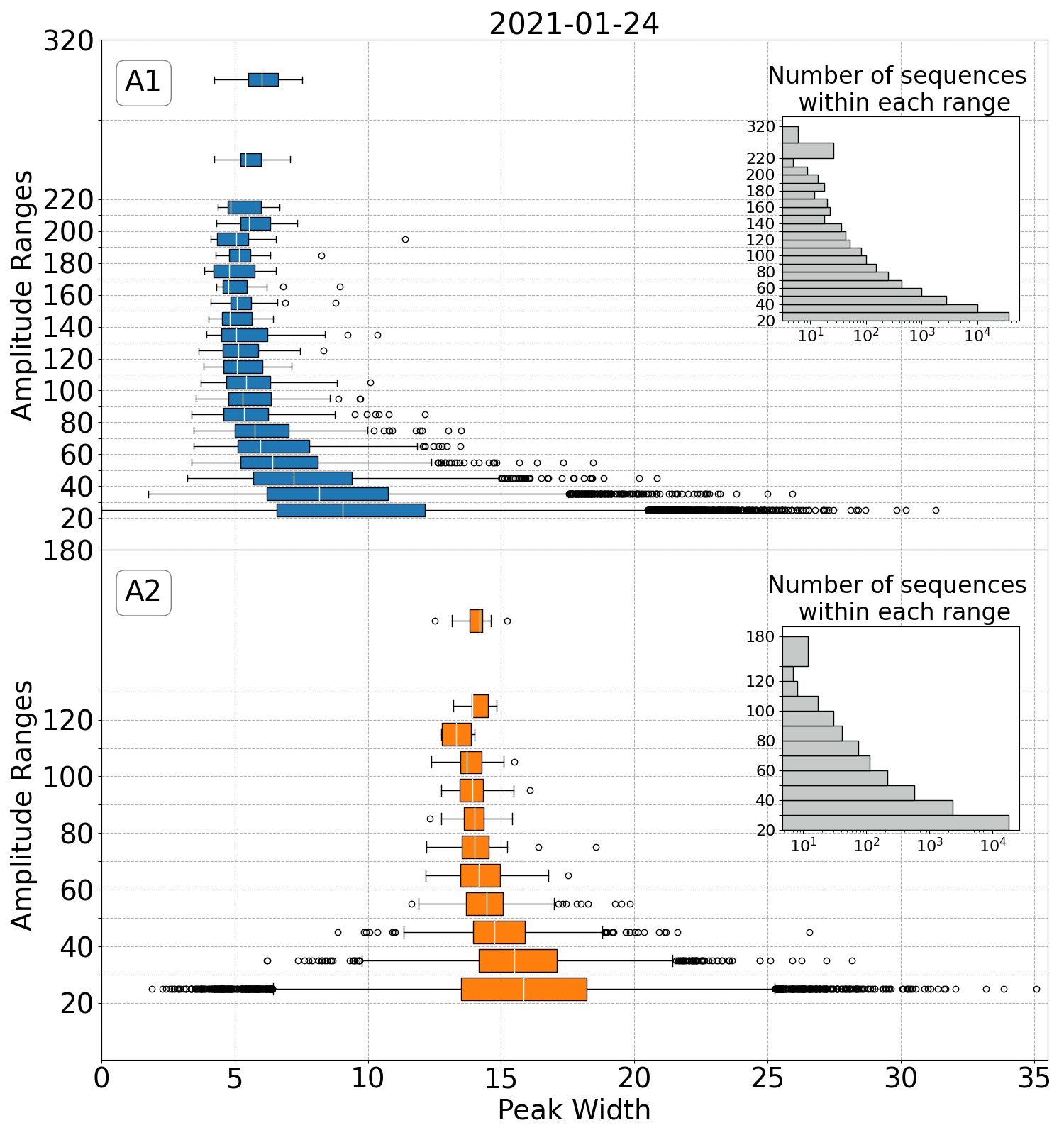}
    \caption{Distribution of the pulses amplitudes versus arrival time as measured by the position of their peaks for the 2021-01-24 observations. On the right the distribution of the pulses amplitudes versus its mean widths for the 2021-01-24 observations. Radio telescope A1 on top and Radio telescope A2 on bottom.}
    \label{fig:56A12}
\end{figure*}

Fig.~\ref{fig:56A12} displays the distribution of the pulses amplitudes versus arrival time as measured by the position of their peaks. The insets show the histograms of pulse amplitudes for each observation. We again observe a trend for peaks with larger amplitudes to appear earlier than pulses with lower amplitude in both radio antennas observations indicating that this is a robust feature of the pulses distributions. On the right
column of Fig.~\ref{fig:56A12} we display the distribution of the pulses amplitudes versus its width as measured by the half amplitude of their peaks. 
The amplitude versus pulse width shows again a preference of narrower high amplitude pulses as seen by the A1 observations, but much less evident in the A2 two polarizations observations, indicating some differences when we observe this features with two polarizations than when we observe them in a single one.

\begin{figure*}
	\includegraphics[width=1.15\columnwidth]{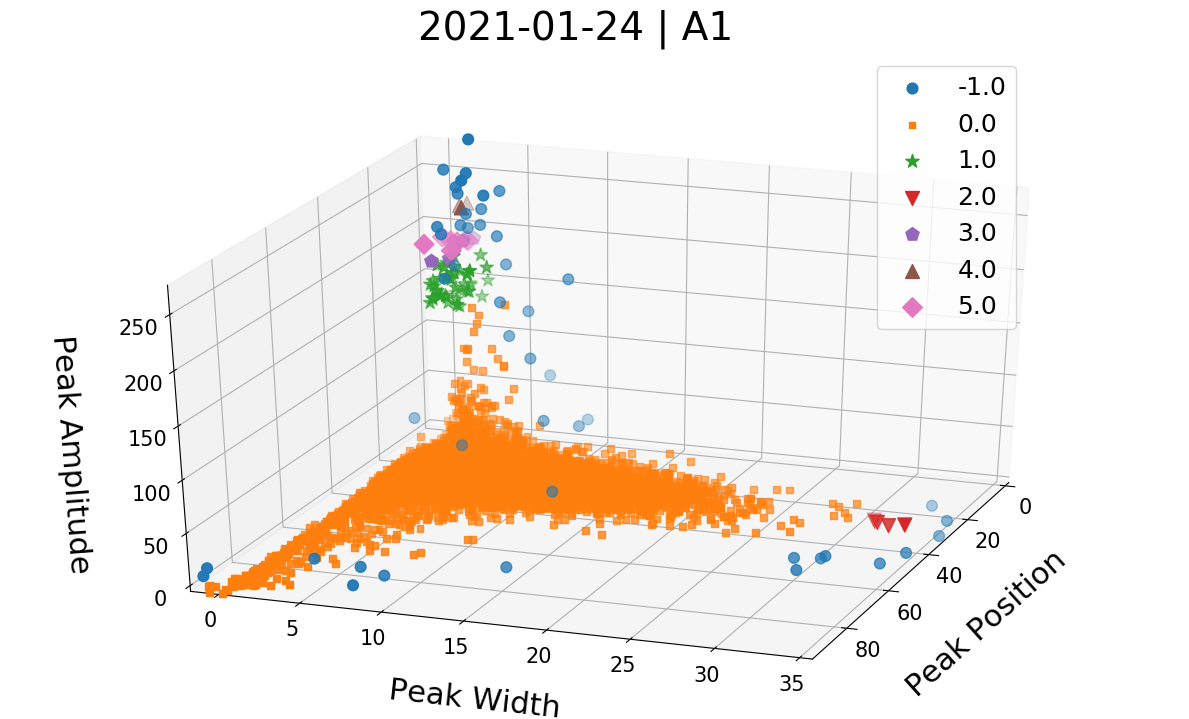}
	\includegraphics[width=0.75\columnwidth]{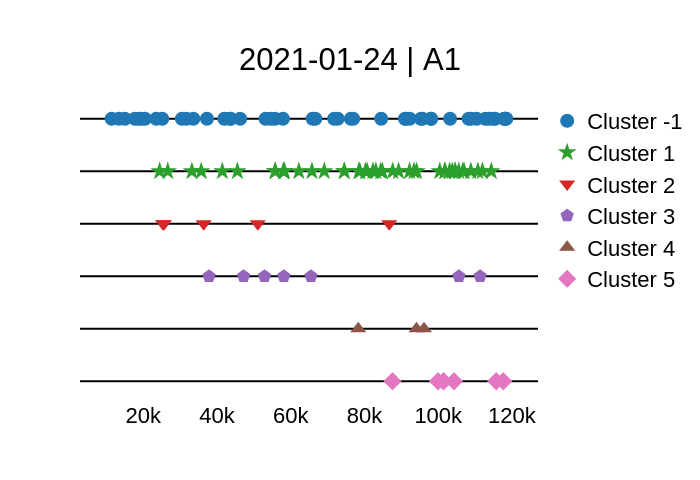}\\
	\includegraphics[width=1.15\columnwidth]{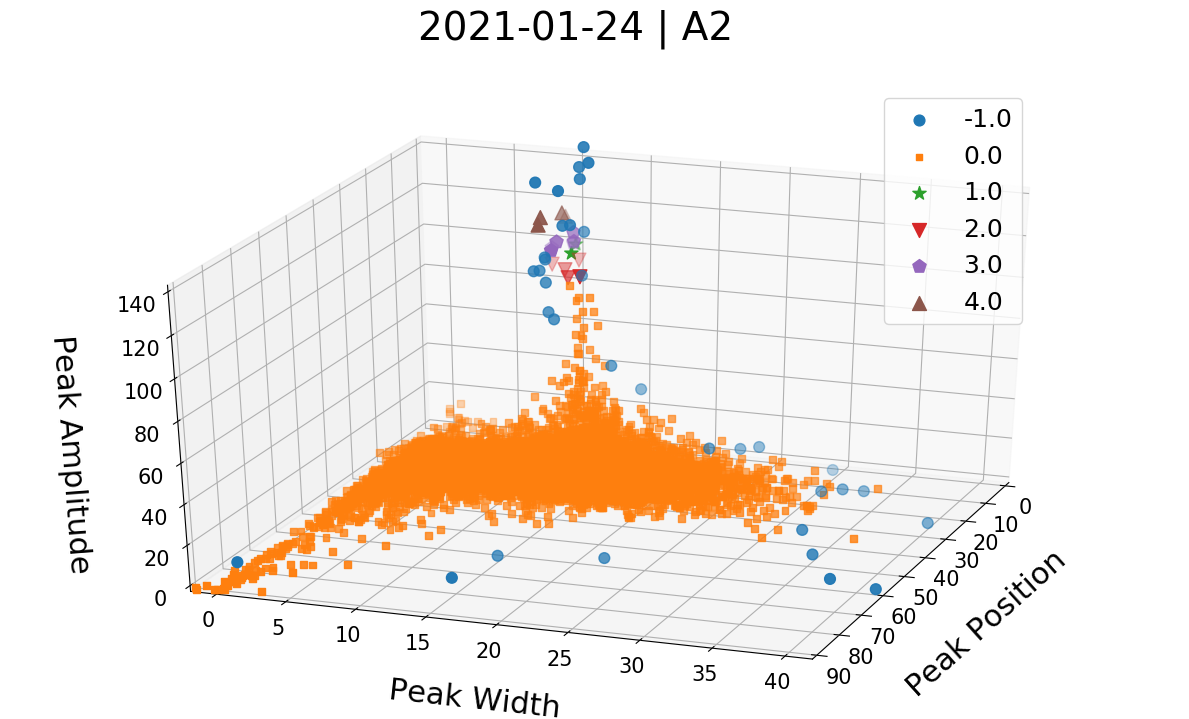}
	\includegraphics[width=0.75\columnwidth]{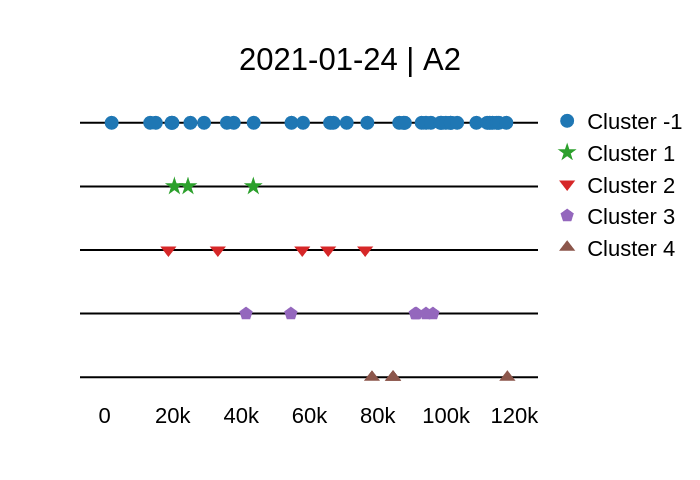}
    \caption{(On the left) 3D distribution of the pulses peak amplitudes, position, and widths for the 2021-01-24 observations. Different colors represent different clusters according to density based scan criteria.
    (On the right) Distribution of the pulses over the duration of the observations 2021-1-24. Upper figures as for Radio telescope A1, lower figures are for Radio telescope A2 on bottom. Different colors represent different clusters according to DBSCAN criteria.}
    \label{fig:7A12}
\end{figure*}

We next performed a Density Based Scan (DBSCAN) analysis of the pulses clustering for each radio telescope's observation as displayed in Fig.~\ref{fig:7A12}. This shows the different clusters by increasing amplitudes and some spreading in peak location and width. It also creates a baseline cluster (in orange labeled as 0) and an enveloping outlier (in light blue, labeled as -1). We also display the detail of each pulse in the top amplitude DBSCAN clusters over the duration of the observation, labeled by the pulse index number. There is a slight preference for the top amplitude pulses to lie in the second part of the observations, but in very small numbers to make a statistically significant trend.

%In Fig.~\ref{fig:2A12pi} we display the detail of each pulse in the top amplitude DBSCAN clusters over the duration of the observation, labeled by the pulse index number.

%\begin{figure}
%	\includegraphics[width=\columnwidth]{2A1pulse_index.png}
%	\includegraphics[width=\columnwidth]{2A2pulse_index.png}
%    \caption{Distribution of the pulses over the duration of the observations 2021-1-24. Upper figures as for Radio telescope A1, lower figures are for %Radio telescope A2 on. Different colors represent different clusters according to density based scan criteria.}
%    \label{fig:2A12pi}
%\end{figure}

%In Fig.~\ref{fig:8A12} we display the detail of each pulse in the top amplitude DBSCAN clusters in comparison with the total average pulse in blue.

%\begin{figure}
	% To include a figure from a file named example.*
	% Allowable file formats are eps or ps if compiling using latex
	% or pdf, png, jpg if compiling using pdflatex
%	\includegraphics[width=\columnwidth]{8A1a.png}
%	\includegraphics[width=\columnwidth]{8A1b.png}
%	\includegraphics[width=\columnwidth]{8A2a.png}
%	\includegraphics[width=\columnwidth]{8A2b.png}
%    \caption{Distribution of the pulses peak amplitudes, position, and widths. Upper figures as for Radio telescope A1, lower figures are for Radio telescope A2 on. Different colors represent different clusters according to density based scan criteria. The total average pulse in depicted in blue.}
%    \label{fig:8A12}
%\end{figure}

\subsubsection{Observations 2021-01-28}\label{sec:2021-01-28}

The third observation of our selected week at the end of January 2021 is the one with the top S/N according to our analysis in Appendix \ref{sec:RFIClean}. 
Fig.~\ref{fig:910A12} displays on the left the distribution of the pulses amplitudes versus arrival time as measured by the position of their peaks. Insets show the histograms of pulse amplitudes for each observation. We observe again, that there is a trend for peaks with larger amplitudes to appear earlier than pulses with lower amplitude indicating its a robust feature over time of observation as well as Radio telescope / polarization. We leave the more quantitative study of this feature for the next two sections, where we analyze the effects of scintillation and consider the SOM clustering.
On the right, Fig.~\ref{fig:910A12} displays the distribution of the pulses amplitudes versus its mean width as measured by the half amplitude of their peaks. There is again a trend for peaks with larger amplitudes to appear narrower than pulses with lower amplitude in Radio telescope A1 observations that is weakly confirmed by the corresponding observation with Radio telescope A2, suggesting again this effect has a dependence on the single polarization observations with A1.
Note also the the magnitude of the amplitudes in the A1 observations nearly doubles that of the A2, again, indicating the higher sensitivity to a wider bandwidth than to the (circular) addition of the two polarizations, although there might be some systematic effects that are better resolved in the two polarization observations, like the width dependence of the pulses on the larger amplitudes.

\begin{figure*}
	% To include a figure from a file named example.*
	% Allowable file formats are eps or ps if compiling using latex
	% or pdf, png, jpg if compiling using pdflatex
	\includegraphics[width=.95\columnwidth]{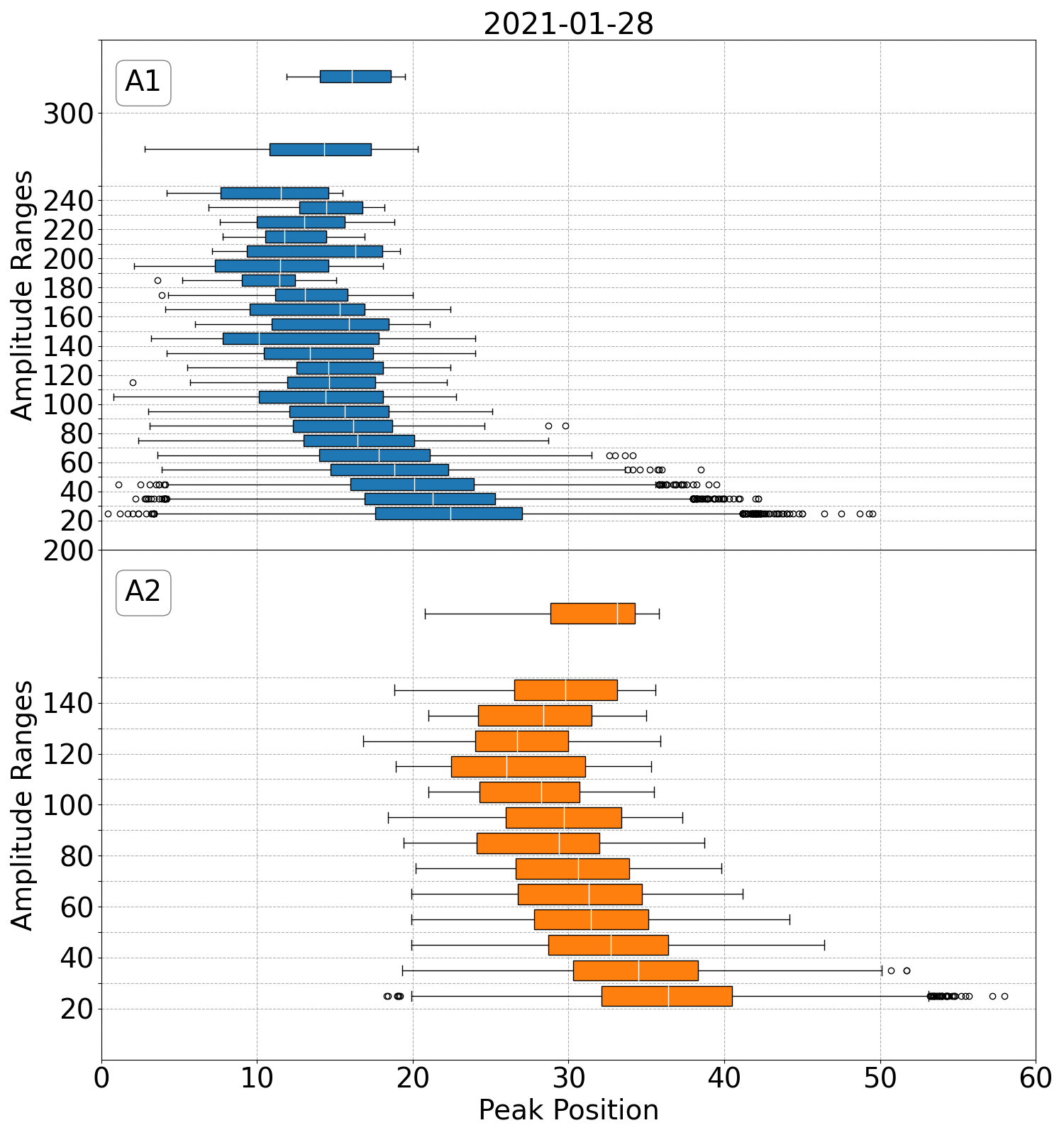}
	\includegraphics[width=.95\columnwidth]{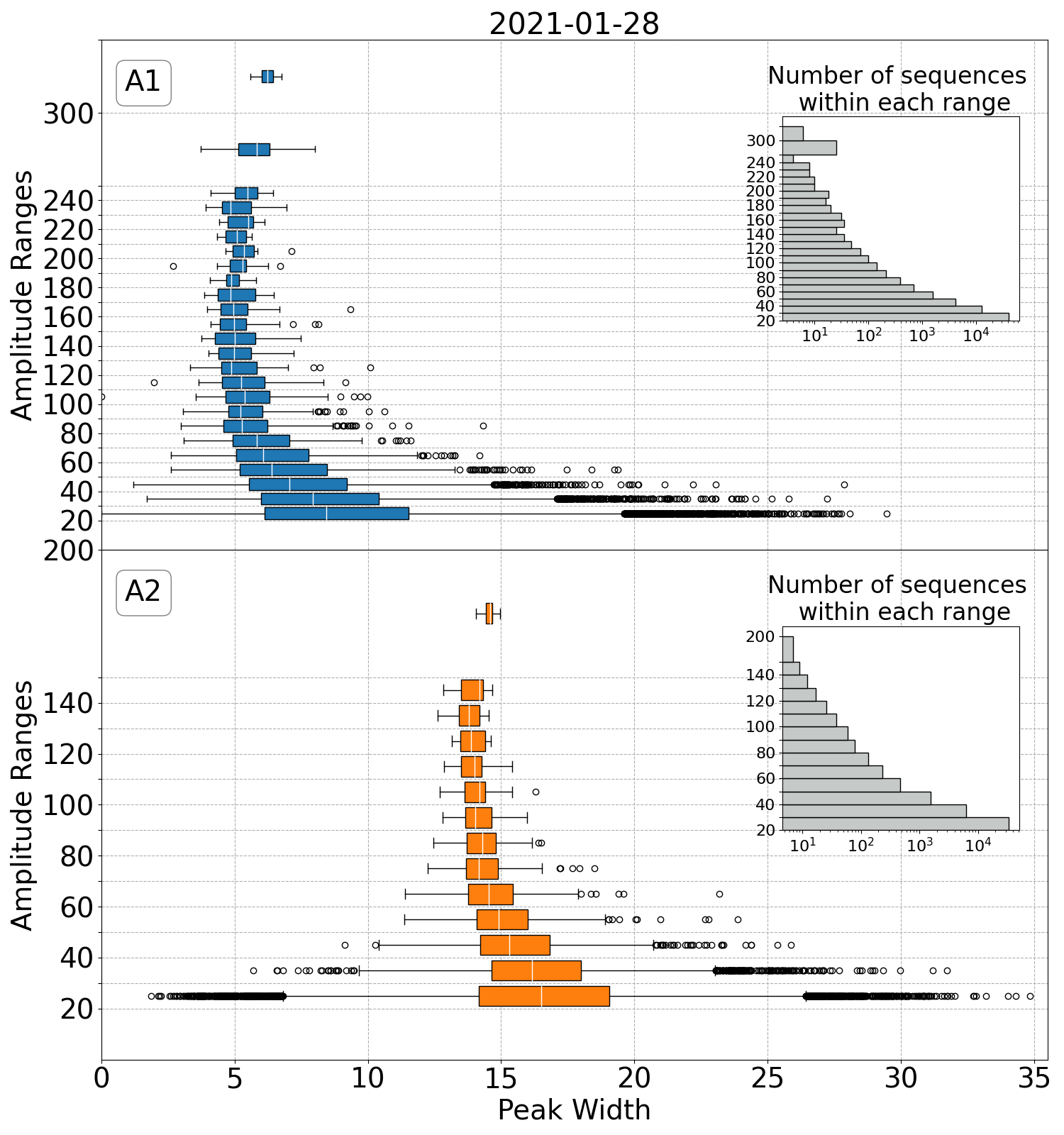}
    \caption{Distribution of the pulses amplitudes versus arrival time as measured by the position of their peaks for the 2021-01-28 observations. Distribution of the pulses amplitudes versus its mean widths for the 2021-01-28 observations. Radio telescope A1 on top and Radio telescope A2 on bottom. The differences in behavior can be attributed to the single polarization (A1) versus two polarizations (A2) observations.}
    \label{fig:910A12}
\end{figure*}

\begin{figure*}
	% To include a figure from a file named example.*
	% Allowable file formats are eps or ps if compiling using latex
	% or pdf, png, jpg if compiling using pdflatex
	\includegraphics[width=1.15\columnwidth]{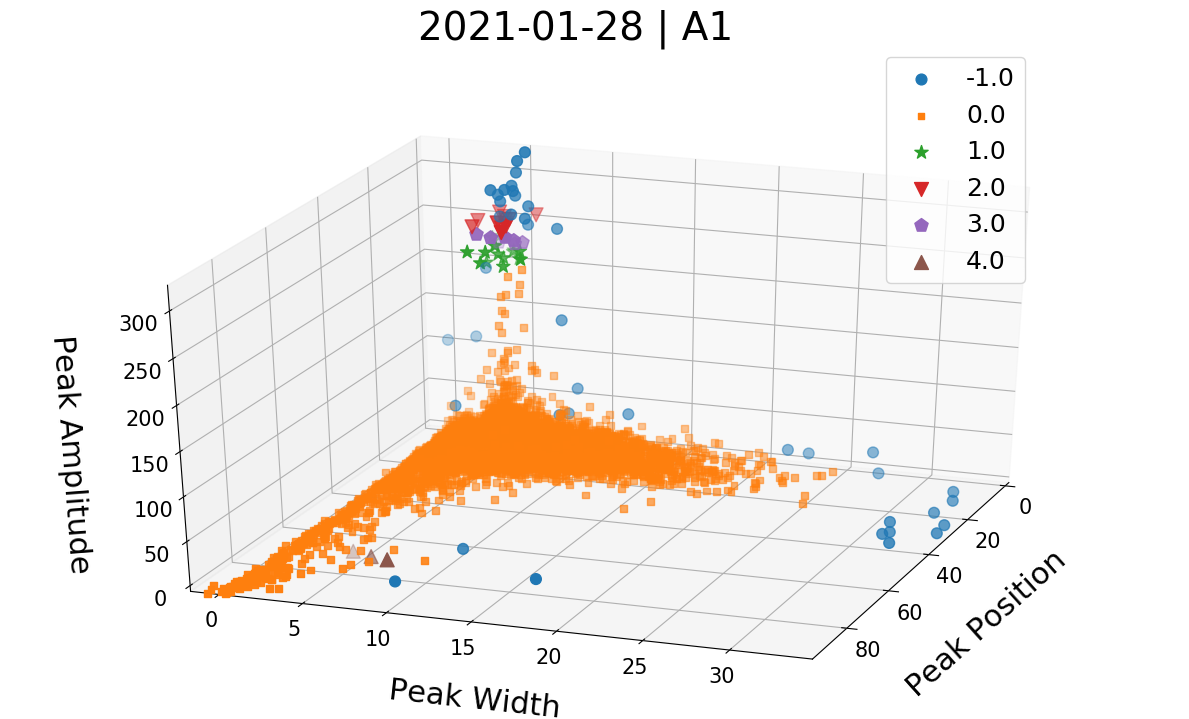}
	\includegraphics[width=0.75\columnwidth]{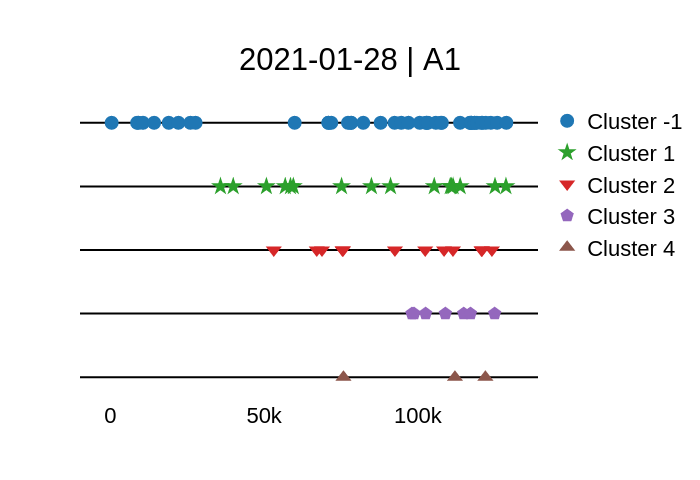}\\
	\includegraphics[width=1.15\columnwidth]{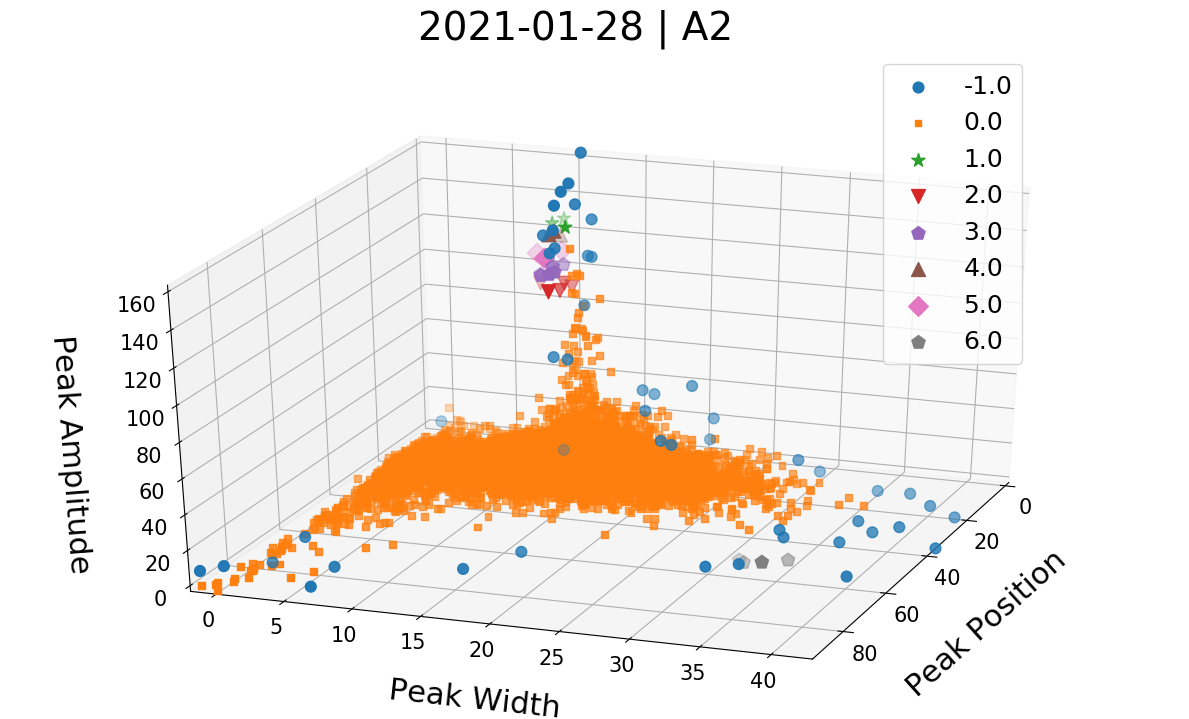}
	\includegraphics[width=0.75\columnwidth]{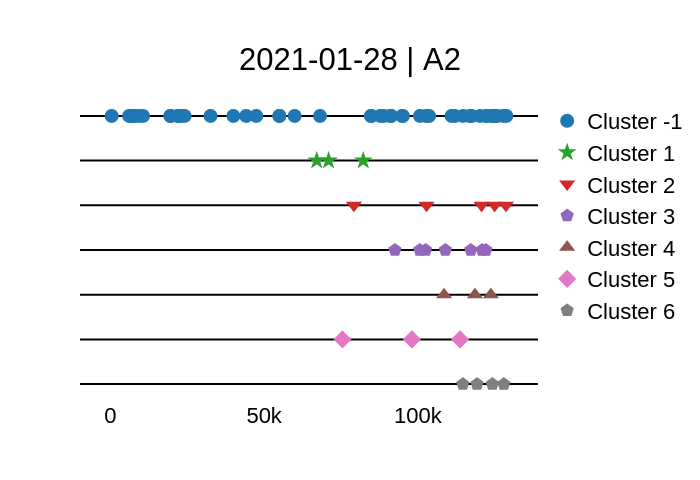}
    \caption{(On the left) 3D distribution of the pulses peak amplitudes, position, and widths for the 2021-01-28 observations. Different colors represent different clusters according to density based scan criteria.
    (On the right) Distribution of the pulses over the duration of the observations 2021-01-28. Upper figures as for Radio telescope A1, lower figures are for Radio telescope A2 on bottom. Different colors represent different clusters according to DBSCAN criteria.}
    \label{fig:11A12}
\end{figure*}

We performed a DBSCAN analysis of the pulses clustering for each Radio telescope's observation as displayed in Fig.~\ref{fig:11A12} which selects the different clusters by increasing amplitudes, also creates a baseline cluster (in orange labeled as 0), and an enveloping outlier (in light blue, labeled as -1). We also display the detail of each pulse in the top amplitude DBSCAN clusters over the duration of the observation, labeled by the pulse index number. Those seem to display a preference to appear in the second part of the observation. This seems to indicate perhaps a region of the local sky, towards west, with less interference noise (RFIs), during this week of January 2021.

%In Fig.~\ref{fig:3A12pi} we display the detail of each pulse in the top amplitude DBSCAN clusters over the duration of the observation, labeled by the pulse index number.
%
%\begin{figure}
%	\includegraphics[width=\columnwidth]{3A1pulse_index.png}
%	\includegraphics[width=\columnwidth]{3A2pulse_index.png}
%    \caption{Distribution of the pulses over the duration of the observations 2021-01-28. Upper figures as for Radio telescope A1, lower figures are for %Radio telescope A2 on. Different colors represent different clusters according to density based scan criteria.}
%    \label{fig:3A12pi}
%\end{figure}

%In Fig.~\ref{fig:12A12} we display the detail of each pulse in the top amplitude DBSCAN clusters in comparison with the total average pulse in blue.

%\begin{figure}
	% To include a figure from a file named example.*
	% Allowable file formats are eps or ps if compiling using latex
	% or pdf, png, jpg if compiling using pdflatex
%	\includegraphics[width=\columnwidth]{12A1a.png}
%	\includegraphics[width=\columnwidth]{12A1b.png}
%	\includegraphics[width=\columnwidth]{12A2a.png}
%	\includegraphics[width=\columnwidth]{12A2b.png}
%    \caption{Distribution of the pulses peak amplitudes, position, and widths. Upper figures as for Radio telescope A1, lower figures are for Radio telescope A2 on. Different colors represent different clusters according to density based scan criteria. The total average pulse in depicted in blue.}
%    \label{fig:12A12}
%\end{figure}

\subsection{Scintillations}\label{sec:scintillations}

%A fit to the above distributions have been done using the beta prime distribution \citep{} with parameters $\alpha$ and $\beta$. \Carlos{Valentina: We can also try the scintillation function we used for the paper on J0437 \cite{Fiscella:2020jey}.}

%\begin{table}
%	\centering
%	\caption{Pulse peak amplitude distributions per day of observation per Radio telescope }
%	\label{tab:beta}
%	\begin{tabular}{llllllll}
%		\hline
%		Day \#& Radio telescope & mean & mode  & variance & Skewness & $\alpha$ & $\beta$\\
%		\hline
%	2021-01-21 & A1 & 1 & 2 & 3 & 4 & 5 & 6 \\
%	2021-01-21 & A2 & 1 & 2 & 3 & 4 & 5 & 6 \\
%	2021-01-24 & A1 & 1 & 2 & 3 & 4 & 5 & 6 \\
%	2021-01-24 & A2 & 1 & 2 & 3 & 4 & 5 & 6 \\
%	2021-01-28 & A1 & 1 & 2 & 3 & 4 & 5 & 6 \\
%	2021-01-28 & A2 & 1 & 2 & 3 & 4 & 5 & 6 \\
%		\hline
%	\end{tabular}
%\end{table}

\begin{figure}
	\includegraphics[width=\columnwidth]{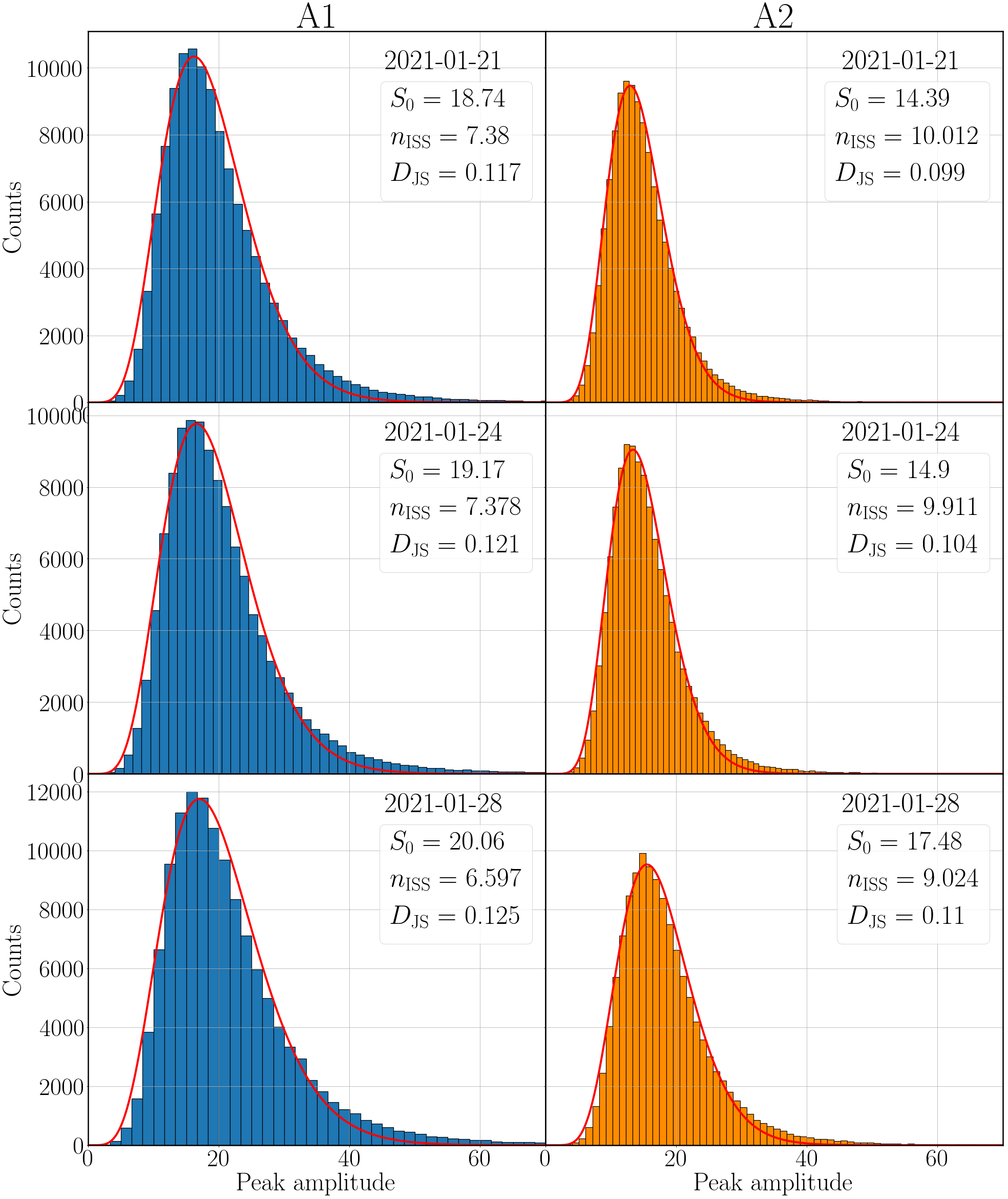}
    \caption{Histograms of projected pulse amplitude for J0835$-$4510 for A1 (left column) and A2 (right column) for the January 2021 observations. The line shows the estimated scintillation distribution from fitting $n_\mathrm{ISS}$ in Eq.~(\ref{eq:scintillation}).}
    \label{fig:scintillations}
\end{figure}

Scintillation due to the interstellar media can change the intensity of the pulses.
Fig.~\ref{fig:scintillations} shows a histogram of the projected pulse S/N for A1 and A2. The line shows the estimated probability density function (PDF) from scintillation \citep{Cordes:1997my}, 
\begin{equation}
    f_S(S|n_\mathrm{ISS}) = \frac{\left(S\, n_\mathrm{ISS}/S_0 \right)^{n_\mathrm{ISS}}}{S\,\Gamma(n_\mathrm{ISS})}\exp{\left( \frac{-S\, n_\mathrm{ISS}}{S_0} \right)} \; \Theta(S),
    \label{eq:scintillation}
\end{equation} 
where $n_\mathrm{ISS}$ is the number of scintles, $S_0$ is the mean value of the signal $S$ (i.e., $S_0 = \langle \mathrm{S/N} \rangle$), and $\Theta$ is the Heaviside step function. Since $S \propto T_{\mathrm{peak}}$ \citep{lorimer2012handbook}, it follows that $T_{\mathrm{peak}}$ also obeys the PDF in Eq.~(\ref{eq:scintillation}).
Here we will explore the possibility of modeling the individual pulses amplitude distribution entirely in terms of a pure interstellar scintillation effect.
We therefore calculate $n_\mathrm{ISS}$ by fitting the observed single pulse peak amplitudes for each Radio telescope. The bin size is determined using the Knuth's rule \citep{Knuth2006} algorithm provided in \texttt{astropy} \citep{astropy2013,astropy2018}. We also normalize the number of observations in each bin by the total number of single pulses in each observation. For each of these fittings we perform a Jensen-Shannon divergence test \citep{JSTest} to quantify the difference between the empirical and the theoretical distribution.
Table~\ref{table:niss_table} gives the scintillation fit parameters 
of the amplitude versus the number of pulses (the insets in Figs.~\ref{fig:12A12},\ref{fig:56A12},\ref{fig:910A12},\ref{fig:1314A12}) for the four days of observation on both radio telescopes.

\begin{table}
    \centering
    \caption{Adjusted values of $n_\mathrm{ISS}$ for each set of observations and the Jensen-Shannon divergence test $D_{\mathrm{JS}}$-value for each fitting. * Was performed in two polarizations at half the bandwidth}\label{table:niss_table}
    \begin{tabular}{cc|ccc}
    \hline \hline
     & Date & $n_\mathrm{ISS}$ & $S_0$ & $D_{\mathrm{JS}}$ \\
    \hline
    \parbox[c]{2mm}{\multirow{4}{*}{\rotatebox[origin=c]{90}{A1}}} & Jan 21 & 7.38 $\pm$ 0.11 & 18.735 $\pm$ 0.068 & 0.117 \\
    & Jan 24 & 7.378 $\pm$ 0.112 & 19.169 $\pm$ 0.07 & 0.121 \\
    & Jan 28 & 6.597 $\pm$ 0.118 & 20.057 $\pm$ 0.092 & 0.125 \\
    & *Mar 29 & 8.186 $\pm$ 0.073 & 7.065 $\pm$ 0.014 & 0.119 \\
    \hline
     \parbox[c]{2mm}{\multirow{4}{*}{\rotatebox[origin=c]{90}{A2}}} & Jan 21 & 10.012 $\pm$ 0.091 & 14.389 $\pm$ 0.026 & 0.099 \\
    & Jan 24 & 9.911 $\pm$ 0.098 & 14.9 $\pm$ 0.03 & 0.104 \\
    & Jan 28 & 9.024 $\pm$ 0.103 & 17.482 $\pm$ 0.043 & 0.11 \\
    & Mar 29 & 9.756 $\pm$ 0.093 & 15.418 $\pm$ 0.03 & 0.106 \\
    \hline \hline
    \end{tabular}
\end{table}

As a result, for Vela, we find $n_\mathrm{ISS} \sim 6.6 - 7.4$ with A1 and $n_\mathrm{ISS} \sim 9 - 10$ for A2. In contrast, in \cite{Fiscella:2020jey}, for J0437$-$4715, we obtained $n_\mathrm{ISS}=2.67 \pm 0.31$ for A1 and $n_\mathrm{ISS}=2.17 \pm 0.25$ for A2 when using $217$~minutes-long observations. 
We also note the large value of the $n_\mathrm{ISS}$ found in comparison to the typical $n_\mathrm{ISS}<2$ found for longer time scales and different radio-frequencies. \cite{2000astro.ph..7233C} found two scintillation scales observing Vela at 2.5~GHz of 15s and 26s. Rescaling those scales to our observing frequency, 1400~MHz, we find time scales of $\Delta t_\mathrm{d,1} = 7.48$~s and $\Delta t_\mathrm{d,2} = 12.97$~s. Likewise we rescale the scintillation bandwidths
to 1400~MHz, and find $\Delta \nu_\mathrm{d,1} = 3.84$~MHz and $\Delta \nu_\mathrm{d,2} = 6.49$~MHz, respectively. We can now compare 
our values of $n_\mathrm{ISS}$ with theoretical estimations via the formula \citep{Cordes:1997my}
\begin{equation}
    n_\mathrm{ISS} \approx \left(1+\eta_t\frac{T}{\Delta t_\mathrm{d}}\right)\left(1+\eta_\nu\frac{BW}{\Delta \nu_\mathrm{d}}\right)
\end{equation}
where $\eta_t$ and $\eta_\nu$ are filling factors $\sim 0.25$. The estimated $n_\mathrm{ISS}$ for $T=0.089$~s A1 (BW~=~112~MHz) and for A2 (BW~=~56~MHz) are
$n_\mathrm{ISS,1}=8.3$ and $n_\mathrm{ISS,1}=4.7$ for A1 and A2 respectively, and $n_\mathrm{ISS,2}=5.3$ for A1 and $n_\mathrm{ISS,2}=3.2$ for A2. 

We then conclude that $n_\mathrm{ISS}$ over a shorter (0.089 seconds) timescale is expected to be smaller than measured for A2 and polarization dependent (values roughly match the one-polarization measures of A1). We also find a relatively good agreement between the observational data and the theoretical PDF, showing that Eq.~(\ref{eq:scintillation}) holds valid even at such short timescales. Nonetheless, we also note an excess in the number of high-amplitude single pulses that cannot be explained solely on the basis of scintillation. Those represent several thousands of pulses, and leave room for the interpretation in terms of the mini-giant pulses discussed in teh previous Sec.~\ref{sec:math}. Besides, the Jensen-Shannon divergence test $D_{\mathrm{JS}}$ values displayed in Table \ref{table:niss_table} are closer to zero than to one (the ideal fit).
 This opens the possibility of a different sort of modeling to the pulse amplitude distribution, requiring a complementary approach to the one provided above that we will describe next.
%This suggest that another pulsar-linked mechanism is largely at play.
%A plausible explanation is that their generation takes place at a potentially upper location in the neutron star magnetosphere.

\subsection{Self-Organizing Map (SOM) techniques and results}\label{sec:CS}

%\Carlos{Description intro/review CS techniques by Ryan, Prahsnna, Linwei, Cameron?}

This section describes a deep neural network-based machine learning method for unsupervised clustering of the individual pulses of Vela. In recent years, deep learning has produced remarkable results across many domains, including computer vision, natural languages, and medical imaging \citep{lecun2015deep}. With deep learning, one tries to learn representations of the data hierarchically (hence the term deep). The underlying representations capture variations of the data explaining different attributes of the data. Often these representations are regressed to the target prediction because of the interest in the prediction model. In our case, we are more interested in the underlying attributes themselves and resort to unsupervised learning. As we show later, these representations can be grouped into different meaningful clusters of the individual pulses.

In deep learning literature, there are various unsupervised approaches to learn or capture the representations of the data. The most common ones include the autoencoder and its variants, a class of deep learning algorithms that take in the input and try to reconstruct the same input by passing it through the low-dimensional bottleneck subjected to different regularizations (e.g., sparsity). In our case, we consider a variational autoencoder (VAE) which is a probabilistic model with stochastic latent space \citep{kingma2014autoencoding}. Formally, given an individual pulse $\mathbf{x_i}$, we estimate the mean $\mu$ and standard deviation $\sigma$ using a deep encoder. We then sample a latent vector $\mathbf{z_i}$ from the approximated Gaussian distribution $q_{\phi}(\mathbf{z_i}|\mathbf{x_i})$ and pass it to the deep decoder to reconstruct the original signal $p_{\theta}(\mathbf{x_i}|\mathbf{z_i})$. The whole model is trained together using evidence lower bound objective:
\begin{equation}
  \label{eq:VAEObj}
  \log{p}(\mathbf{x}) \geq \mathcal{L}  = \mathop{\mathbb{E}_{q_{\phi}(\mathbf{y}|\mathbf{x})}}[\log{p_{\theta}(\mathbf{x}|\mathbf{y})}] - KL(q_{\phi}(\mathbf{y}|\mathbf{x})||p(\mathbf{y}))
\end{equation} 

The first term can be represented as the data reconstruction objective and metrics such as the mean squared error or negative log-likelihood as standard choices \citep{kingma2014autoencoding}. The second penalty term constrains the approximated posterior distribution of our latent variable $\mathbf{z}$ to be similar to a given prior $p(\mathbf{z})$ by minimizing the Kullback-Leibler divergence. 
For the prior $p(\mathbf{z})$, we consider isotropic Gaussian distribution, a standard option in VAE models. We explored other priors as well, which we discuss in Appendix \ref{sec:analysisMethodsAppendix}.
% \Prashnna{Prashnna: should we mention that we tried other priors like Beta-Bernoulli process as prior, but only standard isotropic Gaussian worked best for us?} \Ryan{Maybe it is worth it in the appendix? A brief little bit on what we tried throughout the process?}
% In standard VAE, isotropic Gaussian distribution is considered for the prior $p(\mathbf{z})$. 

After training the VAE, we consider the Self-Organizing Map (SOM) for unsupervised clustering. SOM is a type of neural network that produces a low-dimensional map (2D), a discretized representation of the input samples. In a typical SOM \citep{teuvo1988som} with a total of $M$ nodes, $\mathcal{V}=\{v_1,v_2,...,v_{M}\}$, on a two-dimensional grid, we learn a so-called weight vector $\mathbf{r}^{v}$ for each node $v\in\mathcal{V}$ such as to minimize the distance between each input data and its distance to the closest weight vector (also known as the Best Matching Unit (BMU)). During training, these node weight vectors are updated towards the input data without destroying the topological structure. This is done by dragging neighboring nodes alongside the node being updated. Once trained, the weight vectors on the SOM nodes provide prototypes of the input data, where similar data will be mapped to neighboring prototypes. The SOM can be seen as a generalized case of the K-Means algorithm, where K-Means represents a subclass where the nodes exert no influence on each other during updates.

\begin{figure*}
    	\includegraphics[width=0.5\columnwidth]{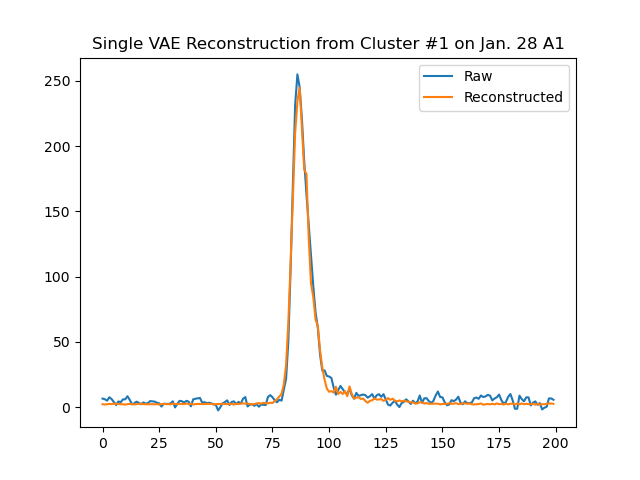}
	\includegraphics[width=0.5\columnwidth]{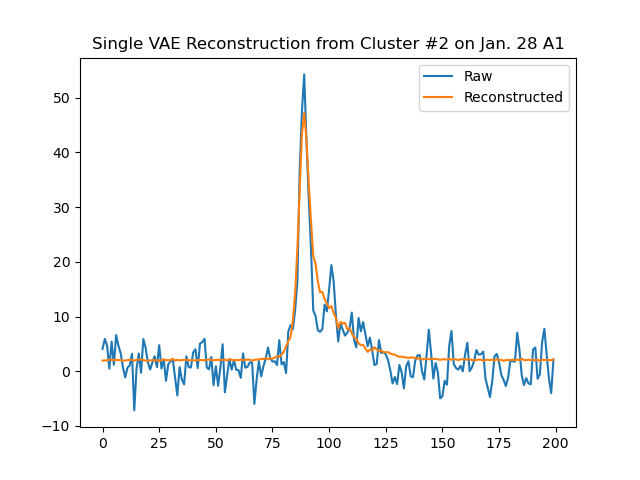}
    	\includegraphics[width=0.5\columnwidth]{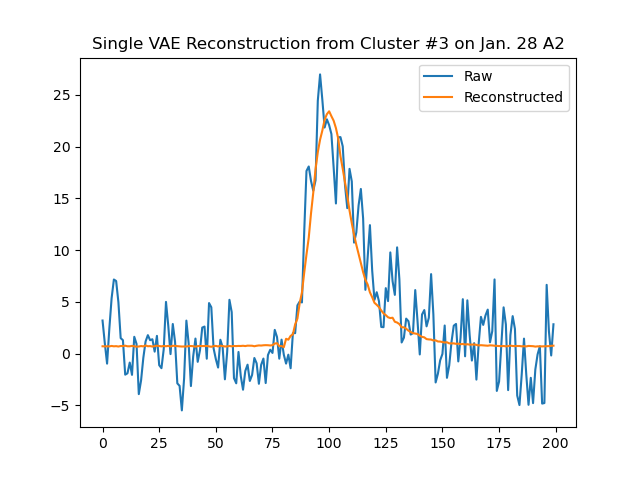}
	\includegraphics[width=0.5\columnwidth]{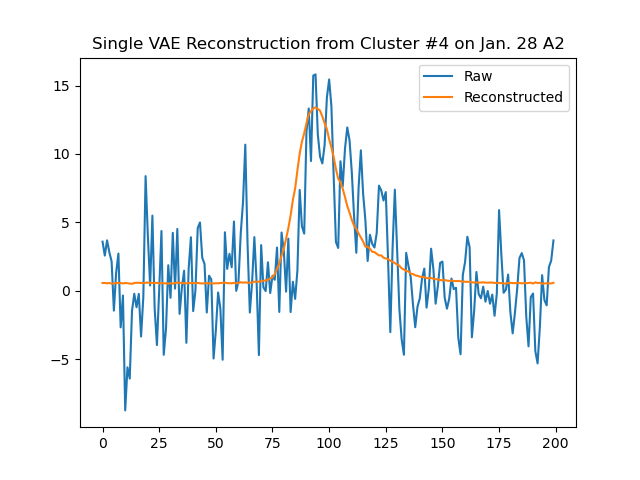}\\
	\includegraphics[width=0.5\columnwidth]{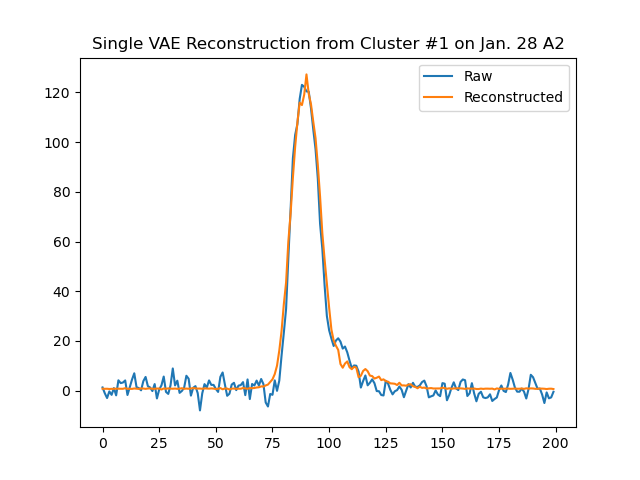}
	\includegraphics[width=0.5\columnwidth]{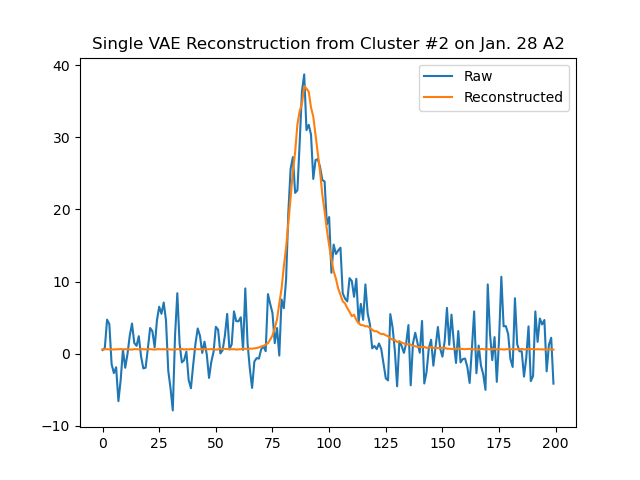}
    	\includegraphics[width=0.5\columnwidth]{signal83364Cluster3.png}
	\includegraphics[width=0.5\columnwidth]{signal65547Cluster4.png}
    \caption{Representative VAE reconstruction individual signals for the observations 2021-01-28 from each of the four SOM clusters.}
    \label{fig:VAEA12}
\end{figure*}

\begin{figure*}
    	\includegraphics[width=0.7\textwidth]{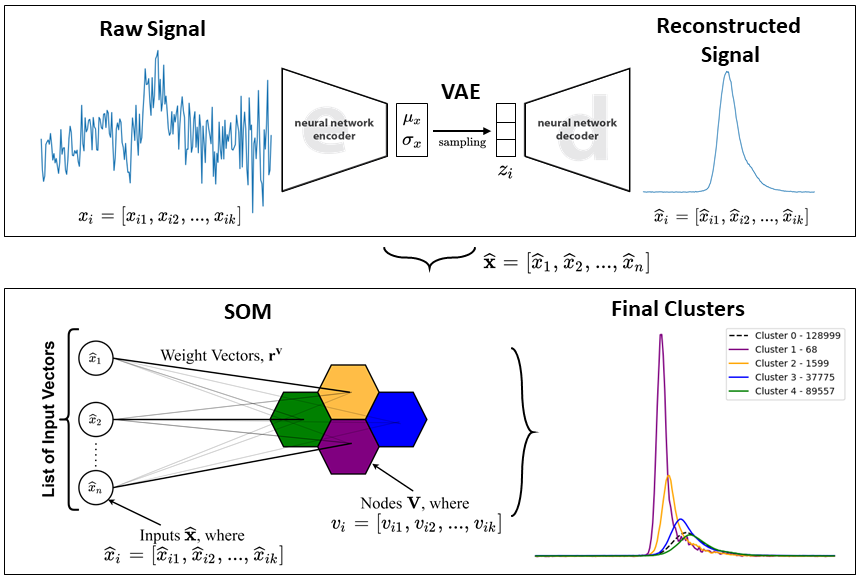}
   \caption{Schematics of the VAE reconstruction of all $N$ individual pulses signals and of SOM clustering.}
    \label{fig:VAESOM}
\end{figure*}

We consider both the latent representation of the signal and the original signal as the input data to the SOM algorithm. Since the original signal is noisy, we consider reconstructed signal $\hat{\mathbf{x_i}}$ $\sim$ $p_{\theta}(\mathbf{x_i}|\mathbf{z_i})$ (some samples are provided in Fig. \ref{fig:VAEA12}) as the input to the SOM, given that they offer a noiseless approximation of the original signal $\mathbf{x_i}$. We present the schematic diagram of VAE and usage of SOM for clustering in Fig. \ref{fig:VAESOM}.

%\Carlos{Description Results; Reconstruction pulses. Clustering. Correlations Width-Peak-Amplitude-Skew}

%\Carlos{This paragraph can also be moved to later in the paper}.
In Fig.~\ref{fig:VAEA12} we display representative original and VAE reconstructed pulses of each of the SOM clusters for the observation of the 2021-01-28 for each antenna. This data has been later used to train the VAE reconstruction on all other days of observation. This process shown robustness and close agreement with the untrained reconstruction of individual days of observation.

\subsubsection{Observations 2021-01-21}\label{sec:2021-01-21CS}

\begin{figure}\begin{center}
	\includegraphics[width=.9\columnwidth]{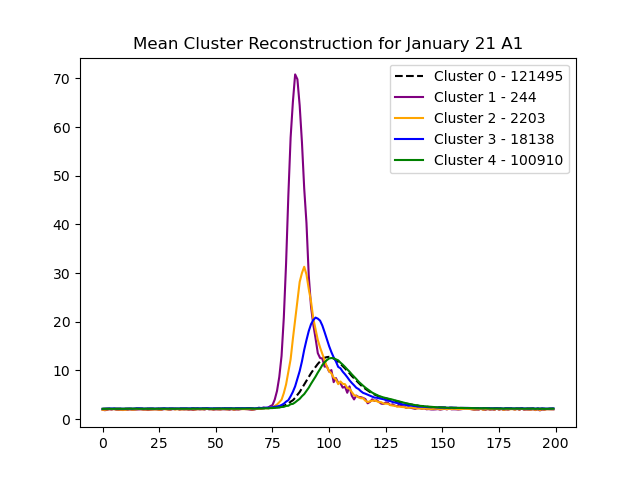}\vskip-10pt
	\includegraphics[width=.9\columnwidth]{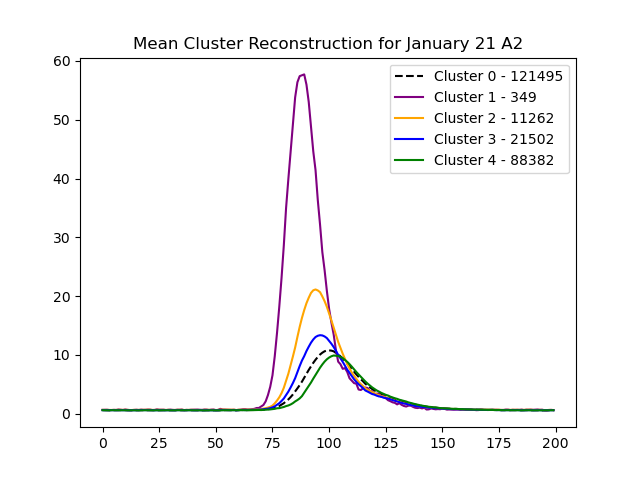}
    \caption{Distribution of the SOM clustering average signals for the observations 2021-01-21. Radio telescope A1 on top and Radio telescope A2 on bottom with respective VAE training performed on the January 28 observation.}
    \label{fig:1A12CS}
    \end{center}
\end{figure}

In Fig.~\ref{fig:1A12CS} we display the average value of the pulses in each SOM cluster for each Radio telescope observation. Those have been obtained by first applying a reconstruction of the raw pulses with the VAE technique, for which we have used the reconstruction of our best day of observation (according to the analysis in Appendix \ref{sec:RFIClean}) as a training case to apply to the rest of the days of observation. This training has been applied for each antenna individually. SOM allows us to specify then the number of clusters we seek to subdivide the whole set. We have studied several possible cases, 4, 6, 10, 25, 100, finding that the simplest four cluster analysis presents the most robust results. In fact we observe in Fig.~\ref{fig:1A12CS} the similarity of the mean clustering between the two antennas observations
(we also studied average vs. centroid's clusters). While the total number of members of each cluster changes, the qualitative mean value of the pulses (average pulse over the entire cluster) seems to be robust. We also note the visual displacement towards earlier times of the center and peak of each cluster average pulse for those with the largest amplitude with respect to each other and with respect to the total average pulse, in the plot denoted by the black dashed curve and labeled as 0. This would be the usual reference pulse we would obtain from the total observation. We also see that largest cluster (labeled as 1), counting a few hundred pulses, is a factor about 5 larger in amplitude for than the average pulse. Both features in qualitative agreement with the statistical analysis of the previous Sec.~\ref{sec:math}.

Table~\ref{tab:2021-01-21A12} give a quantitative account of the results displayed in Fig.~\ref{fig:1A12CS}. 
For each antenna's observation
we provide the number of pulses of each cluster \# pulses; peak location from the index of the maximum value in the pulse sequence;
peak height from the maximum value of the pulse sequence;
peak width done by first finding the maximum value of the sequence, then performing full-width half maximum of peak
(library used for this: \url{https://docs.scipy.org/doc/scipy/reference/generated/scipy.signal.peak_widths.html});
for the peak skew we evaluated the Fisher-Pearson coefficient of skewness 
(using the scipy for this computation \url{https://docs.scipy.org/doc/scipy/reference/generated/scipy.stats.skew.html}).
%; and finally MSE is the standard mean squared error $\sum_{i=1}^N(x_i-\bar{x})^2/N$.
The trend in the peak location towards later times is clear and above the quoted errors in its determination. Note that the cluster 4, the most numerous,
is (necessarily) showing a later arrival than the average total pulse, labeled as 0. A trend is also marginally seen in the width of the pulse, with
narrower values for the higher amplitude clusters while also carrying a higher skewness. Although specific numbers differ in both antennas observations,
the trends seem to be the same.

\begin{table*}
	\centering
	\caption{SOM Clustering for 2021-01-21 observations with Antennas 1 and 2.}
	\label{tab:2021-01-21A12}
	 \begin{tabular}{l|llllll}
	 	 \hline
	 	 &Cluster \#& 0 & 1 & 2 & 3 & 4 \\ 
	 	 \hline 
	 	 &\# pulses & 121495 & 244 & 2203 & 18138 & 100910 \\ 
	 	 &peak loc & $99.28 \pm 5.70$ & $84.99 \pm 1.89$ & $87.75 \pm 2.15$ & $93.28 \pm 3.13$ & $100.64 \pm 5.05$ \\ 
	 A1	 &peak height & $16.64 \pm 10.40$ & $95.24 \pm 49.60$ & $36.47 \pm 22.83$ & $24.70 \pm 14.26$ & $14.57 \pm 6.07$ \\ 
	 	 &peak width & $17.76 \pm 4.53$ & $6.05 \pm 1.13$ & $9.73 \pm 3.00$ & $11.70 \pm 3.32$ & $18.74 \pm 4.00$ \\ 
	 	 &peak skew & $2.49 \pm 0.42$ & $4.68 \pm 0.53$ & $3.51 \pm 0.56$ & $2.90 \pm 0.41$ & $2.38 \pm 0.31$ \\ 
%	 	 &MSE & $0.00011 \pm 0.00019$ & $0.08197 \pm 0.27763$ & $0.00687 \pm 0.01703$ & $0.00077 \pm 0.00142$ & $0.00013 \pm 0.00021$\\ 
	 	 \hline
	 	 &Cluster \#& 0 & 1 & 2 & 3 & 4 \\ 
	 	 \hline 
	 	 &\# pulses & 121495 & 349 & 11262 & 21502 & 88382 \\ 
	 	 &peak loc & $100.31 \pm 4.81$ & $87.75 \pm 2.84$ & $93.55 \pm 3.25$ & $95.30 \pm 2.78$ & $102.44 \pm 3.38$ \\ 
	 A2	 &peak height & $12.40 \pm 6.16$ & $64.91 \pm 23.05$ & $23.17 \pm 7.25$ & $13.90 \pm 2.91$ & $10.45 \pm 3.37$ \\ 
	 	 &peak width & $17.44 \pm 1.25$ & $14.52 \pm 0.55$ & $17.51 \pm 1.00$ & $21.58 \pm 0.97$ & $23.53 \pm 1.83$ \\ 
	 	 &peak skew & $2.09 \pm 0.22$ & $2.98 \pm 0.10$ & $2.46 \pm 0.13$ & $2.17 \pm 0.13$ & $2.01 \pm 0.18$ \\ 
%	 	 &MSE & $0.00009 \pm 0.00013$ & $0.03975 \pm 0.07801$ & $0.00098 \pm 0.00151$ & $0.00050 \pm 0.00074$ & $0.00012 \pm 0.00017$\\ 
		\hline
	\end{tabular}
\end{table*}

\subsubsection{Observations 2021-01-24}\label{sec:2021-01-24CS}

\begin{figure}\begin{center}
	\includegraphics[width=.9\columnwidth]{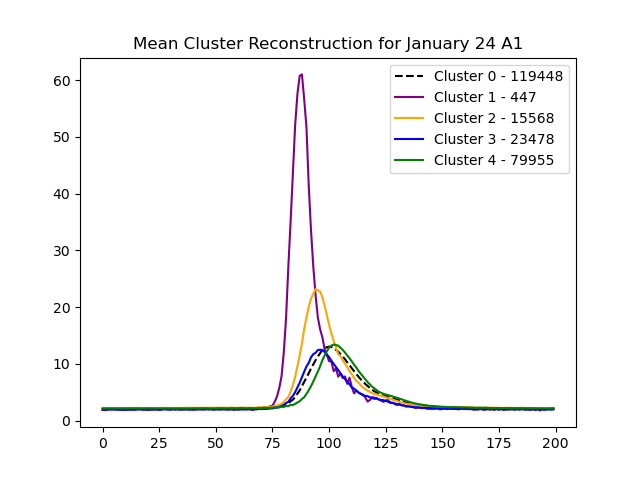}\vskip-10pt
	\includegraphics[width=.9\columnwidth]{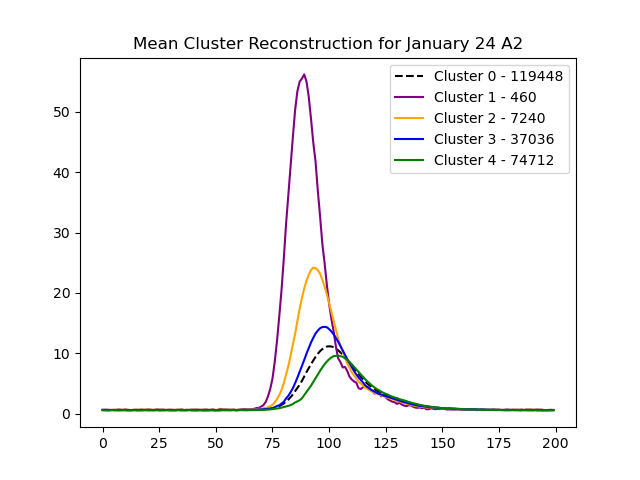}
    \caption{Distribution of the SOM clustering average signals for the observations 2021-01-24. Radio telescope A1 on top and Radio telescope A2 on bottom with respective VAE training performed on the January 28 observation.}
    \label{fig:2A12CS}
    \end{center}
\end{figure}

In Fig.~\ref{fig:2A12CS} we display the average value of the pulses in each SOM cluster for each Radio telescope observation. Those have been obtained by first applying a reconstruction of the raw pulses with the VAE technique, for which we have used the reconstruction of our best day of observation (according to the analysis in Appendix \ref{sec:RFIClean}) as a training case to apply to the rest of the days of observation. This training has been applied for each antenna individually. SOM allows us to specify then the number of clusters we seek to subdivide the whole set. We have studied several possible cases, 4, 6, 10, 25, 100, finding that the simplest four cluster choice represents the most robust results. We observe again in Fig.~\ref{fig:2A12CS} the similarity of the mean clustering between the two antennas observations. While the total number of members of each cluster changes, the qualitative mean value of the pulses (average pulse over the entire cluster) seems to be robust. We also note the visual displacement towards earlier times of the center and peak of each cluster average pulse for those with the largest amplitude with respect to each other and with respect to the total average pulse, in the plot denoted by the black dashed curve and labeled as 0. This is the usual reference pulse we obtain from the total observation. We also see that the largest amplitude cluster (labeled as 1), counting a few hundred pulses, is a factor about 5 larger in amplitude than the average pulse. Both features are in qualitative agreement with the statistical analysis of the previous section~\ref{sec:math}. Another interesting feature of the average pulses for each cluster is particularly evident in the A2 observations: The right shoulder of the pulses superpose (this is also evident in the Jan. 21 observations), indicating a sort of superposition of components at earlier arrival times.

Table~\ref{tab:2021-01-24A12} give a quantitative account of the results displayed in Fig.~\ref{fig:2A12CS}. 
For each antenna's observation on 2021-01-24,
we provide the number of pulses of each cluster \# pulses; peak location from the index of the maximum value in the pulse sequence;
peak height from the maximum value of the pulse sequence;
peak width done by first finding the maximum value of the sequence, then performing full-width half maximum of peak;
%(library used for this: \url{https://docs.scipy.org/doc/scipy/reference/generated/scipy.signal.peak_widths.html});
for the peak skew we evaluated the Fisher-Pearson coefficient of skewness. %; 
%(using the scipy for this computation \url{https://docs.scipy.org/doc/scipy/reference/generated/scipy.stats.skew.html}); 
%and finally MSE is the standard mean squared error. %$\sum_{i=1}^N(x_i-\bar{x})^2/N$.
The trend in the peak location towards later times is clear and above the quoted errors in its determination. Note that the cluster 4, the most numerous,
is (necessarily) showing a later arrival than the average total pulse, labeled as 0. A trend is also marginally seen in the width of the pulse, with
narrower values for the higher amplitude clusters while also carrying a higher skewness. Although specific numbers differ in both antennas observations,
the trends seem to be the same.

\begin{table*}
	\centering
	\caption{SOM Clustering for 2021-01-24 with Antennas 1 and 2.}
	\label{tab:2021-01-24A12}
	 \begin{tabular}{l|llllll}
	 	 \hline
	 	 &Cluster \#& 0 & 1 & 2 & 3 & 4 \\ 
	 	 \hline 
	 	 &\# pulses & 119448 & 447 & 15568 & 23478 & 79955 \\ 
	 	 &peak loc & $99.77 \pm 5.64$ & $86.46 \pm 2.26$ & $93.25 \pm 3.51$ & $94.62 \pm 3.52$ & $102.63 \pm 4.03$ \\ 
	A1 	 &peak height & $17.03 \pm 11.10$ & $86.96 \pm 50.04$ & $29.56 \pm 17.44$ & $13.96 \pm 4.21$ & $15.09 \pm 6.41$ \\ 
	 	 &peak width & $17.39 \pm 4.32$ & $6.07 \pm 1.14$ & $10.84 \pm 3.23$ & $17.33 \pm 3.64$ & $18.43 \pm 3.77$ \\ 
	 	 &peak skew & $2.50 \pm 0.43$ & $4.55 \pm 0.58$ & $3.07 \pm 0.48$ & $2.47 \pm 0.27$ & $2.38 \pm 0.32$ \\ 
%	 	 &MSE & $0.00011 \pm 0.00018$ & $0.04149 \pm 0.13950$ & $0.00090 \pm 0.00176$ & $0.00054 \pm 0.00089$ & $0.00015 \pm 0.00026$\\ 
	 	 \hline 
	 	 &Cluster \#& 0 & 1 & 2 & 3 & 4 \\ 
	 	 \hline 
	 	 &\# pulses & 119448 & 460 & 7240 & 37036 & 74712 \\ 
	 	 &peak loc & $100.53 \pm 4.80$ & $88.15 \pm 2.62$ & $92.90 \pm 2.91$ & $96.91 \pm 3.08$ & $103.14 \pm 3.38$ \\ 
	A2	 &peak height & $12.93 \pm 6.67$ & $62.24 \pm 24.60$ & $26.32 \pm 8.78$ & $15.25 \pm 3.98$ & $10.17 \pm 3.07$ \\ 
	 	 &peak width & $21.92 \pm 1.99$ & $14.76 \pm 0.56$ & $17.22 \pm 0.93$ & $21.61 \pm 1.50$ & $22.73 \pm 1.36$ \\ 
	 	 &peak skew & $2.11 \pm 0.22$ & $2.95 \pm 0.11$ & $2.52 \pm 0.13$ & $2.22 \pm 0.13$ & $2.01 \pm 0.18$ \\ 
%	 	 &MSE & $0.00009 \pm 0.00013$ & $0.03049 \pm 0.05987$ & $0.00155 \pm 0.00241$ & $0.00029 \pm 0.00043$ & $0.00014 \pm 0.00021$\\ 
	 \hline 
	 \end{tabular} 
\end{table*} 

\subsubsection{Observations 2021-01-28}\label{sec:2021-01-28CS}

\begin{figure}\begin{center}
	\includegraphics[width=.9\columnwidth]{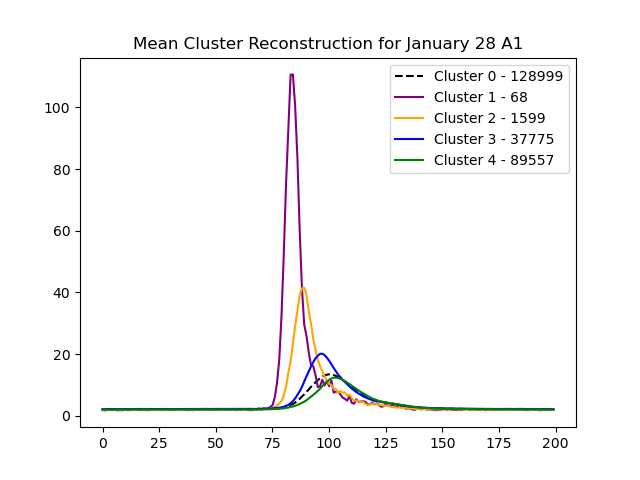}\vskip-10pt
	\includegraphics[width=.9\columnwidth]{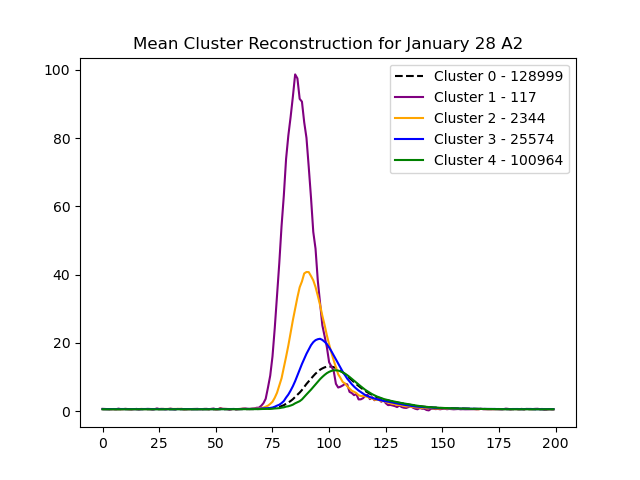}
    \caption{Distribution of the SOM clustering average signals for the observations 2021-01-28. Radio telescope A1 on top and Radio telescope A2 on bottom. VAE training has been performed on these observations.}
    \label{fig:3A12CS}    
\end{center}
\end{figure}

In Fig.~\ref{fig:3A12CS} we display the average value of the pulses in each of the four SOM cluster for each Radio telescope observation.
Those have been obtained by first applying a reconstruction of the raw pulses with the VAE technique.
Accordingly to the analysis in Appendix \ref{sec:RFIClean} this observations of the 2021-01-28 are the best regarding its signal-to-noise ratio
and are the ones used (for each antenna) for VAE pulse reconstruction training for the rest of the observing days.
We observe again in Fig.~\ref{fig:3A12CS} the similarity of the mean clustering between the two antennas observations. 
While the total number of members of each cluster changes, the qualitative mean value of the pulses (average pulse over the cluster) seems to be robust. 
We also note again the visual displacement towards earlier times of the center and peak of each cluster average pulse 
with the largest amplitude with respect to each other and with respect to the total average pulse.
Most notably here is the amplification factor of the cluster 1 with respect to the total average pulse nearing a factor ten, although being less
numerous than during previous observations days. Note also that in the A2 observations there seems to be a superposition of the cluster pulses 
profile at late times. This feature is also true in the previous two days of observations (and the March observation displayed in the Appendix \ref{sec:2021-03-29CS} in A2, clearly displaying the earlier arrival of the higher amplitude components.

Table~\ref{tab:2021-01-28A12} gives a quantitative account of the results displayed in Fig.~\ref{fig:2A12CS}. 
For each antenna's observation on 2021-01-28,
we provide the number of pulses of each cluster \# pulses; peak location from the index of the maximum value in the pulse sequence;
peak height from the maximum value of the pulse sequence;
peak width done by first finding the maximum value of the sequence, then performing full-width half maximum of peak;
%(library used for this: \url{https://docs.scipy.org/doc/scipy/reference/generated/scipy.signal.peak_widths.html});
for the peak skew we evaluated the Fisher-Pearson coefficient of skewness. %; 
%(using the scipy for this computation \url{https://docs.scipy.org/doc/scipy/reference/generated/scipy.stats.skew.html}); 
%and finally MSE is the standard mean squared error. %$\sum_{i=1}^N(x_i-\bar{x})^2/N$.
We observe again that the trend in the peak location towards later times is clear and above the quoted errors in its determination. 
Note that the cluster 4, the most numerous,
is (necessarily) showing a later arrival than the average total pulse, labeled as 0. A trend is also marginally seen in the width of the pulse, with
narrower values for the higher amplitude clusters while also carrying a higher skewness. Although specific numbers differ in both antennas observations,
the trends seem to be the same showing the robustness of the effect. 

\begin{table*}
	\centering
	\caption{SOM Clustering for 2021-01-28 with Antennas 1 and 2. This set was used for A1 and A2 VAE training.}
	\label{tab:2021-01-28A12}
	 \begin{tabular}{l|llllll}
	 	 \hline
	 	 &Cluster \#& 0 & 1 & 2 & 3 & 4 \\ 
	 	 \hline 
	 	 &\# pulses & 128999 & 68 & 1599 & 37775 & 89557 \\ 
	 	 &peak loc & $99.52 \pm 5.82$ & $83.57 \pm 1.47$ & $87.21 \pm 2.24$ & $95.19 \pm 3.90$ & $101.57 \pm 5.22$ \\ 
	 A1	 &peak height & $17.94 \pm 12.39$ & $132.11 \pm 46.02$ & $53.42 \pm 39.95$ & $25.15 \pm 14.62$ & $14.18 \pm 5.70$ \\ 
	 	 &peak width & $17.27 \pm 4.54$ & $5.72 \pm 0.61$ & $6.98 \pm 1.91$ & $12.52 \pm 3.54$ & $18.85 \pm 3.79$ \\ 
	 	 &peak skew & $2.54 \pm 0.44$ & $5.06 \pm 0.41$ & $3.91 \pm 0.65$ & $2.87 \pm 0.42$ & $2.38 \pm 0.30$ \\ 
%	 	 &MSE & $0.00010 \pm 0.00019$ & $0.36319 \pm 1.65015$ & $0.01011 \pm 0.02502$ & $0.00037 \pm 0.00070$ & $0.00015 \pm 0.00025$\\ 
	 	 \hline
	 	 &Cluster \#& 0 & 1 & 2 & 3 & 4 \\ 
	 	 \hline 
	 	 &\# pulses & 128999 & 117 & 2344 & 25574 & 100964 \\ 
	 	 &peak loc & $100.45 \pm 5.04$ & $85.74 \pm 2.53$ & $89.95 \pm 2.37$ & $94.85 \pm 2.95$ & $102.13 \pm 4.07$ \\ 
	 A2	 &peak height & $15.64 \pm 8.54$ & $104.39 \pm 24.05$ & $44.33 \pm 18.04$ & $22.83 \pm 7.61$ & $13.05 \pm 4.71$ \\ 
	 	 &peak width & $21.81 \pm 2.62$ & $14.08 \pm 0.59$ & $15.71 \pm 0.74$ & $17.97 \pm 0.99$ & $22.64 \pm 2.21$ \\ 
	 	 &peak skew & $2.20 \pm 0.23$ & $3.12 \pm 0.08$ & $2.79 \pm 0.12$ & $2.43 \pm 0.14$ & $2.13 \pm 0.19$ \\ 
%	 	 &MSE & $0.00008 \pm 0.00012$ & $0.13918 \pm 0.30824$ & $0.00518 \pm 0.00886$ & $0.00042 \pm 0.00065$ & $0.00010 \pm 0.00016$\\ 
	 	 \hline 
	 \end{tabular} 
\end{table*} 

%\Carlos{Here we discuss the effects of training on one day, clustering 4p vs. 6, 25, and average vs. centroids.}
%\Carlos{Here a unified analysis with the CS and Math approaches}

In the next section we will provide a simple geometrical interpretation of these four components and its properties, in particular the appearance of the earlier high amplitude pulses and its relatively narrower properties, particularly in one polarization.

\subsection{Geometrical modeling of the cluster components}\label{sec:magneto}

\begin{figure}
	\includegraphics[width=.9\columnwidth]{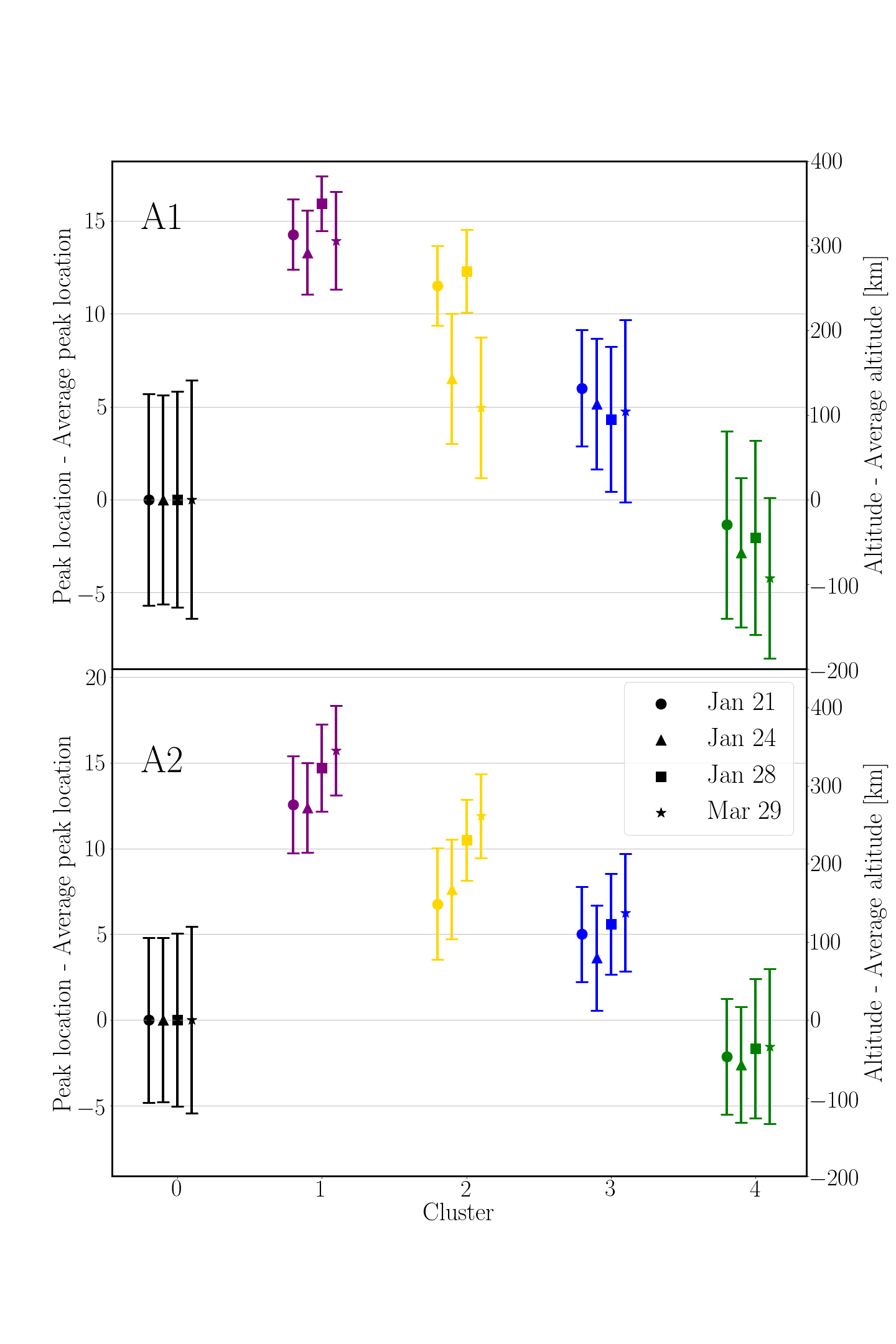}
    \caption{Peak location and magnetosphere altitude, with the corresponding error bars, for each of the pulse clusters presented in Sec.~\ref{sec:CS}.}
    \label{fig:altitudes}
\end{figure}

As presented Sec.~\ref{sec:scintillations}, the excess of high amplitude pulses cannot be explained solely due to effects of scintillation. In Sec.~\ref{sec:CS} we also find that each cluster has a different average peak location, with brighter pulses arriving earlier. Therefore, following the classic work of \cite{Downs1983},
we may attribute these variations in pulse amplitude and location to different altitudes in the neutron star magnetosphere where the pulses of each cluster are emitted. To this end, we measure the displacement of each cluster peak location relative to the average pulse location, and then relate those pulse displacements to differences in the emission altitude by
\begin{equation}
    h - \Bar{h} = \frac{x - \Bar{x}}{n_\mathrm{bins}} c P ,
\end{equation}
\noindent where $x$ and $h$ are the cluster peak location and altitude in the magnetosphere, $\Bar{x}$ and $\Bar{h}$ are the average peak location and the average altitude (corresponding to cluster 0), $n_\mathrm{bins}$ is the number of time bins in each pulse (in our case, $1220$), and $P$ is the pulsar rotational period. 
Fig.\ref{fig:altitudes} displays the results of applying this model to each of the four days of observation for each antenna. The right hand side ordinate gives
the components distances to the average pulse reference height in the pulsar magnetosphere. We note the consistency between the components for each of the four days and for each individual antenna's observations. The four components appear to be almost equidistant (this maybe an effect of the SOM clustering method) and roughly of the order of $\sim100$ kilometers.

More quantitative results are presented in Table.~\ref{table:altitude}. Cluster 4, naturally appears with negative heights with respect to the average pulse location since it has to compensate the positive contributions of the larger amplitude clusters. The shift is relatively small since cluster 4 is the overwhelming most numerous cluster, except of the special configuration of A1 during March 29th observation.

\begin{table}
    \centering
    \caption{Magnetosphere altitude in kilometers for each of the pulse clusters presented in Sec.~\ref{sec:CS}.}\label{table:altitude}
    \begin{tabular}{cc|cccc}
    \hline \hline
     & Cluster & 1 & 2 & 3 & 4 \\
    \hline
    \parbox[c]{0mm}{\multirow{4}{*}{\rotatebox[origin=c]{90}{A1}}}
    & Jan 21 &  313.7$\pm$41.5 & 253.1$\pm$47.2 & 131.7$\pm$68.7 & -29.9$\pm$110.9 \\
    & Jan 24 & 292.2$\pm$49.6 & 143.1$\pm$77.1 & 113.1$\pm$77.3 & -62.8$\pm$88.5 \\
    & Jan 28 & 350.1$\pm$32.3 & 270.2$\pm$49.2 & 95.1$\pm$85.6 & -45.0$\pm$114.6 \\
    & Mar 29 & 306.0$\pm$57.7 & 108.9$\pm$83.0 & 104.7$\pm$108.0 & -92.6$\pm$95.1 \\
    \hline
     \parbox[c]{0mm}{\multirow{4}{*}{\rotatebox[origin=c]{90}{A2}}} 
    & Jan 21 & 275.7$\pm$62.3 & 148.4$\pm$71.3 & 110.0$\pm$61.0 & -46.8$\pm$74.2 \\
    & Jan 24 & 271.8$\pm$57.5 & 167.5$\pm$63.9 & 79.5$\pm$67.6 & -57.3$\pm$74.2 \\
    & Jan 28 & 322.9$\pm$55.5 & 230.5$\pm$52.0 & 122.9$\pm$64.8 & -36.9$\pm$89.3 \\
    & Mar 29 & 345.3$\pm$57.3 & 261.2$\pm$53.8 & 137.4$\pm$75.3 & -33.8$\pm$99.4 \\
    \hline \hline
    \end{tabular}
\end{table}

The location of the four components seems to roughly follow those studied in \cite{Downs1983} (See their Fig. 16), with distances between components ranging from 100 km to 500 km. The coincidence on the scale of distances is perhaps expected from the decomposition of the main pulse into ordered components (See Fig. 9 in \cite{Downs1983}).
Further studies of the Vela radio emission region \cite{1997ApJ...483L..53G} also estimate its size in about 500 km from the modulation of the pulsar's scintillation.
While more recent and detailed studies \cite{2012ApJ...758....7G} of the scintillation properties at 18 cm wavelength describe the early part of the pulse from $-400$~km to the later part of the pulse to $+800$~km with respect to its peak. While using novel scintillation statistics techniques at 760~MHz \cite{2012ApJ...758....8J} find that the radio emission altitude for the Vela pulsar is less than 340~km.

\section{Conclusions}\label{sec:conclusions}

%The last numbered section should briefly summarise what has been done, and describe
%the final conclusions which the authors draw from their work.

In this paper we have performed a prototypical study of individual pulses of Vela using standard machine learning techniques. We have assumed an agnostic approach to the data and applied clustering DBSCAN and SOM techniques independently to analyze the top amplitude pulses and a global dissection of the pulses respectively. The former provided us with a way to stratify those mini-giant pulses that are relatively rare but still appear in each day of observation. The SOM techniques are applied after a VAE reconstruction of the pulses with a training on the best day of observation (based on a S/N criteria). We have found a robustness of the clustering this way over the days of observations and antennas, which represents an important cross-check of our systematics.

The simple statistical distribution of the pulse amplitudes was found to be roughly fitted by the scintillation function (\ref{eq:scintillation}) (a $\chi^2$-distribution) and we have been able to provide effective $n_{iss}$ indices with rather high values, ranging from around 7 to 10. While
this fit is rather good (See Fig.~\ref{fig:scintillations}) it leaves a residual high amplitude set of pulses that cannot be simply explained by scintillation, suggesting there is an intrinsic pulsar emission of those mini-giant pulses. We couple this to the observation of those pulses arriving earlier than the average pulse during the over 3 hours of our daily observations (See Figs.~\ref{fig:12A12},\ref{fig:56A12},\ref{fig:910A12}) to support a simple model based on the height location of the emission regions in the pulsar magnetosphere for those large amplitude pulses. We use the SOM clustering to determine four relevant sets of pulses characteristics that seem to linearly array along the emission regions of the magnetosphere (See Fig.~\ref{fig:altitudes}) separated by roughly $100$km each. 
Those results, based on Machine learning clustering, seem to agree with specific studies of the Vela pulses and its scintillation responses.

This first study provides us with a proof-of-principle technique to be applied more massively to daily observations of Vela. 
Our studies for the distribution of pulses amplitudes, widths and peaks timing have been performed in a relatively calm period of Vela, in between major glitches. Of particular interest is to study if any of these properties suffers a change or provides precursor information of the mayor glitches Vela suffers every 2--3 years. The most recent one took place on July 21st 2021 \cite{2021ATel14806....1S} (while this paper was nearing completion) and the previous one was
on February 1st, 2019 (See \cite{Gancio2020} and references there in).

The fact that high amplitude pulses are also thinner, could have implications for their use for pulsar timing \cite{2015MNRAS.452..607K}. In particular, if this features is also carried out to millisecond pulsars, like PSR J0437$-$4715, this can further improve its timing to detect gravitational waves from supermassive binary black holes \cite{Fiscella:2020jey} and will be the subject of study of a forthcoming paper. Another pulsar of interest for
single pulses study in the southern hemisphere that is being observed at IAR is PSR B1641$-$45\,/\,J1644$-$4559 \cite{2004MNRAS.348.1229J}.

\section*{Acknowledgements}

%The Acknowledgements section is not numbered. Here you can thank helpful colleagues, acknowledge funding agencies, telescopes and facilities used etc. Try to keep it short.

We thank Yogesh Maan for numerous beneficial discussions about the best use of RFIClean,
Cameron Knight and Carolina Negrelli for participating in the early studies involving machine learning techniques, and Adolfo Simaz Bunzel and Ezequiel Zubieta in the observations reduction pipeline.
COL gratefully acknowledge the National Science Foundation (NSF) for financial support from Grants No.\ PHY-1912632 and PHY-2018420. JAC and FG acknowledge support by PIP 0102 (CONICET). JAC is CONICET researcher. This work received financial support from PICT-2017-2865 (ANPCyT).
 JAC was also supported by grant PID2019-105510GB-C32/AEI/10.13039/501100011033 from the Agencia Estatal de Investigaci\'on of the Spanish Ministerio de Ciencia, Innovaci\'on y Universidades, and by Consejer\'{\i}a de Econom\'{\i}a, Innovaci\'on, Ciencia y Empleo of Junta de Andaluc\'{\i}a as research group FQM-322, as well as FEDER funds.

%\Carlos{Here goes other grants acknowledgements.}

%%%%%%%%%%%%%%%%%%%%%%%%%%%%%%%%%%%%%%%%%%%%%%%%%%
\section*{Data Availability}
%The inclusion of a Data Availability Statement is a requirement for articles published in MNRAS. Data Availability Statements provide a standardised format for readers to understand the availability of data underlying the research results described in the article. The statement may refer to original data generated in the course of the study or to third-party data analysed in the article. The statement should describe and provide means of access, where possible, by linking to the data or providing the required accession numbers for the relevant databases or DOIs.

Data generated by our calculations or observations are available from the corresponding authors upon reasonable request.

%%%%%%%%%%%%%%%%%%%% REFERENCES %%%%%%%%%%%%%%%%%%

% The best way to enter references is to use BibTeX:

\bibliographystyle{mnras}
\bibliography{references,biblio} % if your bibtex file is called example.bib

\begin{thebibliography}{}
\makeatletter
\relax
\def\mn@urlcharsother{\let\do\@makeother \do\$\do\&\do\#\do\^\do\_\do\%\do\~}
\def\mn@doi{\begingroup\mn@urlcharsother \@ifnextchar [ {\mn@doi@}
  {\mn@doi@[]}}
\def\mn@doi@[#1]#2{\def\@tempa{#1}\ifx\@tempa\@empty \href
  {http://dx.doi.org/#2} {doi:#2}\else \href {http://dx.doi.org/#2} {#1}\fi
  \endgroup}
\def\mn@eprint#1#2{\mn@eprint@#1:#2::\@nil}
\def\mn@eprint@arXiv#1{\href {http://arxiv.org/abs/#1} {{\tt arXiv:#1}}}
\def\mn@eprint@dblp#1{\href {http://dblp.uni-trier.de/rec/bibtex/#1.xml}
  {dblp:#1}}
\def\mn@eprint@#1:#2:#3:#4\@nil{\def\@tempa {#1}\def\@tempb {#2}\def\@tempc
  {#3}\ifx \@tempc \@empty \let \@tempc \@tempb \let \@tempb \@tempa \fi \ifx
  \@tempb \@empty \def\@tempb {arXiv}\fi \@ifundefined
  {mn@eprint@\@tempb}{\@tempb:\@tempc}{\expandafter \expandafter \csname
  mn@eprint@\@tempb\endcsname \expandafter{\@tempc}}}

\bibitem[\protect\citeauthoryear{{Agarwal}, {Aggarwal}, {Burke-Spolaor},
  {Lorimer}  \& {Garver-Daniels}}{{Agarwal} et~al.}{2020}]{2020MNRAS.497.1661A}
{Agarwal} D.,  {Aggarwal} K.,  {Burke-Spolaor} S.,  {Lorimer} D.~R.,
  {Garver-Daniels} N.,  2020, \mn@doi [\mnras] {10.1093/mnras/staa1856}, \href
  {https://ui.adsabs.harvard.edu/abs/2020MNRAS.497.1661A} {497, 1661}

\bibitem[\protect\citeauthoryear{{Andersson}, {Glampedakis}, {Ho}  \&
  {Espinoza}}{{Andersson} et~al.}{2012}]{Andersson:2012iu}
{Andersson} N.,  {Glampedakis} K.,  {Ho} W.~C.~G.,   {Espinoza} C.~M.,  2012,
  \mn@doi [\prl] {10.1103/PhysRevLett.109.241103}, \href
  {https://ui.adsabs.harvard.edu/abs/2012PhRvL.109x1103A} {109, 241103}

\bibitem[\protect\citeauthoryear{Ankerst, Breunig, Kriegel  \& Sander}{Ankerst
  et~al.}{1999}]{optics99}
Ankerst M.,  Breunig M.~M.,  Kriegel H.-P.,   Sander J.,  1999, in Proc. ACM
  SIGMOD Int. Conf. on Management of Data (SIGMOD'99). Philadelphia, PA, pp
  49--60

\bibitem[\protect\citeauthoryear{{Astropy Collaboration} et~al.,}{{Astropy
  Collaboration} et~al.}{2013}]{astropy2013}
{Astropy Collaboration} et~al., 2013, \mn@doi [\aap]
  {10.1051/0004-6361/201322068}, \href
  {https://ui.adsabs.harvard.edu/abs/2013A&A...558A..33A} {558, A33}

\bibitem[\protect\citeauthoryear{{Astropy Collaboration} et~al.,}{{Astropy
  Collaboration} et~al.}{2018}]{astropy2018}
{Astropy Collaboration} et~al., 2018, \mn@doi [\aj] {10.3847/1538-3881/aabc4f},
  \href {https://ui.adsabs.harvard.edu/abs/2018AJ....156..123A} {156, 123}

\bibitem[\protect\citeauthoryear{{Bethapudi} \& {Desai}}{{Bethapudi} \&
  {Desai}}{2018}]{2018A&C....23...15B}
{Bethapudi} S.,  {Desai} S.,  2018, \mn@doi [Astronomy and Computing]
  {10.1016/j.ascom.2018.02.002}, \href
  {https://ui.adsabs.harvard.edu/abs/2018A&C....23...15B} {23, 15}

\bibitem[\protect\citeauthoryear{{Cairns}, {Johnston}  \& {Das}}{{Cairns}
  et~al.}{2001}]{Cairns+2001}
{Cairns} I.~H.,  {Johnston} S.,   {Das} P.,  2001, \mn@doi [\apjl]
  {10.1086/338359}, \href
  {https://ui.adsabs.harvard.edu/abs/2001ApJ...563L..65C} {563, L65}

\bibitem[\protect\citeauthoryear{{Chamel}}{{Chamel}}{2013}]{Chamel:2012ae}
{Chamel} N.,  2013, \mn@doi [\prl] {10.1103/PhysRevLett.110.011101}, \href
  {https://ui.adsabs.harvard.edu/abs/2013PhRvL.110a1101C} {110, 011101}

\bibitem[\protect\citeauthoryear{{Cordes}}{{Cordes}}{2000}]{2000astro.ph..7233C}
{Cordes} J.~M.,  2000, arXiv e-prints, \href
  {https://ui.adsabs.harvard.edu/abs/2000astro.ph..7233C} {pp
  astro--ph/0007233}

\bibitem[\protect\citeauthoryear{{Cordes} \& {Chernoff}}{{Cordes} \&
  {Chernoff}}{1997}]{Cordes:1997my}
{Cordes} J.~M.,  {Chernoff} D.~F.,  1997, \mn@doi [\apj] {10.1086/304179},
  \href {https://ui.adsabs.harvard.edu/abs/1997ApJ...482..971C} {482, 971}

\bibitem[\protect\citeauthoryear{{Cordes}, {Downs}  \&
  {Krause-Polstorff}}{{Cordes} et~al.}{1988}]{Cordes+1988}
{Cordes} J.~M.,  {Downs} G.~S.,   {Krause-Polstorff} J.,  1988, \mn@doi [\apj]
  {10.1086/166518}, \href
  {https://ui.adsabs.harvard.edu/abs/1988ApJ...330..847C} {330, 847}

\bibitem[\protect\citeauthoryear{{Devine}, {Goseva-Popstojanova}  \&
  {McLaughlin}}{{Devine} et~al.}{2016}]{2016MNRAS.459.1519D}
{Devine} T.~R.,  {Goseva-Popstojanova} K.,   {McLaughlin} M.,  2016, \mn@doi
  [\mnras] {10.1093/mnras/stw655}, \href
  {https://ui.adsabs.harvard.edu/abs/2016MNRAS.459.1519D} {459, 1519}

\bibitem[\protect\citeauthoryear{Dodson, Lewis  \& McCulloch}{Dodson
  et~al.}{2007}]{Dodson+2007}
Dodson R.,  Lewis D.,   McCulloch P.,  2007, \mn@doi [Astrophysics and Space
  Science] {10.1007/s10509-007-9372-4}, 308, 585

\bibitem[\protect\citeauthoryear{{Eatough}, {Keane}  \& {Lyne}}{{Eatough}
  et~al.}{2009}]{Eatough2009}
{Eatough} R.~P.,  {Keane} E.~F.,   {Lyne} A.~G.,  2009, \mn@doi [\mnras]
  {10.1111/j.1365-2966.2009.14524.x}, \href
  {https://ui.adsabs.harvard.edu/abs/2009MNRAS.395..410E} {395, 410}

\bibitem[\protect\citeauthoryear{Ester, Kriegel, Sander  \& Xu}{Ester
  et~al.}{1996}]{ester1996densitybased}
Ester M.,  Kriegel H.-P.,  Sander J.,   Xu X.,  1996, in Proc. of 2nd
  International Conference on Knowledge Discovery and. pp 226--231

\bibitem[\protect\citeauthoryear{{Gancio} et~al.,}{{Gancio}
  et~al.}{2020}]{Gancio2020}
{Gancio} G.,  et~al., 2020, \mn@doi [\aap] {10.1051/0004-6361/201936525}, \href
  {https://ui.adsabs.harvard.edu/abs/2020A&A...633A..84G} {633, A84}

\bibitem[\protect\citeauthoryear{Gold}{Gold}{1968}]{Gold1968}
Gold T.,  1968, Rotating Neutron Stars as the Origin of the Pulsating Radio
  Sources.
Springer US, Boston, MA, pp 74--75, \mn@doi{10.1007/978-1-4899-6387-1_40}, \url
  {https://doi.org/10.1007/978-1-4899-6387-1_40}

\bibitem[\protect\citeauthoryear{{Goldreich} \& {Julian}}{{Goldreich} \&
  {Julian}}{1969}]{Goldreich+1969}
{Goldreich} P.,  {Julian} W.~H.,  1969, \mn@doi [\apj] {10.1086/150119}, \href
  {https://ui.adsabs.harvard.edu/abs/1969ApJ...157..869G} {157, 869}

\bibitem[\protect\citeauthoryear{{Guo} et~al.,}{{Guo}
  et~al.}{2019}]{2019MNRAS.490.5424G}
{Guo} P.,  et~al., 2019, \mn@doi [\mnras] {10.1093/mnras/stz2975}, \href
  {https://ui.adsabs.harvard.edu/abs/2019MNRAS.490.5424G} {490, 5424}

\bibitem[\protect\citeauthoryear{{Gwinn} et~al.,}{{Gwinn}
  et~al.}{1997}]{1997ApJ...483L..53G}
{Gwinn} C.~R.,  et~al., 1997, \mn@doi [\apjl] {10.1086/310734}, \href
  {https://ui.adsabs.harvard.edu/abs/1997ApJ...483L..53G} {483, L53}

\bibitem[\protect\citeauthoryear{{Gwinn} et~al.,}{{Gwinn}
  et~al.}{2012}]{2012ApJ...758....7G}
{Gwinn} C.~R.,  et~al., 2012, \mn@doi [\apj] {10.1088/0004-637X/758/1/7}, \href
  {https://ui.adsabs.harvard.edu/abs/2012ApJ...758....7G} {758, 7}

\bibitem[\protect\citeauthoryear{Gyawali, Li, Knight, Ghimire, Horacek, Sapp
  \& Wang}{Gyawali et~al.}{2019}]{gyawali2019improving}
Gyawali P.,  Li Z.,  Knight C.,  Ghimire S.,  Horacek B.~M.,  Sapp J.,   Wang
  L.,  2019, in 2019 IEEE International Conference on Data Mining (ICDM). pp
  1078--1083

\bibitem[\protect\citeauthoryear{{Haskell} \& {Melatos}}{{Haskell} \&
  {Melatos}}{2015}]{Haskell:2015jra}
{Haskell} B.,  {Melatos} A.,  2015, \mn@doi [International Journal of Modern
  Physics D] {10.1142/S0218271815300086}, \href
  {https://ui.adsabs.harvard.edu/abs/2015IJMPD..2430008H} {24, 1530008}

\bibitem[\protect\citeauthoryear{Hewish, Bell, Pilkington, Scott  \&
  Collins}{Hewish et~al.}{1968}]{Hewish+1968}
Hewish A.,  Bell S.~J.,  Pilkington J. D.~H.,  Scott P.~F.,   Collins R.~A.,
  1968, \mn@doi [Nature] {10.1038/217709a0}, 217, 709

\bibitem[\protect\citeauthoryear{{Johnson}, {Gwinn}  \& {Demorest}}{{Johnson}
  et~al.}{2012}]{2012ApJ...758....8J}
{Johnson} M.~D.,  {Gwinn} C.~R.,   {Demorest} P.,  2012, \mn@doi [\apj]
  {10.1088/0004-637X/758/1/8}, \href
  {https://ui.adsabs.harvard.edu/abs/2012ApJ...758....8J} {758, 8}

\bibitem[\protect\citeauthoryear{{Johnston}}{{Johnston}}{2004}]{2004MNRAS.348.1229J}
{Johnston} S.,  2004, \mn@doi [\mnras] {10.1111/j.1365-2966.2004.07428.x},
  \href {https://ui.adsabs.harvard.edu/abs/2004MNRAS.348.1229J} {348, 1229}

\bibitem[\protect\citeauthoryear{Johnston, van Straten, Kramer  \&
  Bailes}{Johnston et~al.}{2001}]{Johnston+2001}
Johnston S.,  van Straten W.,  Kramer M.,   Bailes M.,  2001, \mn@doi [The
  Astrophysical Journal] {10.1086/319154}, 549, L101

\bibitem[\protect\citeauthoryear{{Kerr}}{{Kerr}}{2015}]{2015MNRAS.452..607K}
{Kerr} M.,  2015, \mn@doi [\mnras] {10.1093/mnras/stv1296}, \href
  {https://ui.adsabs.harvard.edu/abs/2015MNRAS.452..607K} {452, 607}

\bibitem[\protect\citeauthoryear{Kingma \& Welling}{Kingma \&
  Welling}{2014}]{kingma2014autoencoding}
Kingma D.~P.,  Welling M.,  2014, Auto-Encoding Variational Bayes (\mn@eprint
  {arXiv} {1312.6114})

\bibitem[\protect\citeauthoryear{{Knuth}}{{Knuth}}{2006}]{Knuth2006}
{Knuth} K.~H.,  2006, arXiv e-prints, \href
  {https://ui.adsabs.harvard.edu/abs/2006physics...5197K} {p. physics/0605197}

\bibitem[\protect\citeauthoryear{Kohonen}{Kohonen}{1988}]{teuvo1988som}
Kohonen T.,  1988, Self-Organized Formation of Topologically Correct Feature
  Maps.
MIT Press, Cambridge, MA, USA, p. 509–521

\bibitem[\protect\citeauthoryear{Kramer, Johnston  \& van Straten}{Kramer
  et~al.}{2002}]{Kramer:2002us}
Kramer M.,  Johnston S.,   van Straten W.,  2002, \mn@doi [Mon. Not. Roy.
  Astron. Soc.] {10.1046/j.1365-8711.2002.05478.x}, 334, 523

\bibitem[\protect\citeauthoryear{{Krishnamohan} \& {Downs}}{{Krishnamohan} \&
  {Downs}}{1983}]{Downs1983}
{Krishnamohan} S.,  {Downs} G.~S.,  1983, \mn@doi [\apj] {10.1086/160682},
  \href {https://ui.adsabs.harvard.edu/abs/1983ApJ...265..372K} {265, 372}

\bibitem[\protect\citeauthoryear{{Lam}}{{Lam}}{2017}]{PyPulse}
{Lam} M.~T.,  2017, {PyPulse: PSRFITS handler} (\mn@eprint {ascl} {1706.011})

\bibitem[\protect\citeauthoryear{{Large}, {Vaughan}  \& {Mills}}{{Large}
  et~al.}{1968}]{1968Natur.220..340L}
{Large} M.~I.,  {Vaughan} A.~E.,   {Mills} B.~Y.,  1968, \mn@doi [\nat]
  {10.1038/220340a0}, \href
  {https://ui.adsabs.harvard.edu/abs/1968Natur.220..340L} {220, 340}

\bibitem[\protect\citeauthoryear{LeCun, Bengio  \& Hinton}{LeCun
  et~al.}{2015}]{lecun2015deep}
LeCun Y.,  Bengio Y.,   Hinton G.,  2015, nature, 521, 436

\bibitem[\protect\citeauthoryear{{Lin}, {Li}  \& {Luo}}{{Lin}
  et~al.}{2020}]{2020MNRAS.493.1842L}
{Lin} H.,  {Li} X.,   {Luo} Z.,  2020, \mn@doi [\mnras]
  {10.1093/mnras/staa218}, \href
  {https://ui.adsabs.harvard.edu/abs/2020MNRAS.493.1842L} {493, 1842}

\bibitem[\protect\citeauthoryear{{Lopez Armengol} et~al.,}{{Lopez Armengol}
  et~al.}{2019}]{atel_vela}
{Lopez Armengol} F.~G.,  et~al., 2019, The Astronomer's Telegram, \href
  {https://ui.adsabs.harvard.edu/abs/2019ATel12482....1L} {12482, 1}

\bibitem[\protect\citeauthoryear{{Lorimer}}{{Lorimer}}{2008}]{Lorimer2008}
{Lorimer} D.~R.,  2008, \mn@doi [Living Reviews in Relativity]
  {10.12942/lrr-2008-8}, \href
  {https://ui.adsabs.harvard.edu/abs/2008LRR....11....8L} {11, 8}

\bibitem[\protect\citeauthoryear{{Lorimer} \& {Kramer}}{{Lorimer} \&
  {Kramer}}{2012}]{lorimer2012handbook}
{Lorimer} D.~R.,  {Kramer} M.,  2012, {Handbook of Pulsar Astronomy}.
Cambridge University Press

\bibitem[\protect\citeauthoryear{{Maan}, {van Leeuwen}  \& {Vohl}}{{Maan}
  et~al.}{2021}]{Maan2020}
{Maan} Y.,  {van Leeuwen} J.,   {Vohl} D.,  2021, \mn@doi [\aap]
  {10.1051/0004-6361/202040164}, \href
  {https://ui.adsabs.harvard.edu/abs/2021A&A...650A..80M} {650, A80}

\bibitem[\protect\citeauthoryear{{Mateos}, {Riveaud}  \& {Lamberti}}{{Mateos}
  et~al.}{2017}]{JSTest}
{Mateos} D.~M.,  {Riveaud} L.~E.,   {Lamberti} P.~W.,  2017, \mn@doi [Chaos]
  {10.1063/1.4999613}, \href
  {https://ui.adsabs.harvard.edu/abs/2017Chaos..27h3118M} {27, 083118}

\bibitem[\protect\citeauthoryear{{Michilli} et~al.,}{{Michilli}
  et~al.}{2018}]{2018MNRAS.480.3457M}
{Michilli} D.,  et~al., 2018, \mn@doi [\mnras] {10.1093/mnras/sty2072}, \href
  {https://ui.adsabs.harvard.edu/abs/2018MNRAS.480.3457M} {480, 3457}

\bibitem[\protect\citeauthoryear{{Morello}, {Barr}, {Bailes}, {Flynn}, {Keane}
  \& {van Straten}}{{Morello} et~al.}{2014}]{2014MNRAS.443.1651M}
{Morello} V.,  {Barr} E.~D.,  {Bailes} M.,  {Flynn} C.~M.,  {Keane} E.~F.,
  {van Straten} W.,  2014, \mn@doi [\mnras] {10.1093/mnras/stu1188}, \href
  {https://ui.adsabs.harvard.edu/abs/2014MNRAS.443.1651M} {443, 1651}

\bibitem[\protect\citeauthoryear{{Morello} et~al.,}{{Morello}
  et~al.}{2019}]{2019MNRAS.483.3673M}
{Morello} V.,  et~al., 2019, \mn@doi [\mnras] {10.1093/mnras/sty3328}, \href
  {https://ui.adsabs.harvard.edu/abs/2019MNRAS.483.3673M} {483, 3673}

\bibitem[\protect\citeauthoryear{Pacini}{Pacini}{1967}]{Pacini1967}
Pacini F.,  1967, \mn@doi [Nature] {10.1038/216567a0}, 216, 567

\bibitem[\protect\citeauthoryear{{Palfreyman}, {Dickey}, {Hotan}, {Ellingsen}
  \& {van Straten}}{{Palfreyman} et~al.}{2018}]{2018Natur.556..219P}
{Palfreyman} J.,  {Dickey} J.~M.,  {Hotan} A.,  {Ellingsen} S.,   {van Straten}
  W.,  2018, \mn@doi [\nat] {10.1038/s41586-018-0001-x}, \href
  {https://ui.adsabs.harvard.edu/abs/2018Natur.556..219P} {556, 219}

\bibitem[\protect\citeauthoryear{{Pang}, {Goseva-Popstojanova}, {Devine}  \&
  {McLaughlin}}{{Pang} et~al.}{2018}]{2018MNRAS.480.3302P}
{Pang} D.,  {Goseva-Popstojanova} K.,  {Devine} T.,   {McLaughlin} M.,  2018,
  \mn@doi [\mnras] {10.1093/mnras/sty1992}, \href
  {https://ui.adsabs.harvard.edu/abs/2018MNRAS.480.3302P} {480, 3302}

\bibitem[\protect\citeauthoryear{{Piekarewicz}, {Fattoyev}  \&
  {Horowitz}}{{Piekarewicz} et~al.}{2014}]{Piekarewicz:2014lba}
{Piekarewicz} J.,  {Fattoyev} F.~J.,   {Horowitz} C.~J.,  2014, \mn@doi [\prc]
  {10.1103/PhysRevC.90.015803}, \href
  {https://ui.adsabs.harvard.edu/abs/2014PhRvC..90a5803P} {90, 015803}

\bibitem[\protect\citeauthoryear{{Radhakrishnan} \& {Cooke}}{{Radhakrishnan} \&
  {Cooke}}{1969}]{Radhakrishnan+1969b}
{Radhakrishnan} V.,  {Cooke} D.~J.,  1969, \aplett, \href
  {https://ui.adsabs.harvard.edu/abs/1969ApL.....3..225R} {3, 225}

\bibitem[\protect\citeauthoryear{Radhakrishnan \& Manchester}{Radhakrishnan \&
  Manchester}{1969}]{Radhakrishnan+1969c}
Radhakrishnan V.,  Manchester R.~N.,  1969, \mn@doi [Nature]
  {10.1038/222228a0}, 222, 228

\bibitem[\protect\citeauthoryear{Radhakrishnan, Cooke, Komesaroff  \&
  Morris}{Radhakrishnan et~al.}{1969}]{Radhakrishnan+1969a}
Radhakrishnan V.,  Cooke D.~J.,  Komesaroff M.~M.,   Morris D.,  1969, \mn@doi
  [Nature] {10.1038/221443a0}, 221, 443

\bibitem[\protect\citeauthoryear{{Ransom}}{{Ransom}}{2018}]{PRESTO}
{Ransom} S.,  2018, PRESTO - Pulsar Exploration and Search Toolkit, \url
  {https://www.cv.nrao.edu/~sransom/presto/}

\bibitem[\protect\citeauthoryear{Reichley \& Downs}{Reichley \&
  Downs}{1969}]{Reichley+1969}
Reichley P.~E.,  Downs G.~S.,  1969, \mn@doi [Nature] {10.1038/222229a0}, 222,
  229

\bibitem[\protect\citeauthoryear{{Sarkissian}, {Hobbs}, {Reynolds},
  {Palfreyman}  \& {Olney}}{{Sarkissian} et~al.}{2019}]{atel_newglitch_vela}
{Sarkissian} J.,  {Hobbs} G.,  {Reynolds} J.,  {Palfreyman} J.,   {Olney} S.,
  2019, The Astronomer's Telegram, \href
  {https://ui.adsabs.harvard.edu/abs/2019ATel12466....1S} {12466, 1}

\bibitem[\protect\citeauthoryear{Schubert, Sander, Ester, Kriegel  \&
  Xu}{Schubert et~al.}{2017}]{DBSCAN}
Schubert E.,  Sander J.,  Ester M.,  Kriegel H.,   Xu X.,  2017, \mn@doi [ACM
  Transactions on Database Systems] {10.1145/3068335}, 42, 1

\bibitem[\protect\citeauthoryear{Shu}{Shu}{2014}]{gmvae}
Shu R.,  2014, Gaussian Mixture VAE: Lessons in Variational Inference,
  Generative Models, and Deep Nets., \url {http://ruishu.io/2016/12/25/gmvae/}

\bibitem[\protect\citeauthoryear{Sosa~Fiscella et~al.}{Sosa~Fiscella
  et~al.}{2021a}]{Fiscella:2020jey}
Sosa~Fiscella V.,  et~al., 2021a, \mn@doi [Astrophys. J.]
  {10.3847/1538-4357/abceb3}, 908, 158

\bibitem[\protect\citeauthoryear{{Sosa-Fiscella} et~al.,}{{Sosa-Fiscella}
  et~al.}{2021b}]{2021ATel14806....1S}
{Sosa-Fiscella} V.,  et~al., 2021b, The Astronomer's Telegram, \href
  {https://ui.adsabs.harvard.edu/abs/2021ATel14806....1S} {14806, 1}

\bibitem[\protect\citeauthoryear{{Sturrock}}{{Sturrock}}{1971}]{Sturrock1971}
{Sturrock} P.~A.,  1971, \mn@doi [\apj] {10.1086/150865}, \href
  {https://ui.adsabs.harvard.edu/abs/1971ApJ...164..529S} {164, 529}

\bibitem[\protect\citeauthoryear{Van~der Maaten \& Hinton}{Van~der Maaten \&
  Hinton}{2008}]{van2008visualizing}
Van~der Maaten L.,  Hinton G.,  2008, Journal of machine learning research, 9

\bibitem[\protect\citeauthoryear{{Wang} et~al.,}{{Wang}
  et~al.}{2019}]{2019SCPMA..6259507W}
{Wang} H.,  et~al., 2019, \mn@doi [Science China Physics, Mechanics, and
  Astronomy] {10.1007/s11433-018-9388-3}, \href
  {https://ui.adsabs.harvard.edu/abs/2019SCPMA..6259507W} {62, 959507}

\bibitem[\protect\citeauthoryear{{Zhang}, {Gajjar}, {Foster}, {Siemion},
  {Cordes}, {Law}  \& {Wang}}{{Zhang} et~al.}{2018}]{2018ApJ...866..149Z}
{Zhang} Y.~G.,  {Gajjar} V.,  {Foster} G.,  {Siemion} A.,  {Cordes} J.,  {Law}
  C.,   {Wang} Y.,  2018, \mn@doi [\apj] {10.3847/1538-4357/aadf31}, \href
  {https://ui.adsabs.harvard.edu/abs/2018ApJ...866..149Z} {866, 149}

\bibitem[\protect\citeauthoryear{{Zhang} et~al.,}{{Zhang}
  et~al.}{2020}]{2020A&A...642A..26Z}
{Zhang} C.,  et~al., 2020, \mn@doi [\aap] {10.1051/0004-6361/201937234}, \href
  {https://ui.adsabs.harvard.edu/abs/2020A&A...642A..26Z} {642, A26}

\bibitem[\protect\citeauthoryear{{Zhu} et~al.,}{{Zhu}
  et~al.}{2014}]{2014ApJ...781..117Z}
{Zhu} W.~W.,  et~al., 2014, \mn@doi [\apj] {10.1088/0004-637X/781/2/117}, \href
  {https://ui.adsabs.harvard.edu/abs/2014ApJ...781..117Z} {781, 117}

\makeatother
\end{thebibliography}

% Alternatively you could enter them by hand, like this:
% This method is tedious and prone to error if you have lots of references
%\begin{thebibliography}{99}
%\bibitem[\protect\citeauthoryear{Author}{2012}]{Author2012}
%Author A.~N., 2013, Journal of Improbable Astronomy, 1, 1
%\bibitem[\protect\citeauthoryear{Others}{2013}]{Others2013}
%Others S., 2012, Journal of Interesting Stuff, 17, 198
%\end{thebibliography}

%%%%%%%%%%%%%%%%%%%%%%%%%%%%%%%%%%%%%%%%%%%%%%%%%%

%%%%%%%%%%%%%%%%% APPENDICES %%%%%%%%%%%%%%%%%%%%%

\appendix

\section{Single pulses extraction}\label{sec:sp}

While conventional timing techniques involve folding observations in time and frequency to obtain an integrated pulse profile, for our purposes of performing single-pulse analysis we have devised an algorithm to extract the single pulses:

\begin{enumerate}
    \item For starters, each observation is folded using the \texttt{prepfold} routine from the pulsar search and analysis software \texttt{PRESTO} \citep{PRESTO}. As a byproduct, the pulsar period along the observation is modeled as a polynomial function in time, and the corresponding coefficients are saved to a \texttt{polycos} file that we save for later use.
    \item We de-disperse and integrate our data in frequency. We use a script adapted from \texttt{waterfaller.py} in the \texttt{PRESTO} package \citep{PRESTO}. For the dispersion measure, we use the value 67.897/68.053 for the January observations with A1/A2, and 68.057 for the March observations with A1 or A2. After this step, the observation is reduced to a single array in time.
%    \item \FedeLA{Rewritting this item in the item below, please check.} We then used that information to calculate the instantaneous period for each single pulse and divided that period into an integer number of time bins. In this way, the bin length is constant within each single pulse, but it suffers a small but steady increase in time from one pulse to the next. However, the number of bins remains the same for all pulses, and it is equal to 1220. We thereby account for the intrinsic pulsar spin-down, which results in each single pulse having a slightly longer period than the previous one.
    \item We proceed to build a grid in time where each instantaneous period of the pulsar is divided by a fixed number of time bins. In this way, we account for the intrinsic pulsar spin-down during the observation, and the single pulses will arrive approximately at the same phase. We use the file \texttt{polycos} to calculate the period for each single pulse, and divided that period into an fixed number of time bins. We use $1220$ bins per period, approximated by dividing an average period of Vela by our time resolution $89.2 \mathrm{ms} / 73 \mathrm{\mu s} \approx 1220$.
    \item The original observational data was sampled at time intervals that do not necessarily match our time grid. We therefore interpolated the datum that would correspond to each mesh point from the data in the adjacent actual measured points.
    \item We applied this process to observations taken with each of the radio telescopes, and that overlapped for at least 3 hours.
    \item In order to obtain truly simultaneous observations, we cropped the interpolated data to obtain two observations that start and finish at the same MJDs.
    
\end{enumerate}

The resulting \texttt{.fil} files thereby obtained are then cleansed of RFIs using the algorithms described in Appendix~\ref{sec:RFIClean}.
%\Carlos{@Valentina @FLA: Do we do the cleanning before of after the pulse extraction?} \Valentina{@Carlos: We clean before extracting the pulses (in fact it's the very first thing we do).}

\section{Other Analysis methods} \label{sec:analysisMethodsAppendix}
Besides standard isotropic Gaussian as priors for the latent representation of vela pulse, we explored non-parametric alternatives via Indian Buffet Process (IBP) as prior in VAE \citep{gyawali2019improving}. With IBP, we aimed to cluster the latent representation into multiple factors in an unsupervised way. However, in this study, we find standard Gaussian prior to work better. 

Additionally, we tried to apply the Gaussian Mixture VAE (GMVAE) in \citep{gmvae}. We found that the Gaussian mixture consistently collapsed into one or two super-clusters and failed to give meaningful results.

Outside of the SOM, another clustering method tried was Agglomerative Clustering (based on the SciPy implementatin \url{https://scikit-learn.org/stable/modules/clustering.html#hierarchical-clustering}) where observations are continually merged together into larger groupings based on linkage criterion. We found that this produced decent clusters on small datasets, however found that it was computationally prohibitive to apply on increasingly larger sets.

A t-distributed stochastic neighbor embedding (t-SNE) algorithm in \citep{van2008visualizing} was applied on the raw data to see if clusters could be found in an embedded space outside of deep learning. While organization of signals were observed in this 2-dimensional space, it was primarily a result on when the spike happened in the signal. We observed the same computational issues with this method when extending to the larger datasets.

\section{Cleaning of Radio Frequency Interference}\label{sec:RFIClean}

\begin{figure}
	\includegraphics[width=\columnwidth]{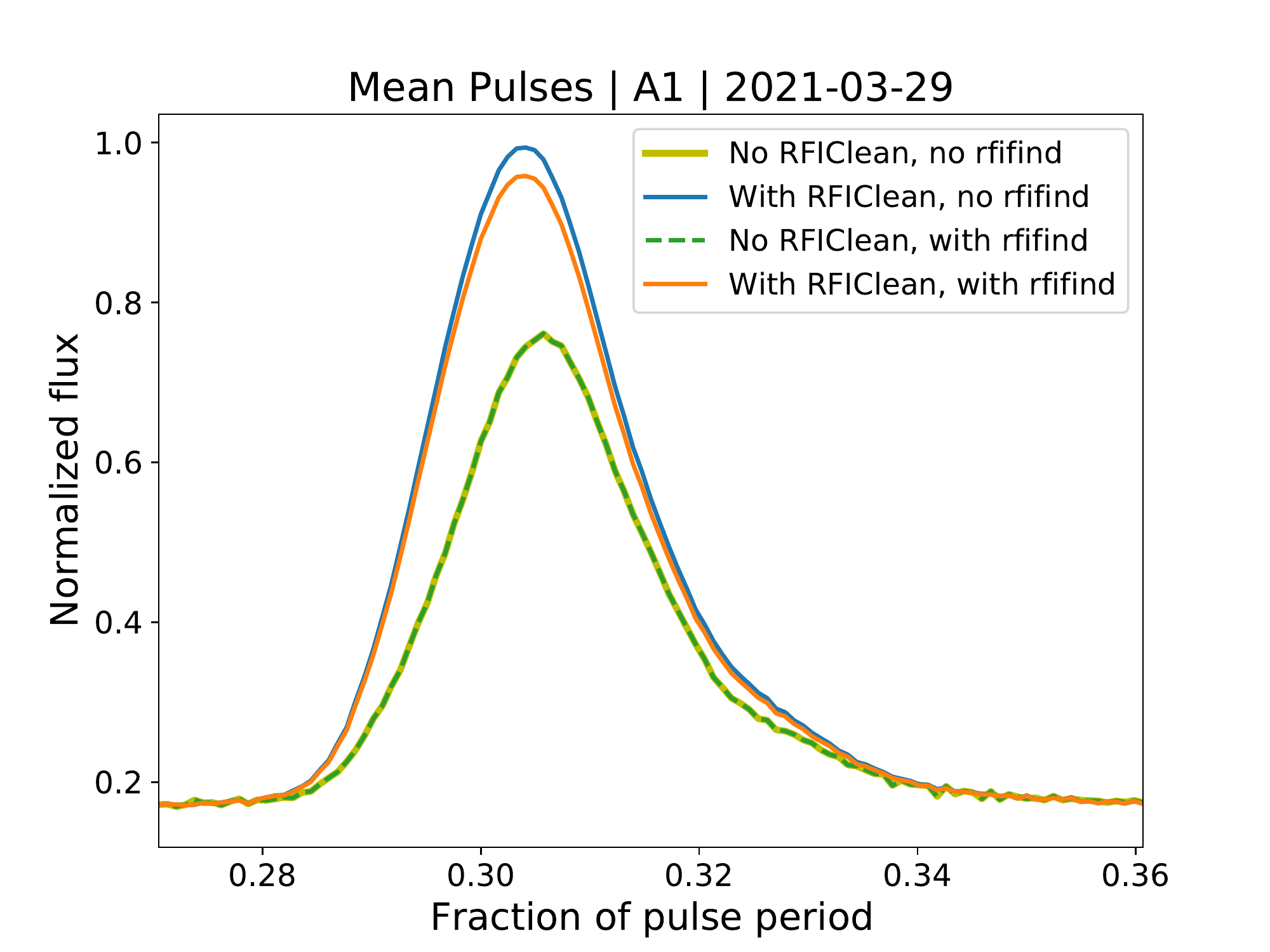}
    \caption{Mean pulses for the March 29th observation using four different RFI-cleaning schemes. The fluxes are uncalibrated and are in arbitrary units. % The SNRs reported in the plots are obtained using the pdmp command (psrstat -c snr pulsar/*.ar). Both epochs show significant improvement in SNR while using RFI mitigation.
    %\Carlos{Note that we have chosen the problematic case/day with A1}
        \label{fig:compare_RFIClean_RFIfind}}
\end{figure}

Time de-dispersion with unfiltered radio frequency interferences (RFIs) results in a modulation of the signal. Therefore, the RFIs must be filtered from the raw data before extracting single pulses. In this work we tested two complementary RFI-cleaning softwares:

\begin{itemize}

    \item \texttt{rfifind} is a task within \texttt{PRESTO} \citep{PRESTO} that searches for narrow-band and short duration broadband RFI by identifying outliers using time domain statistics (mean and standard deviation of each time bloc). Flagged blocks are replaced in subsequent processing by constant data that is chosen based on the median bandpass.

    \item \texttt{RFIClean} \citep{Maan2020} removes \textit{periodic} RFI in each individual frequency channel using Fourier domain analysis. After that, it uses threshold-based techniques to identify time samples as well as frequency channels contaminated by broadband bursts and narrow-band RFI. The identified samples are replaced by mean values in the local regions around the affected samples.

\end{itemize}

Periodic RFI could limit the efficacy of conventional RFI mitigation techniques used by \texttt{rfifind}, since they could potentially result in masking a large fraction of the data. Consequently, we assess the benefits of incorporating \texttt{RFIClean} into our RFI cleaning schemes.

We found that \texttt{RFIClean} gave better results both at mitigating RFIs and providing a higher mean pulse amplitude, as evidenced in Fig.~\ref{fig:compare_RFIClean_RFIfind}. The S/N ratios attained using each cleaning algorithm are presented in Table \ref{tab:snr}. We can readily see that:

\begin{enumerate}
    \item The A1 observations are more heavily affected by RFIs, as evidenced when no RFI cleaning scheme is applied. They also benefit more from using \texttt{RFIClean} than \texttt{rfifind}.
    \item A2 observations do not present a preference for any cleaning algorithm.  Moreover, an over-masking occurs in the January 24th observations when using only one of them.
    \item The quality of the March observations with A1 is severely hindered, possibly due to the different observational setting used in those (two polarizations modes instead of one).
    \item Overall, we see an improvement in S/N when using either RFI mitigation technique. Moreover, we found that using \texttt{rfifind} on the data output from \texttt{RFIClean} further improves the S/N in many cases, as predicted by \cite{Maan2020}.
\end{enumerate}

We therefore ran both programs in all observations. To avoid a possible modification of the periodic signal, we set \texttt{RFIClean} to safeguard the Fourier frequency corresponding to the pulsar's spin period (via the \texttt{-psrf <F0>} modifier). Moreover, given Vela's relatively long spin period, we provided a block size large enough to cover at least 100 of pulsar periods (via \texttt{-t}). Otherwise, if the specified block size is no sufficiently long enough, \texttt{RFIClean} is unable to detect the corresponding spin frequency in the Fourier domain. Including the \texttt{-zerodm} modifier (useful for searching for unknown periodic or transient signals) resulted in a pulse distortion, as noted by \cite{Eatough2009}, and this option was thereby avoided.

\begin{table*}
	{\centering
	\caption{S/N for each observation and each RFI cleaning scheme. These values were obtained using the \texttt{sp.getSN()} class attribute from the \texttt{PyPulse} package \citep{PyPulse}.} % (\texttt{psrstat -c snr *.pfd})}
	\label{tab:snr}
	\begin{tabular}{lrl|rl|rl|rl}
		\hline
		\multirow{2}{*}{}  & 	\multicolumn{2}{c}{No RFIClean} & 	\multicolumn{2}{c}{No RFIClean} & 	\multicolumn{2}{c}{RFIClean} & 	\multicolumn{2}{c}{RFIClean} \\
		 & \multicolumn{2}{c}{No rfifind}  & \multicolumn{2}{c}{rfifind} & \multicolumn{2}{c}{No rfifind} & \multicolumn{2}{c}{rfifind} \\
		\hline
		 & A1 & A2 & A1 & A2 & A1 & A2 & A1 & A2 \\
		Jan21 & \cellcolor[HTML]{f37142}{429.41} & \cellcolor[HTML]{b4e9a9}{2084.61} & \cellcolor[HTML]{fbf7bd}{1936.58} & \cellcolor[HTML]{fbf7bd}{1967.86} & \cellcolor[HTML]{fbf7bd}{1982.96} & \cellcolor[HTML]{fbf7bd}{2038.37} & \cellcolor[HTML]{b4e9a9}{2100.66} & \cellcolor[HTML]{b4e9a9}{2107.03} \\
		Jan24 & \cellcolor[HTML]{ffb84f}{1145.38} & \cellcolor[HTML]{b4e9a9}{2108.17} & \cellcolor[HTML]{fbf7bd}{1894.91} & \cellcolor[HTML]{b4e9a9}{2182.69} & \cellcolor[HTML]{fbf7bd}{1983.3} & \cellcolor[HTML]{b4e9a9}{2142.97} & \cellcolor[HTML]{fbf7bd}{2005.07} & \cellcolor[HTML]{b4e9a9}{2188.26} \\
		Jan28 & \cellcolor[HTML]{f37142}{529.90} & \cellcolor[HTML]{76dd78}{2575.20} & \cellcolor[HTML]{b4e9a9}{2163.26} & \cellcolor[HTML]{76dd78}{2714.59} & \cellcolor[HTML]{b4e9a9}{2188.04} & \cellcolor[HTML]{76dd78}{2676.22} & \cellcolor[HTML]{b4e9a9}{2193.6} & \cellcolor[HTML]{76dd78}{2746.39} \\
		Mar29(*) & \cellcolor[HTML]{f37142}{661.08} & \cellcolor[HTML]{b4e9a9}{2062.26} & \cellcolor[HTML]{ffa001}{1010.64} & \cellcolor[HTML]{b4e9a9}{2121.49} & \cellcolor[HTML]{ffb84f}{1125.62} & \cellcolor[HTML]{b4e9a9}{2192.01} & \cellcolor[HTML]{ffa001}{1115.69} & \cellcolor[HTML]{b4e9a9}{2192.23} \\
		\hline
	\end{tabular}
	\par}
	(*) \textbf{Note.} March 29th observation with A1 used a different configuration with two polarizations.
\end{table*}

\section{Observations 2021-03-29}\label{sec:2021-03-29}

In this section we perform a simultaneous observation with a different A1 configuration. We have chosen to observe in a two polarization, 56MHz+56MHz mode, to mimic more closely the observation mode of A2 and further compare the results of the two radio telescopes as an ultimate cross-check.

\subsection{Scintillation for 2021-03-29}\label{sec:2021-03-29_scintillation}

As described in Sec.~\ref{sec:scintillations}, we also fitted Eq.~(\ref{eq:scintillation}) to the pulse amplitude distribution of the March 29th dataset. In this case, we see that the A1 distribution is tilted towards lower amplitudes, while the A2 distribution is consistent with the previous datasets. We thereby conclude that the A1 observation suffers from a quality loss, possibly due to a malfunction in one of the two polarization modes that were used in this configuration. However, the fitted values of $n_{\mathrm{ISS}}$ for the A2 observation are in good agreement with the corresponding results in Table \ref{table:niss_table}. Therefore, these results are not particular to the January observations but rather part of a more general behaviour when analyzing short timescale scintillation.

\begin{figure}
	\includegraphics[width=\columnwidth]{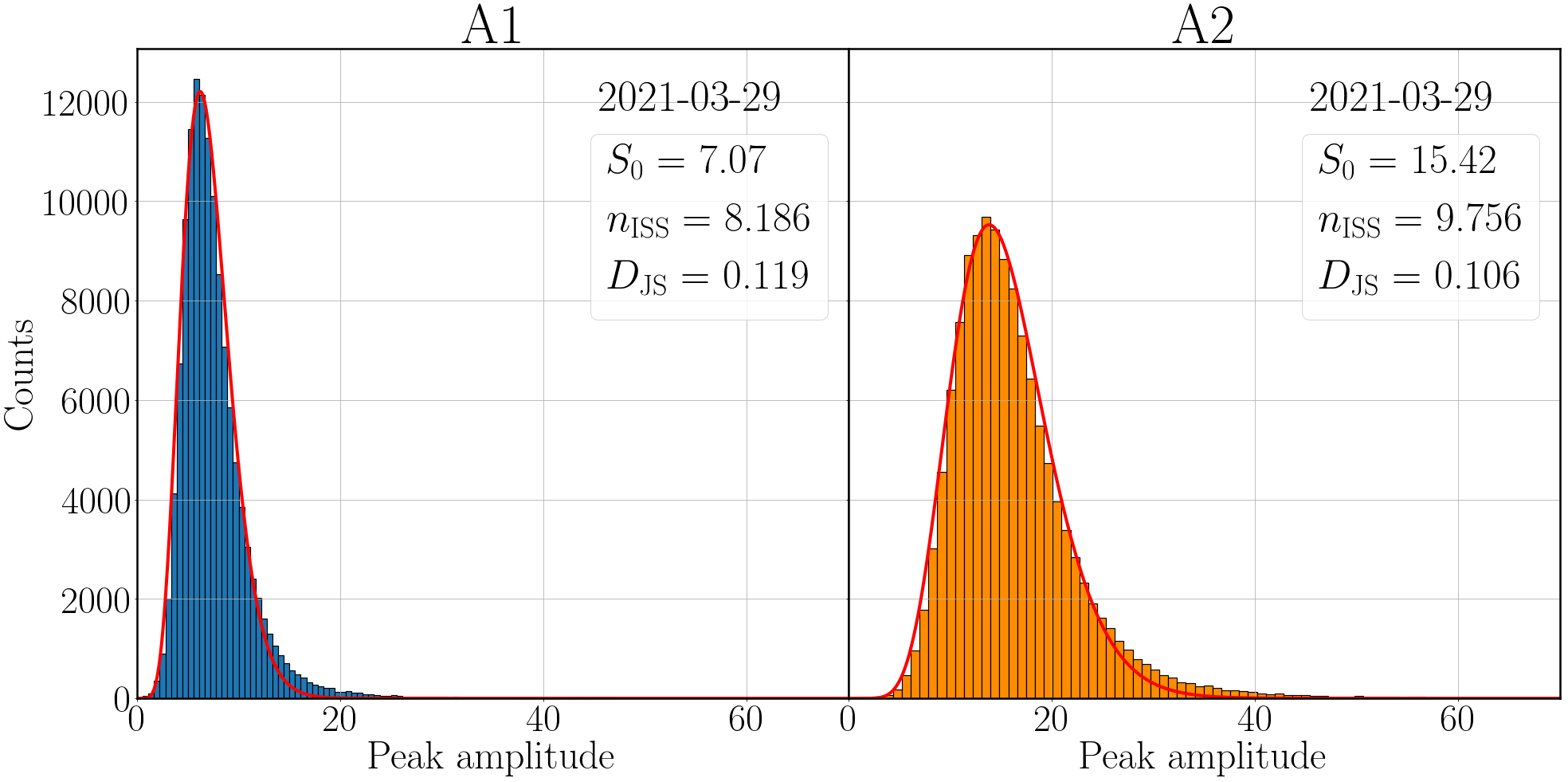}
    \caption{Histograms of projected pulse amplitude for J0835$-$4510 for A1 (left) and A2 (right) for the 2021-03-29 observation. The line shows the estimated scintillation distribution from fitting $n_\mathrm{ISS}$ in Eq.~(\ref{eq:scintillation}).}
    \label{fig:scintillations_march}
\end{figure}

\subsection{Observations 2021-03-29 with DBSCAN analysis}\label{sec:2021-03-29M}

\begin{figure*}
	\includegraphics[width=.95\columnwidth]{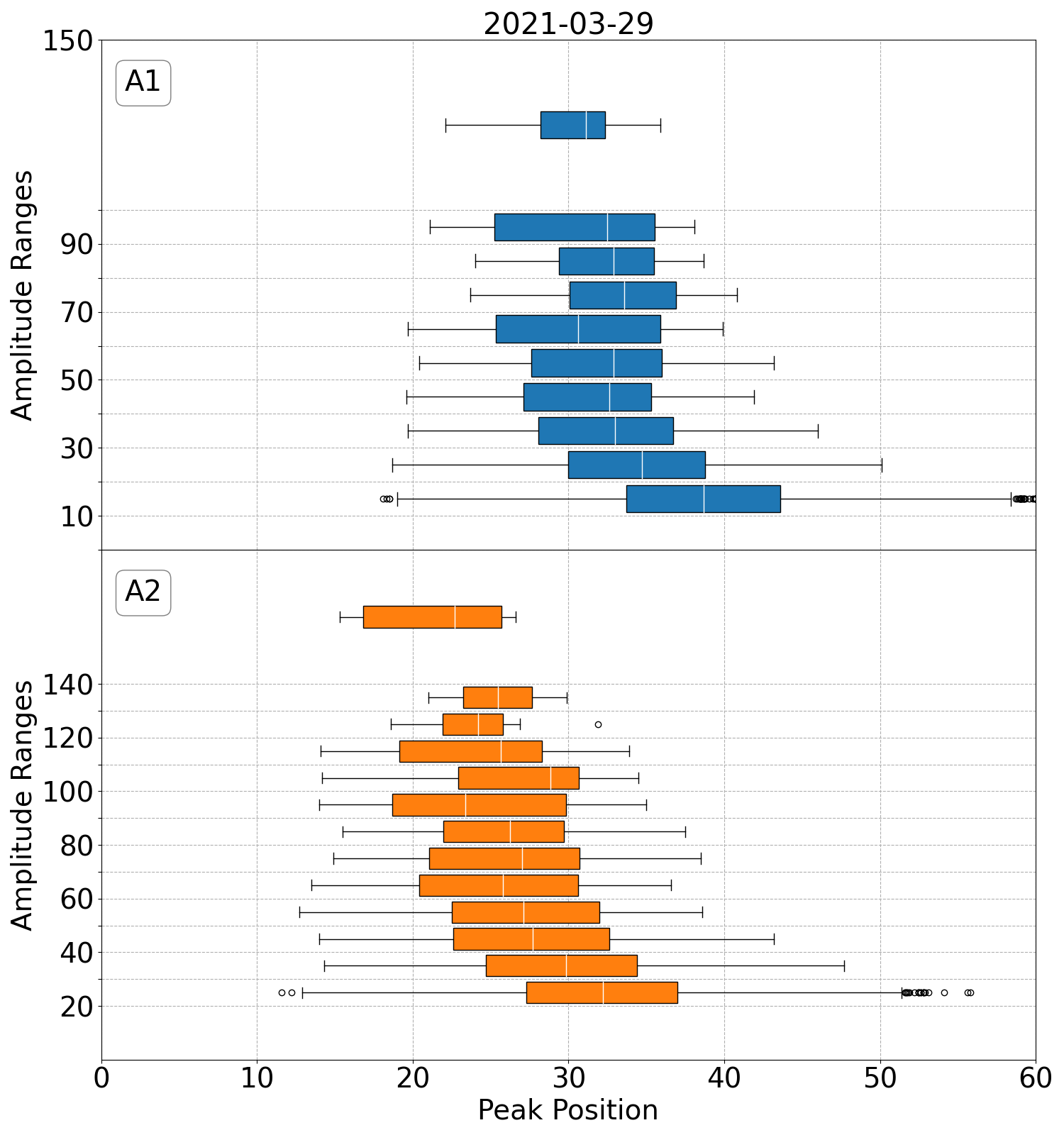}
	\includegraphics[width=.95\columnwidth]{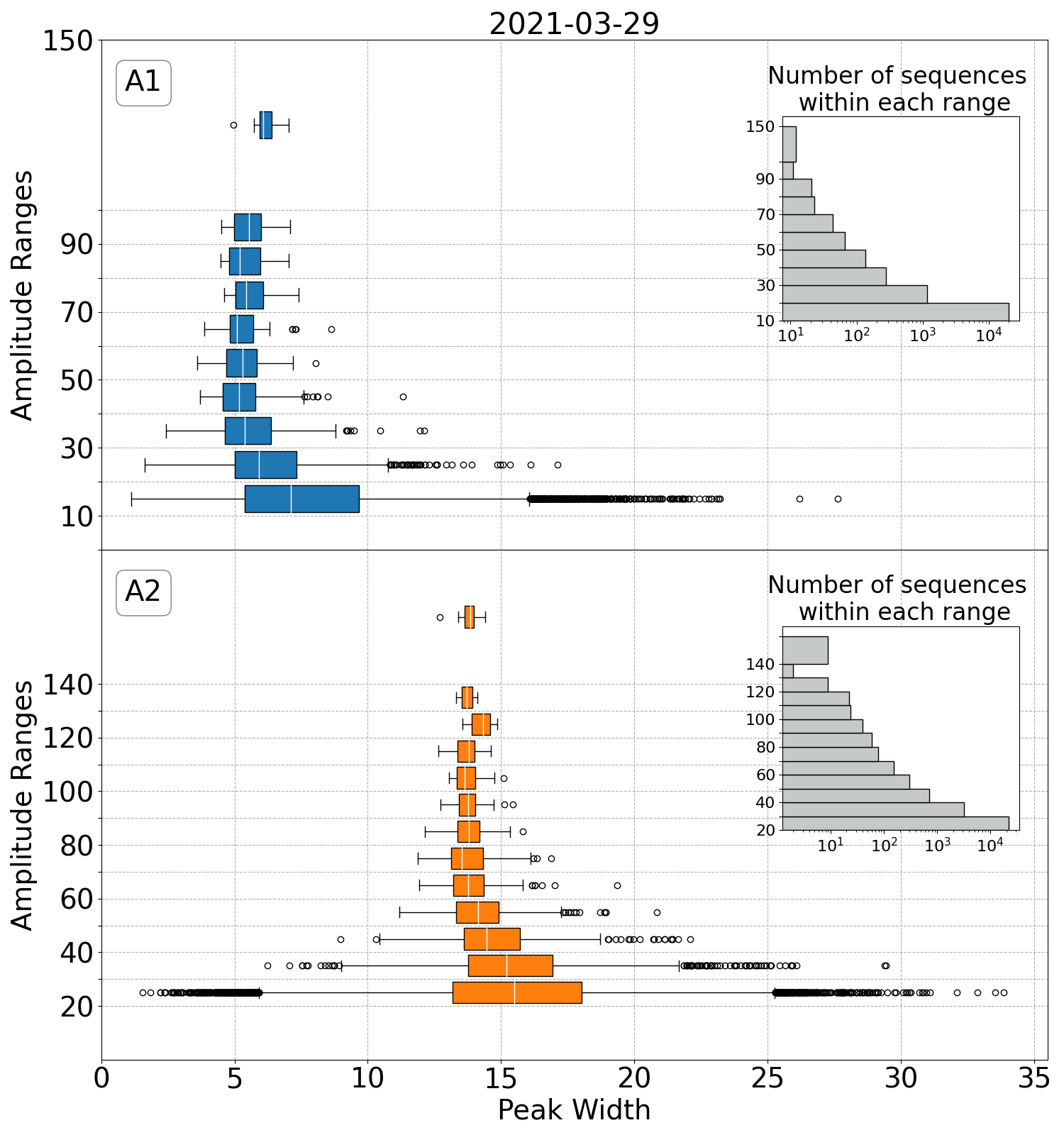}
    \caption{Distribution of the pulses amplitudes versus arrival time as measured by the position of their peaks for the 2021-03-29 observations and distribution of the pulses amplitudes versus its mean widths for the Radio telescope A1 on top and Radio telescope A2 on bottom. The differences in behavior can be attributed to the single polarization versus two polarizations observations with A1 and A2 respectively.}
    \label{fig:1314A12}
\end{figure*}

Fig.~\ref{fig:1314A12} displays on the left column the distribution of the pulses amplitudes versus arrival time as measured by the position of their peaks. Insets show the histograms of pulse amplitudes for each observation. We observe again, that there is a trend for peaks with larger amplitudes to appear earlier that pulses with lower amplitude indicating this is a generic feature. We also note now the similarity of the patterns of A1 and A2 observations with A1 now configured into two polarizations, like A2. 
On the right column, Fig.~\ref{fig:1314A12} displays the distribution of the pulses amplitudes versus its mean width as measured by the half amplitude of their peaks. There is a very weak trend, if at all, for peaks with larger amplitudes to appear narrower than pulses with lower amplitude in Radio telescope A1 observations in this new configuration. This is confirmed by the corresponding observation with Radio telescope A2, suggesting a dependence of this effect on the single polarization observations with A1 during the January runs.

\begin{figure*}
	% To include a figure from a file named example.*
	% Allowable file formats are eps or ps if compiling using latex
	% or pdf, png, jpg if compiling using pdflatex
	\includegraphics[width=1.15\columnwidth]{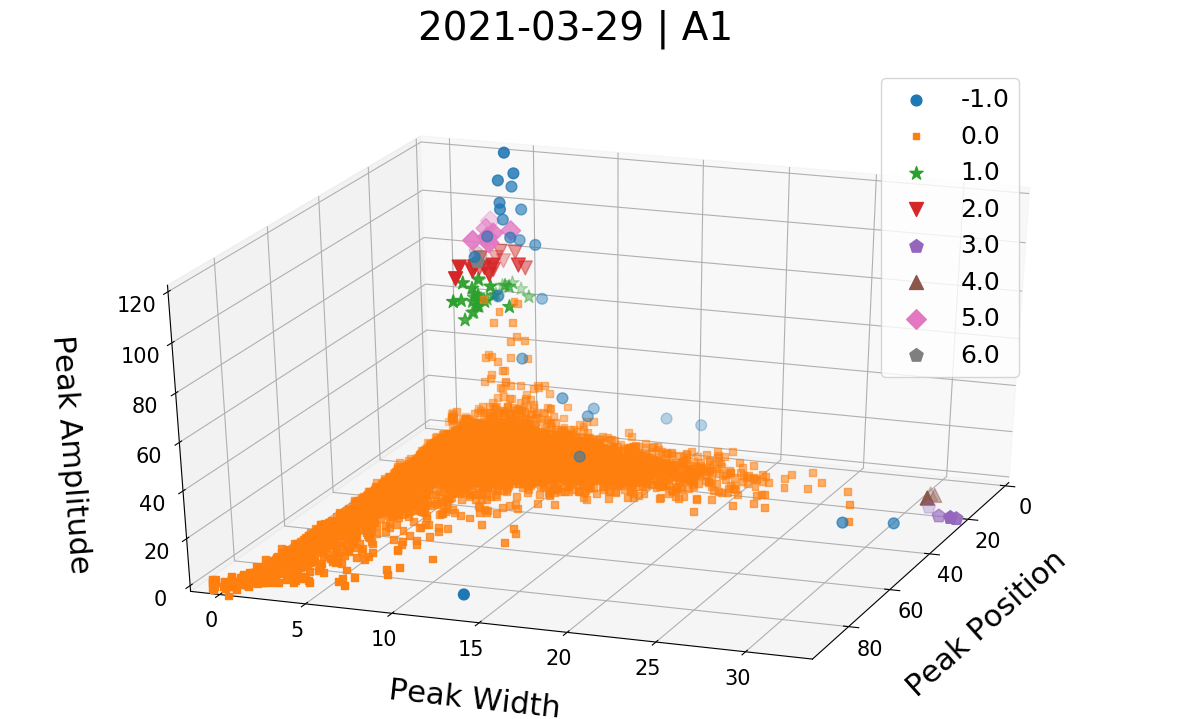}
	\includegraphics[width=0.75\columnwidth]{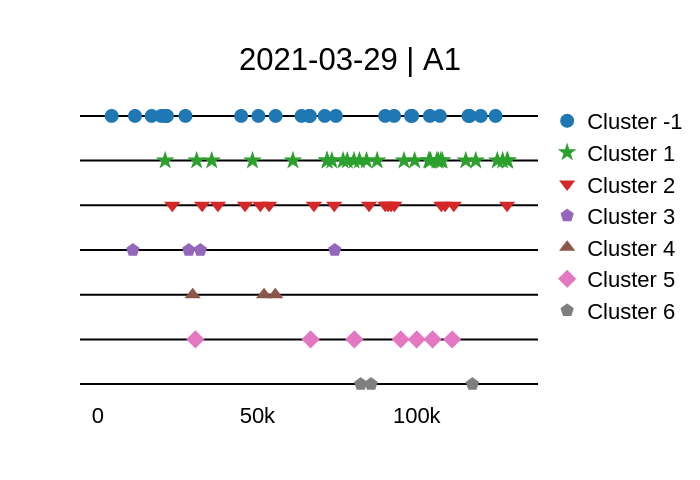}\\
	\includegraphics[width=1.15\columnwidth]{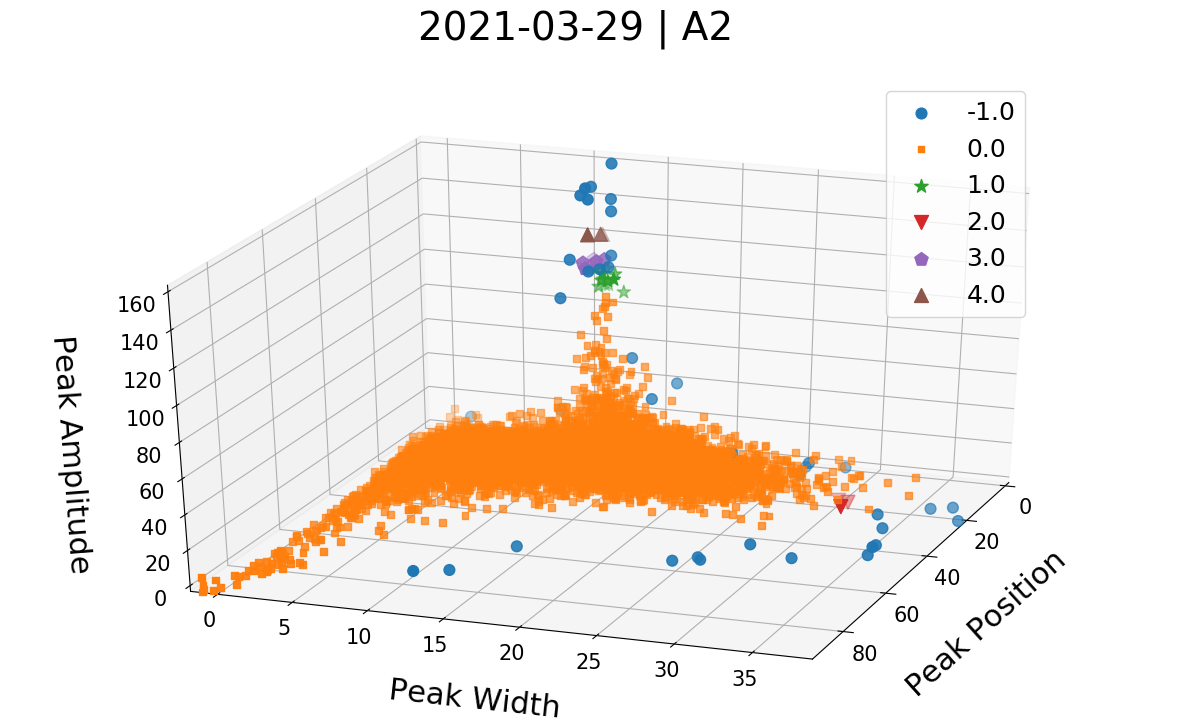}
	\includegraphics[width=0.75\columnwidth]{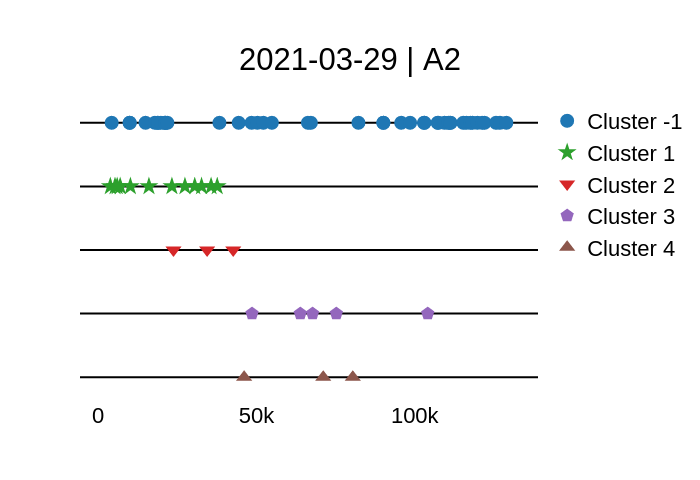}
    \caption{(On the left) 3D distribution of the pulses peak amplitudes, position, and widths for the 2021-03-29 observations. Radio telescope A1 on top and Radio telescope A2 on bottom. Different colors represent different clusters according to density based scan criteria.
    (On the right) Distribution of the pulses over the duration of the observations 2021-03-29. Upper figures as for Radio telescope A1, lower figures are for Radio telescope A2 on. Different colors represent different clusters according to density based scan criteria.}
    \label{fig:15A12}
\end{figure*}

We next performed a Density Based Scan (DBSCAN) analysis of the pulses clustering for each Radio telescope's observation as displayed in Fig.~\ref{fig:15A12} which selects the different clusters by increasing amplitudes, but also creates a baseline cluster (in orange labeled as 0) and an enveloping outlier (in light blue, labeled as -1). We also display the detail of each pulse in the top amplitude DBSCAN clusters over the duration of the observation, labeled by the pulse index number. It is interesting to see here no particular preference of the large amplitude pulses toward a time of the duration of the observation, indicating that during this March observation there seems not to be a preferential direction of the local RFIs.

%In Fig.~\ref{fig:4A12pi} we display the detail of each pulse in the top amplitude DBSCAN clusters over the duration of the observation, labeled by the pulse index number.

%\begin{figure}
%	\includegraphics[width=\columnwidth]{4A1pulse_index.png}
%	\includegraphics[width=\columnwidth]{4A2pulse_index.png}
%    \caption{Distribution of the pulses over the duration of the observations 2021-03-29. Upper figures as for Radio telescope A1, lower figures are for %Radio telescope A2 on. Different colors represent different clusters according to density based scan criteria.}
%    \label{fig:4A12pi}
%\end{figure}

%%%%%%%%%%%%%%%%%%%%%%%

\subsection{Observations 2021-03-29 with VAE/SOM analysis}\label{sec:2021-03-29CS}

\begin{figure*}\begin{center}
	\includegraphics[width=.9\columnwidth]{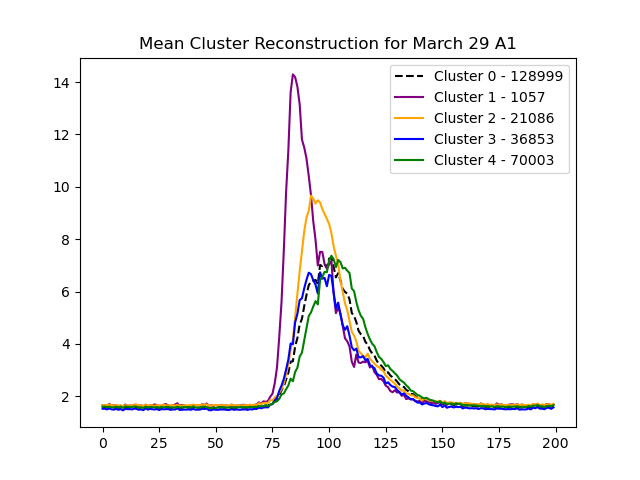}
	\includegraphics[width=.9\columnwidth]{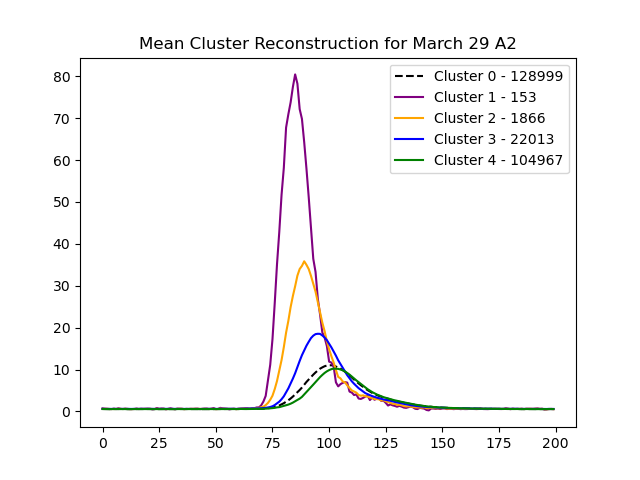}
    \caption{Distribution of the SOM clustering average signals for the observations 2021-03-29. Radio telescope A1 on left and Radio telescope A2 on right panels, with respective VAE training performed on the January 28 observation. Note that A1 was changed to two-polarizations configuration here.}
    \label{fig:4A12CS}    
\end{center}
\end{figure*}

In Fig.~\ref{fig:4A12CS} we display the average value of the pulses in each SOM cluster for each Radio telescope observation.
Those have been obtained by first applying a reconstruction of the raw pulses with the VAE technique, for which we have used the reconstruction of our best day of observation (according to the analysis in Appendix \ref{sec:RFIClean}) as a training case to apply to the rest of the days of observation. This training has been applied for each antenna individually. SOM allows us to specify then the number of clusters we seek to subdivide the whole set. We have studied several possible cases, 4, 6, 10, 25, 100, finding that the simplest four cluster choice represents the most robust results. We observe in Fig.~\ref{fig:4A12CS} the degradation of the A1 observation while we also see the similarity of the mean clustering between the A2 and the two antennas previous observations. While the total number of members of each cluster changes, the qualitative mean value of the pulses (average pulse over the entire cluster) seems to be robust. We also note the visual displacement towards earlier times of the center and peak of each cluster average pulse for those with the largest amplitude with respect to each other and with respect to the total average pulse, in the plot denoted by the black dashed curve and labeled as 0. This is the usual reference pulse we obtain from the total observation. We also see that the largest amplitude cluster (labeled as 1), counting a few hundred pulses, is a factor about 7 larger in amplitude than the average pulse. Both features are in qualitative agreement with the statistical analysis of the previous section~\ref{sec:math}.

Table~\ref{tab:2021-03-29A12} give a quantitative account of the results displayed in Fig.~\ref{fig:4A12CS}. 
For each antenna's observation on 2021-03-29,
we provide the number of pulses of each cluster \# pulses; peak location from the index of the maximum value in the pulse sequence;
peak height from the maximum value of the pulse sequence;
peak width done by first finding the maximum value of the sequence, then performing full-width half maximum of peak;
(library used for this: \url{https://docs.scipy.org/doc/scipy/reference/generated/scipy.signal.peak_widths.html});
for the peak skew we evaluated the Fisher-Pearson coefficient of skewness; 
(using the scipy for this computation \url{https://docs.scipy.org/doc/scipy/reference/generated/scipy.stats.skew.html}); 
and finally MSE is the standard mean squared error. $\sum_{i=1}^N(x_i-\bar{x})^2/N$.
The trend in the peak location towards later times is clear and above the quoted errors in its determination. Note that the cluster 4, the most numerous,
is (necessarily) showing a later arrival than the average total pulse, labeled as 0. A trend is also marginally seen in the width of the pulse, with
narrower values for the higher amplitude clusters while also carrying a higher skewness. Although specific numbers differ in both antennas observations,
the trends seem to be the same, including the particular case of the A1 configuration during this observation.

\begin{table*}
	\centering
	\caption{SOM Clustering for 2021-03-29 with Antennas 1 and 2. A1 configuration has been changed here to two polarizations mode.}
	\label{tab:2021-03-29A12}
	 \begin{tabular}{l|llllll}
	 	 \hline
	 	 &Cluster \#& 0 & 1 & 2 & 3 & 4 \\ 
	 	 \hline 
	 	 &\# pulses & 128999 & 1057 & 21086 & 36853 & 70003 \\ 
	 	 &peak loc & $98.72 \pm 6.42$ & $84.78 \pm 2.63$ & $93.76 \pm 3.78$ & $93.95 \pm 4.92$ & $102.94 \pm 4.33$ \\ 
	 A1	 &peak height & $8.46 \pm 2.97$ & $17.76 \pm 13.30$ & $10.99 \pm 4.68$ & $7.45 \pm 0.97$ & $8.09 \pm 1.59$ \\ 
	 	 &peak width & $21.51 \pm 4.34$ & $11.04 \pm 4.98$ & $18.88 \pm 3.52$ & $19.41 \pm 4.36$ & $22.43 \pm 4.00$ \\ 
	 	 &peak skew & $2.06 \pm 0.29$ & $2.83 \pm 0.70$ & $2.38 \pm 0.28$ & $2.00 \pm 0.21$ & $1.99 \pm 0.22$ \\ 
	 	 &MSE & $0.00005 \pm 0.00009$ & $0.00781 \pm 0.02791$ & $0.00034 \pm 0.00067$ & $0.00016 \pm 0.00027$ & $0.00009 \pm 0.00015$\\ 
	 	 \hline
	 	 &Cluster \#& 0 & 1 & 2 & 3 & 4 \\ 
	 	 \hline 
	 	 &\# pulses & 128999 & 153 & 1866 & 22013 & 104967 \\ 
	 	 &peak loc & $100.49 \pm 5.46$ & $84.76 \pm 2.61$ & $88.59 \pm 2.45$ & $94.23 \pm 3.43$ & $102.03 \pm 4.53$ \\ 
	 A2	 &peak height & $13.31 \pm 7.12$ & $86.35 \pm 27.48$ & $38.66 \pm 14.61$ & $20.43 \pm 6.75$ & $11.27 \pm 3.71$ \\ 
	 	 &peak width & $19.54 \pm 1.54$ & $13.91 \pm 0.98$ & $15.58 \pm 0.77$ & $19.41 \pm 1.30$ & $21.93 \pm 1.32$ \\ 
	 	 &peak skew & $2.12 \pm 0.23$ & $3.14 \pm 0.11$ & $2.76 \pm 0.13$ & $2.38 \pm 0.13$ & $2.06 \pm 0.19$ \\ 
	 	 &MSE & $0.00009 \pm 0.00013$ & $0.10855 \pm 0.26202$ & $0.00676 \pm 0.01160$ & $0.00052 \pm 0.00080$ & $0.00011 \pm 0.00016$\\ 
	 	 \hline 
	 \end{tabular} 
\end{table*}

% Don't change these lines
\bsp	% typesetting comment
\label{lastpage}
\end{document}